\documentclass{article}
\usepackage[utf8]{inputenc}
\usepackage{macros}

\usepackage{graphicx}
\graphicspath{ {./images/} }

\newcommand{\AT}{\mathsf{AT}}
\newcommand{\Res}{\mathsf{Res}}
\newcommand{\reg}{\mathsf{reg}}
\newcommand{\mon}{\mathsf{mon}}
\newcommand{\lin}{\mathsf{lin}}
\newcommand{\ati}{\mathsf{ati}}
\newcommand{\OPT}{\mathsf{OPT}}

\newcommand{\OL}{\mathsf{OL}}

\DeclareMathOperator{\conv}{conv}
\DeclareMathOperator{\sgn}{sign}

\DeclareMathOperator{\diam}{diam}

\DeclareMathOperator{\range}{range}
\DeclareMathOperator{\ranges}{ranges}

\title{New Subset Selection Algorithms for Low Rank Approximation: Offline and Online}
\author{
David P. Woodruff \\ Carnegie Mellon University \\ \texttt{dwoodruf@cs.cmu.edu} \and Taisuke Yasuda \\ Carnegie Mellon University \\ \texttt{taisukey@cs.cmu.edu}
}

\begin{document}

\maketitle

\thispagestyle{empty}
\begin{abstract}
Subset selection for the rank $k$ approximation of an $n\times d$ matrix $\mathbf A$ offers improvements in the interpretability of matrices, as well as a variety of computational savings. This problem is well-understood when the error measure is the Frobenius norm, with various tight algorithms known even in challenging models such as the \emph{online} model, where an algorithm must select the column subset irrevocably when the columns arrive one by one. In sharp contrast, when the error measure is replaced by other matrix losses, optimal trade-offs between the subset size and approximation quality have not been settled, even in the standard offline setting. We give a number of results towards closing these gaps.

In the offline setting, we achieve nearly optimal bicriteria algorithms in two settings. First, we remove a $\sqrt k$ factor from a result of \cite{SWZ2019} when the loss function is \emph{any} entrywise loss with an approximate triangle inequality and at least linear growth, which includes, e.g., the Huber loss. Our result is tight when applied to the $\ell_1$ loss. We give a similar improvement for the entrywise $\ell_p$ loss for $p>2$, improving a previous distortion of $\tilde O(k^{1-1/p})$ to $O(k^{1/2-1/p})$. We show this is tight for $p = \infty$, while for $2<p<\infty$, we give the first bicriteria algorithms for $(1+\varepsilon)$-approximate entrywise $\ell_p$ low rank approximation. Our results come from a general technique which improves distortions by replacing the use of a well-conditioned \emph{basis} with a slightly larger \emph{spanning set} for which any vector can be expressed as a linear combination with small Euclidean norm. This idea  may be of independent interest and we show, for example, that it also gives the first oblivious $\ell_p$ subspace embeddings for $1\leq p < 2$ with $\tilde O(d^{1/p})$ distortion, which is nearly optimal and improves the previously best known $\tilde O(d)$ \cite{WW2022} and closes a long line of work. 

In the online setting, we give the first online subset selection algorithm for $\ell_p$ subspace approximation and entrywise $\ell_p$ low rank approximation by showing how to implement the classical sensitivity sampling algorithm online, which is challenging due to the sequential nature of sensitivity sampling. Our main technique is an online algorithm for detecting when an approximately optimal subspace changes substantially. We also give new related results for the online setting, including online coresets for Euclidean $(k,p)$ clustering as well as an online active regression algorithm making $\tilde\Theta(d^{p/2}/\varepsilon^{p-1})$ queries, answering open questions of \cite{MMWY2022, CLS2022}.
\end{abstract}


\clearpage
\setcounter{page}{1}


\newpage


\section{Introduction}

When one needs to efficiently represent an $n\times d$ matrix $\bfA$, it is possible to store $\bfA$ using just $O((n + d)k)$ real numbers if $\bfA$ is rank $k$, which leads to significant computational savings when $k$ is small. While $\bfA$ may not always have rank $k$, it is often useful to \emph{approximate} $\bfA$ by a surrogate rank $k$ matrix $\hat\bfA$. The problem of finding such a matrix is known as \emph{low rank approximation}. Depending on the cost function used to measure the ``quality'' of the approximation, this leads to a variety of problems which have been studied extensively in computer science, and has applications both in theory and in practice to efficient data analysis, machine learning, and computational geometry. In this work, we study two versions of this problem:

\begin{Definition}[Low Rank Approximation Problems]
Let $\bfA\in\mathbb R^{n\times d}$, $p\geq 1$ a constant, and $k\geq 1$.
\begin{itemize}
\item \textbf{Entrywise $\ell_p$ Low Rank Approximation.} We seek a rank $k$ matrix $\hat\bfA\in\mathbb R^{n\times d}$ which approximately minimizes
    \[
    \norm{\bfA - \hat\bfA}_{p,p} \coloneqq \bracks*{\sum_{i=1}^n \norm{\bfa_i - \hat\bfa_i}_p^p}^{1/p} = \bracks*{\sum_{i=1}^n \sum_{j=1}^d \abs{\bfA_{i,j} - \hat\bfA_{i,j}}^p}^{1/p},
    \]
    where $\bfa_i = \bfe_i^\top\bfA$ and $\hat\bfa_i = \bfe_i^\top\hat\bfA$. For $p = 2$, we write $\norm*{\cdot}_F \coloneqq \norm*{\cdot}_{2,2}$ for the Frobenius norm. For $p = \infty$, we write $\norm*{\cdot}_{\infty,\infty}$ for the norm defined by $\norm*{\bfA}_{\infty,\infty} = \max_{(i,j)\in[n]\times[d]} \abs{\bfA_{i,j}}$.

\item \textbf{$\ell_p$ Subspace Approximation.} Let $\mathcal F_k$ denote the set of subspaces $F\subseteq\mathbb R^d$ of rank at most $k$. We seek a rank $k$ subspace $F\in\mathcal F_k$ which approximately minimizes
\[
    \norm{\bfA - \bfA\bfP_F}_{p,2} \coloneqq \bracks*{\sum_{i=1}^n \norm{\bfa_i - \bfP_F\bfa_i}_2^p}^{1/p} = \bracks*{\sum_{i=1}^n \min_{\bfx\in F}\norm{\bfa_i - \bfx}_2^p}^{1/p},
\]
where $\bfa_i = \bfe_i^\top\bfA$ and $\bfP_F$ is the orthogonal projection matrix onto $F$. For $p = \infty$, we write $\norm*{\cdot}_{\infty,2}$ for the norm defined by $\norm*{\bfA}_{\infty,2} = \max_{i\in[n]}\norm*{\bfa_i}_2$.
\end{itemize}
\end{Definition}

One can also consider generalizations of these problems which replace the $\ell_p$ loss function by an arbitrary function $g$, which gives the $g$-norm subspace approximation problem \cite{CW2015b} and entrywise $g$-norm low rank approximation problem \cite{SWZ2019}, respectively. These generalizations allow for a more flexible choice of loss function, for example the popular Huber loss from the robust statistics literature \cite{CW2015a}.

For $p = 2$, both the entrywise $\ell_p$ low rank approximation problem and the $\ell_p$ subspace approximation problem are exactly equivalent to the classical principle component analysis (PCA) problem, and can be solved exactly in polynomial time via the singular value decomposition (SVD). However, it is also desirable to solve the problem for other values of $p$, with $p < 2$ generally allowing for a more \emph{robust} cost function which is less sensitive to outliers, while $p > 2$ is useful for capturing the \emph{extent} of datasets. While the $p = 2$ case is well-understood in a variety of settings, the $p\neq 2$ is much more difficult, and we center our discussion of results and previous work around this case. In particular, the subspace approximation problem is NP-hard to approximate for any $p\neq 2$ \cite{DTV2011, GRSW2012, CW2015b}, and for entrywise $\ell_p$ low rank approximation, a variety of hardness of approximation results are known \cite{Mie2009, GV2018, DHJWZ2018, BBBKLW2019, MW2021}.

\paragraph{Subset Selection and Coresets for Low Rank Approximation.}

In this work, we continue a long line of work which studies low rank approximation based on \emph{subset selection} and \emph{coresets}\footnote{Note that various notions of subset selection and coreset guarantees have been considered in the literature, with subtle differences and somewhat inconsistent use of terminology. We will generally use the terms ``subset selection'' and ``coreset'' interchangably, and precisely state our approximation guarantee whenever necessary.}. In such approaches, we take the low rank factorization $\hat\bfA\in\mathbb R^{n\times d}$ of the original matrix $\bfA\in\mathbb R^{n\times d}$ to be factorized as $\hat\bfA = \bfU\bfV^\top$, where $\bfU$ is formed from a subset of the columns of $\bfA$, or $\bfV$ is formed from a subset of the rows of $\bfA$. Such an approach has numerous advantages over other alternatives: by using the original columns or rows of $\bfA$, one preserves structural properties of $\bfA$ such as sparsity, and also gives better \emph{interpretability} of the resulting factorization. When the columns of $\bfA$ correspond to \emph{features} of a training dataset, then such a result can also be thought of as an unsupervised feature selection result \cite{ABFMRZ2016}, which is of interest in machine learning. Furthermore, in some cases, subset selection algorithms are in fact the \emph{best known} algorithms for low rank approximation.

\subsection{Offline Subset Selection for Entrywise Low Rank Approximation}

We first present our results for the entrywise low rank approximation problem, for both the entrywise $g$-norm and entrywise $\ell_p$ norm. In this setting, we study bicriteria approximation guarantees of the following form:

\begin{Definition}[Bicriteria Coreset for Low Rank Approximation]
\label{def:bicriteria}
    Let $\bfA\in\mathbb R^{n\times d}$, let $k$ be a rank parameter, and let $\norm*{\cdot}$ be any loss function. Let $S\subseteq[d]$ be a subset of columns, and write $\bfA\vert^S$ for the $n\times S$ matrix\footnote{We allow for indexing matrices and vectors by arbitrary sets. For example, $\mathbb R^S$ is the set of vectors with entries indexed by elements $s$ of $S$, and $\mathbb R^{S\times d}$ is the set of matrices with rows indexed by elements of $S$ and columns indexed by $[d]$.} formed by the columns of $\bfA$ indexed by $S$. Then, $S$ is a \emph{bicriteria coreset} with distortion $\kappa\geq 1$ if
    \[
    \min_{\bfX\in\mathbb R^{S\times d}}\norm{\bfA - \bfA\vert^S\bfX} \leq \kappa \min_{\rank(\hat\bfA)\leq k} \norm{\bfA - \hat\bfA}.
    \]
\end{Definition}

\subsubsection{Entrywise \texorpdfstring{$g$}{g}-Norm Low Rank Approximation}

We begin by presenting our result on entrywise $g$-norm low rank approximation, which is low rank approximation with the following loss function, first considered by \cite{SWZ2019}:

\begin{Definition}[\cite{SWZ2019}]
Let $g:\mathbb R\to\mathbb R_{\geq0}$ be a nonnegative scalar cost function. Then for a matrix $\bfA\in\mathbb R^{n\times d}$, we define the \emph{entrywise $g$-norm} $\norm*{\cdot}_g$ as
\[
    \norm*{\bfA}_g \coloneqq \sum_{i=1}^n \sum_{j=1}^d g(\bfA_{i,j})
\]
While we denote this loss as a norm in a standard abuse of notation  \cite{CW2015b, CW2015a, SWZ2019, MMWY2022}, it may not necessarily satisfy the properties of a norm. This definition extends naturally to vectors.
\end{Definition}

We recall several natural properties of $g$, which have been considered in previous work \cite{CW2015b, CW2015a, SWZ2019, MMWY2022} for obtaining provable guarantees for a broad class of loss functions:

\begin{Definition}
    Let $g:\mathbb R \to \mathbb R_{\geq0}$. Then:
    \begin{itemize}
        \item $g$ satisfies the \emph{$\ati_{g,t}$-approximate triangle inequality} if for any $x_1, x_2, \dots, x_t$, $g(\sum x_i) \leq \ati_{g,t}\cdot \sum_i g(x_i)$.
        \item $g$ is \emph{$\mon_g$-monotone} if for any $0\leq\abs{x}\leq\abs{y}$, $g(x) \leq \mon_g \cdot g(y)$.
        \item $g$ has at least \emph{$\lin_g$-linear growth} if for any $0<\abs{x}\leq\abs{y}$, $g(y)/g(x) \geq \lin_g \cdot \abs{y}/\abs{x}$.
    \end{itemize}
\end{Definition}

For example, popular functions that satisfy these bounds include the Huber loss, Fair loss, Cauchy loss, $\ell_1$-$\ell_2$ loss, and the quantile loss \cite{SWZ2019}. While the $\lin_g$-linear growth bound excludes the Tukey loss, which grows quadratically near the origin and stays constant away from the origin, it allows for a modification of the Tukey loss where the constant away from the origin is replaced by an arbitrarily slow linear growth \cite{CW2015a}.

\cite{SWZ2019} showed that, given an algorithm for solving linear regression in the $g$-norm with relative error $\reg_g$\footnote{This parameter can depend on the input matrix, but we take it to be a parameter depending only on $g$ for now, for simplicity.}, it is possible to compute a set of $O(k\log d)$ columns achieving an approximation ratio of
\[
O(k\log k)\cdot\reg_g \cdot \mon_g\cdot \ati_{g,2k+1}.
\]
for $g$ satisfying the $\mon_g$-monotone and $\ati_{g,t}$-approximate triangle inequality properties. We show that for the slightly restricted family of $g$ of at least $\lin_g$-linear growth, which for example includes all convex $g$ \cite{CW2015a}, we obtain an improved approximation ratio of
\[
O(\sqrt{k\log\log k})\cdot\frac{\reg_g \cdot \ati_{g,2s+1}}{\lin_g}.
\]
Our guarantee matches, and in fact improves a log factor, of the $\ell_1$ column subset selection guarantee of \cite{MW2021}, despite being a more general result. Furthermore, our bound is tight, in the sense that the $\sqrt k$ cannot be improved to a smaller polynomial due to a matching lower bound for $\ell_1$ column subset selection \cite{SWZ2017}. Our technique for removing the $\log k$ factor in the distortion is general, and can be used to improve prior results for $\ell_p$ column subset selection as well \cite{CGKLPW2017, DWZZR2019, MW2021}. 

\begin{Theorem}[Improved Guarantees for \cite{SWZ2019}]\label{thm:improved-swz2019}
    Let $\bfA\in\mathbb R^{n\times d}$ and let $k\geq 1$. Let $s = O(k\log\log k)$. Let $g:\mathbb R\to\mathbb R_{\geq0}$ be a loss function satisfying the $\ati_{g,t}$-approximate triangle inequality for $t = s+1$ and the $\lin_g$-linear growth property. Furthermore, suppose that there is an algorithm outputting $\tilde\bfx$ such that
    \[
    \norm*{\bfB\tilde\bfx - \bfb}_g \leq \reg_{g,s} \cdot \min_{\bfx\in\mathbb R^s}\norm*{\bfB\tilde\bfx - \bfb}_g
    \]
    for any $\bfB\in\mathbb R^{n\times s}$ and $\bfb\in\mathbb R^n$. Then, there is an algorithm, Algorithm \ref{alg:swz2019}, which outputs a subset $S\subseteq[d]$ of $\abs{S} = O(k(\log\log k)(\log d)^2)$ columns and $\bfX\in\mathbb R^{t\times d}$ such that
    \begin{align*}
        \norm*{\bfA - \bfA\vert^S\bfX}_g &\leq O(\sqrt s)\frac{\reg_{g,O(s\log d)}\cdot\ati_{g,2s+1}}{\lin_g} \min_{\rank(\hat\bfA) \leq k}\norm{\bfA-\hat\bfA}_g.
    \end{align*}
\end{Theorem}

Our proof is given in Section \ref{sec:m-css}. For the important case of the Huber loss, given by
\[
    H(x) = \begin{cases}
    \abs{x}^2 / 2 & \text{if $\abs{x}\leq 1$} \\
    \abs{x} - 1/2 & \text{if $\abs{x} > 1$}
    \end{cases},
\]
we specialize our technique to give the following optimized result, proven in Section \ref{sec:huber}:

\begin{restatable}[Entrywise Huber Low Rank Approximation]{Theorem}{HuberCSS}\label{thm:huber-css}
    Let $\bfA\in\mathbb R^{n\times d}$ and let $k\geq 1$. There is an algorithm which outputs a subset $S\subseteq[d]$ of $\abs{S} = O(k(\log\log k)\log d)$ columns and $\bfX\in\mathbb R^{S\times d}$ such that
    \begin{align*}
        \norm*{\bfA - \bfA\vert^S\bfX}_{H} &\leq O(k)\min_{\rank(\hat\bfA) \leq k}\norm{\bfA-\hat\bfA}_{H},
    \end{align*}
    where $\norm*{\cdot}_H$ denotes the entrywise Huber loss.
\end{restatable}

The previous best known bound \cite{SWZ2019} gave a distortion of $\tilde O(k^2)$ for the same number of columns.

\paragraph{Well-Conditioned Spanning Sets.}

Our improvements stem from a new technique which replaces the use of a well-conditioned \emph{linear basis} by a slightly larger \emph{spanning set} which satisfies a much stronger well-conditioning guarantee. Consider a set of $n \geq d$ vectors $\{\bfa_i\}_{i=1}^n\subseteq\mathbb R^d$, where the associated $n\times d$ matrix $\bfA$ has rank $d$. It is well-known that a subset $S\subseteq[n]$ of $d$ vectors chosen to maximize the determinant of the associated matrix $\bfA\vert_S\in\mathbb R^{d\times d}$ has an $\ell_\infty$ well-conditioning property, in the sense that for any $i\in[n]$,
\begin{equation}\label{eq:well-cond-subset-infty}
    \norm*{(\bfA\vert_S)^{-\top}\bfa_i}_\infty \leq 1,
\end{equation}
where we let $\bfM^{-\top}$ denote the pseudoinverse (or inverse) transpose of a matrix $\bfM$. One can also take the original set of vectors to be a subspace rather than a finite set, in which case this is known as an \emph{Auerbach basis} from the functional analysis literature \cite{Aue1930, JL2001}. This property has been used crucially in several works in theoretical computer science to obtain various matrix approximation guarantees \cite{SW2011, MM2013, WZ2013, WW2019, WW2022}, including works on low rank approximation with entrywise losses \cite{CGKLPW2017, SWZ2019, BRW2021}. The property in \eqref{eq:well-cond-subset-infty} is rather weak, in the sense that $\ell_\infty$ is the smallest $\ell_p$ norm, and one could ask whether a similar property holds for other $\ell_p$ norms. Unfortunately, a simple construction shows that there exist $d+1$ vectors such that for any $d$ vectors selected, the remaining vector is always written by a linear combination where the coefficients all have absolute value $1$ (see Theorem \ref{thm:lower-bound-spanning-set}).

To improve upon this, we make the following crucial observation: if we relax our notion of a well-conditioned basis to be a \emph{spanning set} which is allowed to consist of more than $d$ vectors, then we can in fact replace the $\ell_\infty$ norm in \eqref{eq:well-cond-subset-infty} by the much stronger $\ell_2$ norm, with only a small increase in the size of the set. This result is formalized in the following theorem:

\begin{Theorem}[Informal Restatement of Corollary \ref{cor:l2-well-cond}]\label{thm:informal-l2-well-cond}
Let $\bfA\in\mathbb R^{n\times d}$. There exists a subset $S\subseteq[n]$ of size at most $\abs{S} \leq O(d\log\log d)$ such that for every $i\in[n]$, there exists a vector $\bfx\in\mathbb R^S$ such that $\bfa_i = \bfA\vert_S^\top \bfx$ and
\[
    \norm*{\bfx}_2 \leq O(1).
\]
\end{Theorem}

In fact, the construction of $S$ is nothing more than a coreset for a L\"owner--John ellipsoid around the set $\{\pm\bfa_i\}_{i=1}^n$, which has been studied extensively in prior work \cite{KY2005, Tod2016}\footnote{While a coreset of size $O(d\log\log d)$ is the best result we are aware of, we note that improvements to constructions of coresets for L\"owner--John ellipsoids immediately imply improvements to Theorem \ref{thm:informal-l2-well-cond}.}. However, to our knowledge, our work is the first to explicitly connect coresets for L\"owner--John ellipsoids to well-conditioned spanning sets in this way. We also provide a faster algorithm for constructing such a coreset for L\"owner--John ellipsoids by using leverage score sampling, at a cost of a coreset that is larger by a factor of $\log d$ (Theorem \ref{thm:linf-embedding-LJ}). We discuss our results for $\ell_2$-well-conditioned spanning sets in Section \ref{sec:well-cond}. Other applications of our well-conditioned spanning sets technique can be found in Section \ref{sec:subspace-embeddings-large-dist}, including subspace embeddings for the average top $k$ norm and cascaded matrix norms. 

\paragraph{Entrywise $g$-Norm Low Rank Approximation.} Next, we discuss how we apply our well-conditioned spanning sets to obtain sharper bounds for entrywise $g$-norm low rank approximation. Our main improvement comes from an improved structural result on uniform sampling. Let $\bfA = \bfA_* + \bfDelta$, where $\bfA_*$ is an optimal rank $k$ solution and $\bfDelta$ is the error matrix. The overall idea of the \cite{SWZ2019} algorithm follows \cite{CGKLPW2017}, and is based on noting that a random subset $H\subseteq[d]$ of $2k$ columns fits each column $i\in[d]\setminus H$ with constant probability. This suffices for the final algorithmic guarantee, since we will then fit a constant fraction of columns by averaging, and repeating for $O(\log d)$ rounds fits all $d$ columns and selects only $O(k\log d)$ columns.

By using a well-conditioned basis given by maximum determinant subsets, \cite{SWZ2019} show that for a random subset $H\subseteq[d]$ of $2k$ columns, a random column $i\in[d]\setminus H$ outside of $H$ can be written as $\bfa_*^i = \bfA_*\vert^H\bfx_*$ where $\norm{\bfx_*}_\infty \leq 1$, with constant probability. It then follows that
\[
    \min_{\bfx\in\mathbb R^d}\norm*{\bfA\vert^H\bfx-\bfa^i}_g \leq \norm*{\bfA\vert^H\bfx_*-\bfa^i}_g = \norm*{\bfDelta\vert^H\bfx_*-\bfdelta^i}_g \leq \ati_{g,2k+1}\cdot\mon_g\cdot \sum_{j\in H\cup\{i\}}\norm*{\bfdelta^j}_g
\]
with constant probability, using approximate monotonicity and approximate triangle inequality. The latter sum is $O(k/d)\norm*{\bfDelta}_g$ on average, which translates to a final error bound of $\ati_{g,2k+1}\cdot\mon_g\cdot O(k)\norm*{\bfDelta}_g$. 

To improve this argument, we now instead take $H$ to be a random subset of $2s = O(k\log\log k)$ columns, and use our well-conditioned spanning set to argue that with constant probability, we can write a random $i\in[d]\setminus H$ as $\bfa_*^i = \bfA_*\vert^H\bfx_*$ with $\norm{\bfx_*}_2 \leq O(1)$. Then, we can replace the earlier argument by
\[
    \min_{\bfx\in\mathbb R^d}\norm*{\bfA\vert^H\bfx-\bfa^i}_g^2 \leq \norm*{\bfA\vert^H\bfx_*-\bfa^i}_g^2 = \norm*{\bfDelta\vert^H\bfx_*-\bfdelta^i}_g^2 \leq 2\ati_{g,2s+1}^2\parens[\Big]{\norm*{\bfdelta^i}_g + \sum_{j\in H}\norm*{(\bfx_*)_j\bfdelta^j}_g}^2
\]
Now, using at least linear growth and then Cauchy--Schwarz, we can bound the latter sum by
\[
\frac1{\lin_g^2}\parens[\Big]{\sum_{j\in H}\abs{(\bfx_*)_j}\norm*{\bfdelta^j}_g}^2 \leq \frac1{\lin_g^2}\norm*{\bfx_*}_2^2\sum_{j\in H}\norm*{\bfdelta^j}_g^2 \leq \frac1{\lin_g^2}O(1)\sum_{j\in H}\norm*{\bfdelta^j}_g^2
\]
Then if the $\bfdelta^i$ all have similar $g$-norms, then this means that the cost of $\bfa^i$ when fit on $\bfA\vert^H$ is only $O(\sqrt s) = O(\sqrt{k\log\log k})$ times the average cost, rather than $k$. To formalize the assumption about the columns having similar $g$-norms, we conduct our analysis by splitting the columns into roughly $O(\log d)$ groups of columns, such that columns $j$ within each group have costs $\norm*{\bfdelta^j}_g$ which are within a constant factor of each other. Then, applying the previous argument on each group of columns only increases the number of columns by a factor of $O(\log d)$, and we are able to obtain the cost improvement from $O(k)$ to $O(\sqrt s)$, as claimed.

\subsubsection{Nearly Optimal Oblivious Subspace Embeddings}

We take a brief detour from our low rank approximation results to note that our result on well-conditioned spanning sets resolves a long-standing problem on \emph{oblivious $\ell_p$ subspace embeddings}, or \emph{$\ell_p$ OSEs}:

\begin{Definition}[Oblivious $\ell_p$ Subspace Embedding]
Let $p\geq1$ and $\kappa \geq 1$ be parameters. Let $\mathcal D$ be a distribution over matrices $\bfS\in\mathbb R^{r\times d}$. Then, $\mathcal D$ is an \emph{oblivious $\ell_p$ subspace embedding} if for \emph{any} $\bfA\in\mathbb R^{n\times d}$, 
\[
    \Pr\braces*{\mbox{for all $\bfx\in\mathbb R^d$, }\norm*{\bfA\bfx}_p \leq \norm*{\bfS\bfA\bfx}_p \leq \kappa \norm*{\bfA\bfx}_p} \geq \frac{99}{100}.
\]
\end{Definition}

In the above definition, the distribution $\mathcal D$ does not depend on $\bfA$, hence the name ``oblivious''. For $p = 2$, OSEs can be obtained with $\kappa = 1+\eps$, and have found many applications \cite{Woo2014}. For $p\in[1,2)$, a line of work has studied the problem of obtaining $\ell_p$ OSEs with $\kappa = \poly(d)$ distortion, as $1+\eps$ approximations require $r$ to be exponential in $d$ \cite{WW2019, LWY2021}. The first such result was given by \cite{SW2011}, who gave a construction with $r,\kappa = O(d\log d)$ for $p = 1$, using Auerbach bases as a crucial ingredient. Analogous results were later obtained for $p\in(1,2)$ \cite{MM2013, WZ2013, WW2019, WW2022} using the existence of an \emph{$(\alpha,\beta,p)$-well-conditioned basis}:

\begin{Definition}[$(\alpha,\beta,p)$-Well-Conditioned Basis, Definition 3, \cite{DDHKM2009}]
Let $\bfA\in\mathbb R^{n\times d}$ be a rank $d$ matrix, let $p\geq 1$, and let $q$ be the H\"older dual of $p$. Then, $\bfU\in\mathbb R^{n\times d}$ is an \emph{$(\alpha,\beta,p)$-well-conditioned basis} if (1) $\norm*{\bfU}_{p,p}\leq\alpha$ and (2) for any $\bfz\in\mathbb R^d$, $\norm*{\bfz}_q \leq \beta\norm*{\bfU\bfz}_p$.
\end{Definition}

It was claimed in \cite{MM2013} and used by \cite{MM2013, WZ2013, WW2019} that $(d^{1/p},1,p)$-well-conditioned bases exist for any $p\in[1,2)$, which in turn gave $\ell_p$ OSEs with $\kappa = O(d\log d)^{1/p}$ and $r = O(d\log d)$, which is nearly optimal \cite{WW2019}. However, an error in the claim of existence of well-conditioned bases with these parameters was later found \cite{WW2022}. Thus, the best known result is to use Auerbach bases instead, which only gives an $\tilde O(d)$ approximation; obtaining tight bounds for $\ell_p$ OSEs thus became an important open problem again:

\begin{Question}[\cite{WW2022}]\label{qn:oblivious-ose}
\it Do there exist oblivious $\ell_p$ subspace embeddings with $\kappa = \tilde O(d^{1/p})$ distortion?
\end{Question}

To address Question \ref{qn:oblivious-ose}, we apply our well-conditioned spanning sets to circumvent the construction of well-conditioned bases. More specifically, we note that the proofs of \cite{MM2013, WZ2013, WW2019} only require that for any $\bfA\bfx$, there exists $\bfz$ with $\bfA\bfx = \bfU\bfz$ and $\norm*{\bfz}_q \leq O(1)$. As we show in Section \ref{sec:ose}, by applying Theorem \ref{thm:informal-l2-well-cond} to the set of $\ell_p$ unit vectors in the column space of $\bfA$, we can in fact prove a stronger result which bounds the $\ell_2$ norm of the coefficient vector $\bfz$, rather than the $\ell_q$ norm (note that $q > 2$ for $p < 2$). 

\begin{Theorem}[$\ell_p$ Well-Conditioned Spanning Sets]\label{thm:well-cond-for-ose}
Let $\bfA\in\mathbb R^{n\times d}$, let $p\geq 1$, and let $q$ be the H\"older dual of $p$. Then, there exists $\bfU\in\mathbb R^{n\times s}$ for $s = O(d\log\log d)$ such that (1) $\norm*{\bfU}_{p,p} \leq s^{1/p}$ and (2) for any $\bfx\in\mathbb R^d$, there exists $\bfz\in\mathbb R^s$ such that $\bfA\bfx = \bfU\bfz$ and $\norm*{\bfz}_2 \leq O(1)\norm{\bfU\bfz}_p$. 
\end{Theorem}

Combining Theorem \ref{thm:well-cond-for-ose} with \cite{MM2013, WZ2013, WW2019} affirmatively answers Question \ref{qn:oblivious-ose}:

\begin{Corollary}[Nearly Optimal Oblivious $\ell_p$ Subspace Embeddings]\label{cor:lp-ose}
There exists an oblivious $\ell_p$ subspace embedding distribution $\mathcal D$ with a distortion of $\kappa = O(d(\log d)(\log\log d))^{1/p} = \tilde O(d^{1/p})$ and $r = O(d\log d)$. 
\end{Corollary}

We give a simple proof of Corollary \ref{cor:lp-ose} in Section \ref{sec:proof-lp-ose}, based on the idea of \cite{SW2011} of taking the sketching matrix $\bfS$ to be an appropriate scaling of a dense $r\times n$ matrix of i.i.d.\ \emph{$p$-stable random variables} \cite{Nol2005}, which are random variables drawn from a distribution $\mathcal D$ with the property that for $\bfs\sim\mathcal D^n$ and any vector $\bfy\in\mathbb R^n$, $\angle*{\bfs,\bfy}$ is a random variable distributed as $\norm{\bfy}_p Y$ for some $Y\sim\mathcal D$. While more sophisticated constructions of $\bfS$ are known which admit faster running time for applying $\bfS$ to $\bfA$ \cite{WZ2013, MM2013, CDMMMW2016, WW2019, WW2022}, the same ideas immediately apply, and thus we opt for a simpler proof for sake of a cleaner presentation. 

To prove Corollary \ref{cor:lp-ose}, we need to prove that (1) $\norm*{\bfS\bfA\bfx}_p$ is never smaller than $\norm*{\bfA\bfx}_p$, and (2) $\norm*{\bfS\bfA\bfx}_p$ never grows larger than $\norm*{\bfA\bfx}_p$ by more than a factor of $\tilde O(d^{1/p})$. The first item (1) follows straightforwardly from a standard combination of a concentration inequality and a net argument. However, such an argument does not work for the second item (2), due to the fact that $p$-stable random variables are \emph{heavy-tailed}. This is where we crucially use our construction of well-conditioned spanning sets $\bfU$ for $\bfA$. We first show that $\norm*{\bfS\bfU}_{p,p} = \tilde O(\norm*{\bfU}_{p,p}) = \tilde O(d^{1/p})$ with high probability, which does not need a net argument. Then, we write any $\bfA\bfx$ as $\bfA\bfx = \bfU\bfz$ for $\bfz\in\mathbb R^s$, and bound
\[
    \norm*{\bfS\bfA\bfx}_p = \norm*{\bfS\bfU\bfz}_p \leq \norm*{\bfS\bfU}_{p,p}\norm{\bfz}_q \leq \tilde O(\norm{\bfU}_{p,p})\norm*{\bfU\bfz}_p = \tilde O(d^{1/p})\norm*{\bfA\bfx}_p
\]
using H\"older's inequality and the guarantee of our well-conditioned spanning sets. This shows (2) and thus Corollary \ref{cor:lp-ose}.

A related result we obtain is the following low rank decomposition result, which improves \cite[Lemma 9]{BRW2021} and may be of independent interest. We prove this result in Section \ref{sec:ose} as well. 
\begin{restatable}[$\ell_p$ Well-Conditioned Matrix Decomposition]{Theorem}{MatrixDecomp}\label{thm:well-cond-decomp}
Let $\bfL\in\mathbb R^{n\times d}$ be a rank $k$ matrix and let $p\geq 1$. Then, there is $s = O(k\log\log k)$ and a decomposition $\bfL = \bfU\bfV^\top$ into $n\times s$ and $s\times d$ matrices such that (1) $\norm*{\bfU\bfe_i}_p \leq 1$ for each $i\in[s]$ and (2) $\norm*{\bfV\bfe_j}_2 \leq O(1)\norm*{\bfL\bfe_j}_p$ for each $j\in[d]$.
\end{restatable}

We use this result to improve the additive error $\ell_p$ low rank approximation result of \cite{BRW2021} in Theorem \ref{thm:lp-lra-additive}  and apply it to give the first $(1+\eps)$ relative error result in Theorem \ref{thm:1+eps-rel-error}.

Various notions of well-conditioned bases are central to many results in theoretical computer science and mathematics, especially in the study of embeddings, and we hope that our techniques, and in particular the idea of relaxing well-conditioned bases to well-conditioned spanning sets, finds further applications. 

\subsubsection{Entrywise \texorpdfstring{$\ell_p$}{lp} Low Rank Approximation}
\label{sec:intro:entrywise-lp}

We return to studying the entrywise $\ell_p$ low rank approximation problem. For $p = 2$, the problem of column subset selection for the Frobenius norm has been studied extensively \cite{FKV2004, DV2006, DKM2006, DMM2008, BW2017, CMM2017}. For $p\neq 2$, efficient bicriteria approximations were obtained in a line of work initiated by \cite{SWZ2017}, who studied the case of $p = 1$. For other $p\neq 2$, \cite{CGKLPW2017, DWZZR2019} gave algorithms selecting $O(k\log d)$ columns achieving a distortion of $\tilde O(k^{1/p})$ for $p < 2$ and $\tilde O(k^{1-1/p})$ for $p > 2$, and a hardness result showing that any approximation spanned by $k$ columns must have distortion at least
\begin{equation}\label{eq:lp-css-lb}
    \Omega(k^{1-1/p})
\end{equation}
Perhaps surprisingly, \cite{MW2021} then showed that the lower bound of \eqref{eq:lp-css-lb} could be circumvented when $p < 2$, by giving an algorithm which selected $\tilde O(k\log d)$ columns and achieved a distortion of $\tilde O(k^{1/p - 1/2})$. Note that this does not contradict the lower bound, since the hardness result of \eqref{eq:lp-css-lb} applies only when \emph{exactly} $k$ columns are selected. It was also shown that this result was optimal for such bicriteria algorithms, with a lower bound ruling out $k^{1/p-1/2-o(1)}$ approximations for any algorithm selecting $\tilde O(k)$ columns, based on a result of \cite{SWZ2017} which ruled out $k^{1/2-o(1)}$ approximations for any set of $\poly(k)$ columns for $p = 1$. 

Unfortunately, the algorithmic result of \cite{MW2021} uses $p$-stable random variables \cite{Nol2005} which only exist for $p \leq 2$, and similar improvements were not given for $p > 2$. Similarly, the hardness results also rely on specific properties of $p < 2$, and do not apply to $p > 2$. This motivates the following question:

\begin{Question}\label{qn:lp-lra}
\it What distortions are possible for entrywise $\ell_p$ low rank approximation, if $O(k\log d)$ columns can be selected?
\end{Question}

Our main result for entrywise $\ell_p$ low rank approximation is an algorithm which achieves the natural analogue of the algorithmic result of \cite{MW2021}, which circumvents \eqref{eq:lp-css-lb}:

\begin{Theorem}[Informal Restatement of Theorems \ref{thm:lp-css} and \ref{thm:linf-css}]\label{thm:informal-lp-css}
Let $p\in[2,\infty]$, let $\bfA\in\mathbb R^{n\times d}$, and let $k\geq 1$. There is an algorithm which outputs a subset $S\subseteq[d]$ of $O(k\log d)$ columns and $\bfX\in\mathbb R^{S\times d}$ such that
    \begin{align*}
        \norm*{\bfA - \bfA\vert^S\bfX}_{p,p} &\leq O(k^{1/2-1/p})\min_{\rank(\hat\bfA) \leq k}\norm{\bfA-\hat\bfA}_{p,p}.
    \end{align*}
\end{Theorem}

For this result, our well-conditioned spanning sets are not tight enough, as they do not use the special structure of $\ell_p$ norms. However, our well-conditioned spanning sets are based on a novel use of L\"owner--John ellipsoids, which suggests the use of ellipsoids which approximate $\ell_p$ norms of vectors in a subspace in a better way. One such tool is given by \emph{Lewis ellipsoids} \cite{Lew1978}, and we use these to prove Theorem \ref{thm:informal-lp-css}. In particular, if we knew the optimal rank $k$ factorization $\bfU\bfV$ of $\bfA$, then we can approximate the $\ell_p$ norm of all vectors $\bfy\in\mathbb R^d$ in the row span of $\bfV\in\mathbb R^{k\times d}$ by the $\ell_2$ norm of the vector $\bfW^{1/2-1/p}\bfy$, where $\bfW$ is the diagonal matrix consisting of the so-called \emph{Lewis weights} \cite{Lew1978}, up to a factor of $k^{1/2-1/p}$. This essentially allows us to translate a problem dealing with $\ell_p$ norms to one dealing with $\ell_2$ norms, which can be handled by prior work \cite{CW2015b}, up to a factor of $k^{1/2-1/p}$. While this argument requires the knowledge of the optimal factorization and thus only gives an \emph{existential} result, it has been shown in prior work how to turn such a statement into an algorithmic result \cite{SWZ2017, CGKLPW2017, SWZ2019, MW2021}.

For $p = \infty$, we show that Theorem \ref{thm:informal-lp-css} is tight by showing in Theorem \ref{thm:linf-css-lb} that any set of at most $\poly(k)$ columns cannot achieve a distortion better than $k^{1/2-o(1)}$. 

\begin{table}[ht]
\caption{Results for $\ell_p$ column subset selection. The distortion and number of columns hides constant factors.}
\label{tab:css-prior}
\centering
\begin{tabular}{ c c c c c c c } 
\toprule 
& $p$ & Distortion & Number of Columns & Work \\
\midrule
Upper Bound & $(2,\infty]$ & $k\log k$ & $k\log d$ & \cite{CGKLPW2017} + \cite{SWZ2019} \\
& $(2,\infty]$ & $(k\log k)^{1-1/p}$ & $k\log d$ & \cite{DWZZR2019} + \cite{SWZ2019} \\
& $[1, 2)$ & $(k\log k)^{1/p-1/2}$ & $k(\log k)\log d$ & \cite{MW2021} \\
\rowcolor{blue!15} & $(2,\infty]$ & $k^{1/2-1/p}$ & $k\log d$ & Our work, Theorems \ref{thm:lp-css}, \ref{thm:linf-css} \\
\midrule
Lower Bound & $[1,\infty]$ & $k^{1-1/p}$ & exactly $k$ & \cite{DWZZR2019} \\
 & $1$ & $k^{1/2-o(1)}$ & $k^{\Theta(1)}$ & \cite{SWZ2017} \\
 & $(1,2)$ & $k^{1/p-1/2-o(1)}$ & $k(\log k)^{\Theta(1)}$ & \cite{MW2021} \\
\rowcolor{blue!15} & $\infty$ & $k^{1/2-o(1)}$ & $k^{\Theta(1)}$ & Our work, Theorem \ref{thm:linf-css-lb} \\
\bottomrule
\end{tabular}
\end{table}

While we do not have lower bounds for $p < \infty$, we use recent additive error low rank approximation results of \cite{BRW2021} along with our relative error algorithms to obtain the first $(1+\eps)$ bicriteria approximation:

\begin{restatable}[Relative Error $(1+\eps)$ Approximation]{Theorem}{RelErr}\label{thm:1+eps-rel-error}
Let $\bfA\in\mathbb R^{n\times d}$, let $2 < p < \infty$, and let $k\geq 1$. There exists an efficient algorithm that outputs a matrix $\bfL'$ of rank at most
\[
    O\parens*{\frac{k^{(p/2-1)(1+2/p) + 1} (\log\log k + \log\log\log d) (\log d)}{\eps^{1+2/p}}} = \tilde O\parens*{\frac{k^{p/2 - 2/p + 1} \log d}{\eps^{1+2/p}}}
\]
such that
\[
    \norm*{\bfA - \bfL'}_{p,p}^p \leq (1+\eps)\min_{\rank(\hat\bfA)\leq k} \norm{\bfA - \hat\bfA}_{p,p}^p
\]
\end{restatable}

\subsection{Online Subset Selection for \texorpdfstring{$\ell_p$}{lp} Low Rank Approximation}

Next, we discuss our results on online subset selection algorithms for $\ell_p$ subspace approximation and entrywise $\ell_p$ low rank approximation. We will initially focus on the $\ell_p$ subspace approximation problem, which admits a $(1+\eps)$ approximation in this setting, and then later show that this algorithm can be used to obtain entrywise $\ell_p$ low rank approximation results as well, based on \cite{JLLMW2021}. We also switch our convention from selecting columns to selecting rows in this section, in order to conform to previous work on this problem.

\subsubsection{\texorpdfstring{$\ell_p$}{lp} Subspace Approximation}

\paragraph{Coresets for Subspace Approximation.}

In the literature of subset selection for $\ell_p$ subspace approximation, many works have studied guarantees which are slightly stronger than the bicriteria guarantees of Definition \ref{def:bicriteria}. In particular, the work of \cite{DV2007} showed that one can select a subset $S\subseteq[n]$ of $\abs{S} = \poly(k/\eps)$ rows which contains a $(1+\eps)$-approximately optimal rank $k$ subspace in its span. Thus, using this subset, it is possible to further reduce the rank of the approximate solution by computing the best rank $k$ solution spanned by this subset, rather than using the subset itself as a bicriteria rank solution\footnote{For $p\leq 2$, it is possible to translate such guarantees for $\ell_p$ subspace approximation into guarantees for the entrywise $\ell_p$ low rank approximation problem \cite{JLLMW2021}.}. Similar guarantees for more general loss functions, based on similar techniques, were obtained in \cite{CW2015b, MRWZ2020, MMWY2022}.

In fact, even stronger guarantees are possible for the $\ell_p$ subspace approximation problem. In particular, rather than spanning an approximately optimal rank $k$ subspace, one could ask for a subset of rows which \emph{approximates the cost of every rank $k$ subspace}. This is possible if we associate weights $\bfw_i$ with the rows such that the weighted cost of the subset of rows approximates the cost of all rows, known as a \emph{strong coreset}:

\begin{Definition}[Strong Coreset]\label{def:strong-coreset}
Let $\bfA\in\mathbb R^{n\times d}$, let $p\geq 1$, and let $k\geq 1$ be a rank parameter. Then, a subset $S\subseteq[n]$ together with weights $\bfw\in\mathbb R^S$ is a \emph{strong coreset} if
\begin{equation}\label{eq:strong-coreset}
    \mbox{for all $F\in\mathcal F_k$,}\qquad\sum_{i=1}^n \norm*{\bfa_i - \bfP_F\bfa_i}_2^p = (1\pm\eps)\sum_{i\in S} \bfw_i \norm*{\bfa_i - \bfP_F\bfa_i}_2^p.
\end{equation}
\end{Definition}

A slightly weaker guarantee is a \emph{weak coreset}, which only approximates the cost of the optimal solution:

\begin{Definition}[Weak Coreset]\label{def:weak-coreset}
Let $\bfA\in\mathbb R^{n\times d}$, let $p\geq 1$, and let $k\geq 1$ be a rank parameter. Then, a subset $S\subseteq[n]$ together with weights $\bfw\in\mathbb R^S$ is a \emph{weak coreset} if
\begin{equation}\label{eq:weak-coreset}
    \min_{F\in\mathcal F_k}\sum_{i=1}^n \norm*{\bfa_i - \bfP_F\bfa_i}_2^p = (1\pm\eps)\min_{F\in\mathcal F_k}\sum_{i\in S} \bfw_i \norm*{\bfa_i - \bfP_F\bfa_i}_2^p.
\end{equation}
\end{Definition}
A weak coreset not only restricts a $(1+\eps)$-approximate solution to be in the span of a few points, but also states that this solution can be found by optimizing the approximated objective function using the weights $\bfw$.

It is known that a strong coreset can be computed efficiently \cite{FMSW2010, FL2011, VX2012, SW2018, HV2020, FKW2021}. This stronger guarantee is useful, for example, when one wishes to solve a \emph{constrained} version of the subspace approximation problem. For example, in applications to algorithms for clustering or projective clustering, preserving the minimum cost over all $F\in\mathcal F_k$ is not sufficient to solve the original problem. 

\paragraph{Streaming Algorithms for Subspace Approximation.}

In practical large data applications, one does not have the luxury of storing the entire dataset in memory, or even having random access to points in a dataset. In these scenarios, the \emph{streaming model} of computation is a more appropriate theoretical model, in which the rows of our dataset $\{\bfa_i\}_{i=1}^n$ arrive one at a time in adversarial order in one pass, and one seeks to minimize the space complexity of the algorithm. However, computing coresets for subspace approximation in the streaming model is difficult. This is because most (perhaps all) known coreset algorithms proceed by either an \emph{adaptive sampling} \cite{DV2007} or a \emph{sensitivity sampling} \cite{VX2012, HV2020} approach, both of which are naturally sequential procedures. In the former, one first computes a subspace $\tilde F$ achieving a crude approximation, and samples additional rows proportional to the residual cost of the points. In the latter, one first computes \emph{sensitivity scores} by again computing a crude approximation $\tilde F$, and then sampling rows proportional to a sensitivity score formed from combining the residual cost and the projection cost onto $\tilde F$. 

\cite{MRWZ2020} considered circumventing this problem by using oblivious sketching techniques to form a coreset. However, their techniques are limited to $p \leq 2$, and only output noisy rows, rather than actual rows of the dataset. \cite{DP2022} obtained a streaming coreset algorithm for all $p\geq 1$, but their error guarantee is a weaker additive error guarantee. The authors of \cite{DP2022} pose the following as their main open question:

\begin{Question}[\cite{DP2022}]\label{qn:one-pass-streaming}
\it Is there a one-pass streaming coreset algorithm for $\ell_p$ subspace approximation with multiplicative error for every $p\geq 1$?\footnote{Note that the work of \cite{DP2022} studies guarantees which only require the subset of rows to contain a nearly optimal solution, without a guarantee on how this can be found. Similar questions can be asked for coresets with stronger guarantees, such as our definition of weak/strong coresets in Definitions \ref{def:weak-coreset} and \ref{def:strong-coreset}.}
\end{Question}

\paragraph{Online Coresets.}

In fact, one answer to Question \ref{qn:one-pass-streaming} is already known; one can use offline constructions of strong coresets for $\ell_p$ subspace approximation \cite{SW2018, HV2020} and ``compose'' them using a \emph{merge-and-reduce} strategy \cite{BDMMUWZ2020, CLS2022}. This yields a coreset algorithm, even in the one pass streaming model, with the same size as the offline construction, up to a polylogarithmic loss in the size of the coreset \cite{JLLMW2021}. However, this does not address the question of whether adaptive sampling or sensitivity sampling can be ``directly'' implemented in the streaming setting or not. To formalize and address this question, we initiate the study of $\ell_p$ subspace approximation in the \emph{online coreset model}.

The \emph{online model} is a challenging variation on the streaming model, which refers to settings which require decisions to be made on the spot and irrevocably. When instantiated for the problem of computing coresets, the online coreset model studies the setting where the rows of a dataset arrive one by one, and for each row, one must irrevocably decide whether to include the row in the coreset or not. We allow for storing ``side information'', which is a small amount of memory typically comparable to the size of the coreset.

The online coreset setting has been studied extensively for problems arising in data analysis, for example for spectral approximation \cite{CMP2020}, principal component analysis \cite{BLVZ2019, BDMMUWZ2020}, $\ell_p$ linear regression \cite{BDMMUWZ2020, CLS2022, WY2022b}, and computational geometry \cite{WY2022a}. Note that any online coreset algorithm gives a one-pass streaming algorithm. Thus, we ask whether there exist coreset algorithms for $\ell_p$ subspace approximation in this stronger model:

\begin{Question}\label{qn:online-coreset}
\it Is there an online coreset algorithm for $\ell_p$ subspace approximation with multiplicative error for every $p\geq 1$?
\end{Question}

We answer both Questions \ref{qn:one-pass-streaming} and \ref{qn:online-coreset} by designing the first relative error online coreset algorithm for $\ell_p$ subspace approximation for all $p\in[1, \infty)\setminus\{2\}$. 

\begin{Theorem}[Strong Online Coreset for Real-Valued Inputs, Informal Restatement of Theorem \ref{thm:main-real}]\label{thm:informal-real}
Let $\bfA\in\mathbb R^{n\times d}$ have online condition number $\kappa^\OL \coloneqq \norm{\bfA}_2\max_{i=1}^n \norm{\bfA_i^-}_2$\footnote{Here, $\bfA_i^-$ is the pseudoinverse of the first $i$ rows of $\bfA$.}, $\eps\in(0,1)$, $p\geq 1$ a constant, and let $k$ be a rank. There is an online coreset algorithm, Algorithm \ref{alg:online-lp-subspace}, which, with probability at least $0.99$, stores a weighted subset of rows $S$ with weights $\bfw\in\mathbb R^S$ satisfying \eqref{eq:strong-coreset} such that, for $\eps' = \eps^{(p+3)\cdot(1\lor(2/p))}$ we have\footnote{For $a,b,\in\mathbb R$, we denote $\max(a,b)$ by $a\lor b$ and $\min(a,b)$ by $a\land b$.}
\[
    \abs*{S} = \begin{dcases}
    O\parens*{k^2\parens*{\eps'^{-2} + \eps^{-2}\eps'^{-1}k^2}}\log(n\kappa^\OL)^{O(1)} & \text{if $p<2$} \\
    O\parens*{k^p\parens*{k^{p/2 + 1} + \eps'^{-2} + \eps^{-2}\eps'^{-1}k^2}}\log(n\kappa^\OL)^{O(1)} & \text{if $2<p<4$} \\
    O\parens*{k^{p}\parens*{k^3 + \eps'^{-2} + \eps^{-2}\eps'^{-1}k^2}}\log(n\kappa^\OL)^{O(p)} & \text{if $p>4$} 
    \end{dcases}
\]
\end{Theorem}

The condition number dependence is typical for results in online coresets for matrix approximation \cite{BLVZ2019, CMP2020, BDMMUWZ2020, WY2022a, CLS2022, WY2022b}, and is necessary \cite{CMP2020, WY2022a}.

Our theorem for real-valued matrices is a corollary of our result for integer matrices which we apply by rounding the input, as it turns out that we are able to prove much stronger guarantees for integer matrices, in the spirit of \cite{BDMMUWZ2020, WY2022a}. This is due to refined control over condition numbers that we can achieve over integer matrices. This is in contrast to much of the previous work on online coresets, which places an emphasis on real-valued inputs first \cite{BLVZ2019, CMP2020, CLS2022, WY2022b}. 

\begin{Theorem}[Strong Online Coreset for Integer-Valued Inputs, Informal Restatement of Theorem \ref{thm:main-int}]\label{thm:informal-int}
Let $\bfA\in\mathbb Z^{n\times d}$ have entries bounded by $\poly(n,\Delta)$ for a parameter $\Delta$, $\eps\in(0,1)$, $p\geq 1$ a constant, and let $k$ be a rank. There is an online coreset algorithm, Algorithm \ref{alg:online-lp-subspace}, which, with probability at least $0.99$, stores a weighted subset of rows $S$ with weights $\bfw\in\mathbb R^S$ satisfying \eqref{eq:strong-coreset} such that, for $\eps' = \eps^{(p+3)\cdot(1\lor(2/p))}$,
\[
    \abs*{S} = \begin{dcases}
    O\parens*{k^2\parens*{\eps'^{-2} + \eps^{-2}\eps'^{-1}k^2}}\log(n\Delta)^{O(1)} & \text{if $p<2$} \\
    O\parens*{k^p\parens*{k^{p/2 + 1} + \eps'^{-2} + \eps^{-2}\eps'^{-1}k^2}}\log(n\Delta)^{O(1)} & \text{if $2<p<4$} \\
    O\parens*{k^{p}\parens*{k^3 + \eps'^{-2} + \eps^{-2}\eps'^{-1}k^2}}\log (n\Delta)^{O(p)} & \text{if $p>4$} 
    \end{dcases}
\]
\end{Theorem}


\begin{table}[ht]
\caption{Coreset sizes for $\ell_p$ subspace approximation. We suppress $\log(n\kappa^\OL)^{O(1)}$ factors. We have slightly weakened our dependence on $k$ here for simplicity; the $3$ can be replaced by $(1\lor(p/2)) + 1$ for $p < 4$.}
\centering
\begin{tabular}{ c c c }
\toprule
 & Coreset size & Model \\
\midrule
\cite{SW2018} & $k^{1\lor(p/2)}\eps^{-O(p)}$ & Offline, Exponential Time \\
\cite{HV2020} & $k^{1\lor(p/2) + 3}\eps^{-O(p)}$ & Offline, Polynomial Time \\
\cite{DP2022} & $k^p\eps^{-p}$ & Streaming, Additive Error \\
\rowcolor{blue!15} This work & $k^{2\lor p + 3}\eps^{-O(p)}$ & Online, Relative Error \\
\bottomrule
\end{tabular}
\end{table}

\begin{Remark}
As is standard for online coreset results, our algorithms assume the knowledge of a good upper bound on $\kappa^\OL$ and the length $n$ of the stream. While this is not without loss of generality, it is not a limiting assumption in practice, since our bounds depend only logarithmically on these quantities. 
\end{Remark}

\subsubsection{Our Techniques for \texorpdfstring{$\ell_p$}{lp} Subspace Approximation}\label{sec:techniques}

We now discuss failed attempts and challenges in obtaining the results of Theorems \ref{thm:informal-real} and \ref{thm:informal-int}, and how we overcome them. For the rest of this paper, we will write $\bfA\in\mathbb R^{n\times d}$ for the matrix which contains the $n$ input points $\{\bfa_i\}_{i=1}^n\subseteq\mathbb R^d$ in its $n$ rows. We write $\bfA_i\in\mathbb R^{i\times d}$ for the first $i$ rows of $\bfA$.

\paragraph{Sensitivity Sampling for $\ell_p$ Subspace Approximation.}

We start with a discussion of the offline sensitivity sampling technique \cite{LS2010, FL2011}, which is a general technique for designing coreset algorithms. The main idea is to use non-uniform sampling to obtain coresets, by sampling rows proportional to their \emph{sensitivities}, which upper bound the fraction of the total cost that a given row can occupy. More concretely, when specialized to the $\ell_p$ subspace approximation problem, the sensitivity of a row $i\in[n]$ is given by
\begin{equation}\label{eq:sensitivity}
    \bfsigma_i(\bfA) \coloneqq \sup_{F\in\mathcal F_k}\frac{\norm*{\bfa_i^\top(\bfI - \bfP_F)}_2^p}{\sum_{j=1}^n \norm*{\bfa_j^\top(\bfI - \bfP_F)}_2^p}.
\end{equation}
It can be shown that sampling proportionally to these scores approximates the cost of any fixed subspace $F\in\mathcal F_k$, and a union bound over a net of $k$-dimensional subspaces shows that the sampling process approximates the cost of all subspaces $F\in\mathcal F_k$ simultaneously. While it is not clear that $\bfsigma_i(\bfA)$ can be computed efficiently, upper bounds to $\bfsigma_i(\bfA)$ suffice, and these can often be computed efficiently. 

One of the main challenges in applying sensitivity sampling is in bounding the \emph{total sensitivity} $\mathfrak S(\bfA) \coloneqq \sum_{i=1}^n \bfsigma_i(\bfA)$, which is the sum of the sensitivities of the individual rows. Indeed, if we sample each row proportionally to its sensitivity score, then the expected number of rows sampled is proportional to the total sensitivity. For $\ell_p$ subspace approximation, it can be shown that the total sensitivity is at most $\poly(k)$, with sensitivity upper bounds which can be computed as follows \cite{VX2012}:
\begin{enumerate}
\itemsep0em 
\item Compute a constant factor approximation $\tilde F\in\mathcal F_k$ (which can be done efficiently \cite{DTV2011}).
\item Project the input points $\{\bfa_i\}_{i=1}^n$ onto $\tilde F$ to get points $\{\bfa_i'\}_{i=1}^n$.
\item Compute the sensitivity of $\bfa_i$ within $\tilde F$, i.e., the sensitivity of the projected points $\{\bfa_i'\}_{i=1}^n$.
\item Output the sensitivity of row $i$ by combining the cost of $\bfa_i$ for $\tilde F$ and the sensitivity of $\bfa_i$ within $\tilde F$. 
\end{enumerate}
In particular, the projection onto a $k$-dimensional subspace is crucial for removing a dependence on $d$ from the total sensitivity to get a bound of $\poly(k)$. However, this poses a problem for an online algorithm, since this algorithm requires a sequential procedure; we must first compute a constant factor approximation for all the rows, and then project the rows onto this subspace, which naturally requires two passes through the input stream. Furthermore, note that our algorithm must work for all prefix subsets $\{\bfa_j\}_{j=1}^i$ for each $i\in[n]$. Thus, it is not clear that the same subspace $\tilde F$ works for all of these prefix subsets.

If one is willing to accept an inefficient algorithm, then one possibility is the following\footnote{See also Remark \ref{rem:random-order} for a simpler and sharper argument in random order streams.}. First, note that the optimal cost on the prefix subset $\{\bfa_j\}_{j=1}^i$ is increasing in $i$, and can only double a small number of times as $i$ ranges over $[n]$. Indeed, we can relate the optimal cost $\OPT_{p,k}(\bfA_i)$ to the optimal cost $\OPT_{2,k}(\bfA_i)$ up to a factor of $\poly(n)$ by the equivalence of $\ell_p$ norms. This quantity in turn is at most $\norm*{\bfA}_2^2$ and at least $\min_{i=1}^n \norm*{\bfA_i^-}_2^{-2}$, which means we can bound the relative change in the optimal cost by $\poly(n, \kappa^\OL)$. Thus, the optimal cost can only double at most $O(\log(n\kappa^\OL))$ times. We can then partition the set $[n]$ into $O(\log(n\kappa^\OL))$ consecutive groups, such that for every pair of indices $i_1, i_2$ in the group, we have $\OPT_{p,k}(\bfA_{i_1}) = \Theta(\OPT_{p,k}(\bfA_{i_2}))$. Let this group $i_{\text{start}}, i_{\text{start}} + 1, i_{\text{start}} + 2, \dots, i_{\text{end}}$. Then, note that a constant factor approximation $\tilde F$ for $i_{\text{end}}$ is also a constant factor approximation for any $i$ with $i_{\text{start}} \leq i \leq i_{\text{end}}$, since
\begin{align*}
    \sum_{j=1}^i \norm*{\bfa_j^\top(\bfI-\bfP_{\tilde F})}_2^p \leq \sum_{j=1}^{i_{\text{end}}} \norm*{\bfa_j^\top(\bfI-\bfP_{\tilde F})}_2^p &\leq O(1)\min_{F\in\mathcal F_k}\sum_{j=1}^{i_{\text{end}}} \norm*{\bfa_j^\top(\bfI-\bfP_{F})}_2^p && \text{$\tilde F$ is a constant factor solution} \\
    &\leq O(1)\min_{F\in\mathcal F_k}\sum_{j=1}^{i} \norm*{\bfa_j^\top(\bfI-\bfP_{F})}_2^p && \text{$i$ is in the same group as $i_{\text{end}}$}
\end{align*}
Now, we can apply the same subspace $\tilde F$ to each of the $O(\log(n\kappa^\OL))$ groups, and then multiply the resulting bound by $O(\log(n\kappa^\OL))$. As for the sensitivities within the subspace, it is not hard to show that the online Lewis weights \cite{WY2022b} give an online algorithm for computing good sensitivities.

The above argument proves that one can efficiently bound the sum of sensitivities in a way that works for all prefixes $\bfA_i$ of the stream. Thus, by estimating the sensitivity \eqref{eq:sensitivity} up to relative error, we can get an algorithm which samples a small number of rows. The challenge, however, is to design an \emph{efficient} algorithm which achieves a similar guarantee. In particular, the argument above uses the knowledge of a subspace $\tilde F$ which is a good approximation for \emph{future} rows and thus we cannot algorithmically make use of this subspace. 

\paragraph{Online Coreset Algorithms for $p=2$: Ridge Leverage Scores.}

Next, we discuss the existing online algorithms for $p = 2$. Given the above challenges, how do existing algorithms for $p = 2$ proceed? We will discuss the online ridge leverage score sampling algorithm, which gives a nearly optimal bound of $\tilde O(\eps^{-2}k(\log n)(\log^2\kappa))$ for the $\ell_2$ subspace approximation problem \cite{BDMMUWZ2020}.

The key ingredient for getting online coresets for $p = 2$ are the \emph{ridge leverage scores}. The offline ridge leverage scores were first introduced by \cite{AM2015}, defined as $\bftau_i^{\lambda}(\bfA) \coloneqq \bfa_i^\top(\bfA^\top\bfA + \lambda\bfI_d)^{-1}\bfa_i$. Subsequently, \cite{CMM2017} applied these scores to give an extremely efficient offline algorithm for sampling a coreset for $\ell_2$ subspace approximation by using these ridge leverage scores with $\lambda = \norm*{\bfA - \bfA_k}_F^2 / k$. It can be shown that these scores upper bound the $\ell_2$ subspace approximation sensitivities, and that they sum to $O(k)$.

Notably, the ridge leverage scores do not depend on a fixed constant factor approximation $\tilde F$ as discussed in the previous section. Thus, the ridge leverage scores circumvent the problem of having to compute approximate solutions, and instead depend directly on the optimal value, which does not change too frequently as described earlier. These characteristics of the ridge leverage scores allow them to be effectively adapted to the online model, as \cite{BDMMUWZ2020} show. However, for $p\neq 2$, there is no known analogue of ridge leverage scores. In particular, the proof that the ridge leverage scores sum to at most $O(k)$ crucially makes use of the singular value decomposition, which gives $\ell_2$ a very special algebraic structure that is not available to $p\neq 2$. 

\paragraph{Our Solution.}
Our approach is to tackle the online implementation of the sensitivity sampling algorithm. The full discussion is in Section \ref{sec:online-lp-subspace-apx}. 

As discussed earlier, the biggest challenge is to compute a constant factor approximate subspace $\tilde F$ online. The problem was that we wanted to use a good subspace for future rows as a good subspace for a current row, but we could not obtain such a subspace algorithmically. A natural idea is to try to argue that a constant factor solution at time $i$ is also a constant factor solution for many future rows as well. Intuitively, one could expect a subspace to stay a good solution as long as no significantly different directions are added to the optimal solution, which should only occur about $k$ times, since the optimal solution is only $k$-dimensional. 

We formalize this intuition as follows. We first recall an algorithm of \cite{CW2015b, FKW2021} for computing a constant factor solution for $\ell_p$ subspace approximation. This algorithm first projects the $\bfa_i$ onto a random $O(k)$-dimensional subspace, and then computes an \emph{$\ell_p$ subspace embedding} coreset of the randomly projected points of size $O(k^{1\lor(p/2)})$. This is shown to be sufficient for a constant factor approximation \cite[Lemma B.4]{FKW2021}. Furthermore, \cite{WY2022b} show that the $\ell_p$ subspace embedding coreset can be implemented in the online coreset model, where for an $O(k)$-dimensional subspace, the online coreset has size at most roughly $O(k\log(n\kappa^\OL))^{1\lor(p/2)}$. In particular, this means that the online coreset can change at most $O(k\log(n\kappa^\OL))^{1\lor(p/2)}$ times, so the constant approximation subspace also changes only this many times. Now, we can \emph{algorithmically} partition the stream into only $O(k\log(n\kappa^\OL))^{1\lor(p/2)}$ groups, and then compute online sensitivities within the groups by projecting onto the constant factor approximation and proceed as before.

In addition to computing the constant factor approximation, a number of other obstructions remain. One is that the \cite{FKW2021} algorithm requires solving a regression problem, whose solution may not have a good condition number bound, even if the input matrices do. This is a problem, as online coresets have a condition number dependence in their size guarantees. We address this by rounding the input to an integer matrix, and then using sharper condition number bounds for integer matrices. We also show that sensitivity sampling works without replacement, to support an online sampling algorithm. This may be of interest more broadly, for example for implementing the streaming coresets for logistic and $p$-probit regression in \cite{WY2022b} online.

\subsubsection{Entrywise \texorpdfstring{$\ell_p$}{lp} Low Rank Approximation}

As a corollary of our strong online coresets for $\ell_p$ subspace approximation, we obtain the first \emph{weak} online coresets for \emph{entrywise $\ell_p$ low rank approximation problem} for $p\in[1, 2)$, via a reduction shown by \cite{JLLMW2021}.

\begin{restatable}[Weak Online Coreset for Entrywise Low Rank Approximation]{Corollary}{Entrywise}\label{cor:entrywise}
Let $\bfA\in\mathbb R^{n\times d}$ have online condition number $\kappa^\OL$, $p\in[1, 2)$ be a constant, and let $k$ be a rank parameter. There is an online coreset algorithm which, with probability at least $0.99$, stores a weighted subset of rows $S$ with weights $\bfw\in\mathbb R^S$ such that
\[
    \abs*{S} = O(k^4)\log(n\kappa^\OL)^{O(1)}
\]
and
\[
    \min_{\rank(\bfV) \leq k} \norm*{\bfV\bfS\bfA - \bfA}_{p,p} \leq O(k^{4(\frac1p - \frac12)})\log(n\kappa^\OL)^{O(1)}\min_{\rank(\hat\bfA) \leq k} \norm{\bfA - \hat\bfA}_{p,p}
\]
where $\bfS\in\mathbb R^{S\times n}$ is the sampling matrix associated with $S$ and $\bfw$.
\end{restatable}

Our proof in fact improves \cite{JLLMW2021} in the offline setting by removing a factor of $k^{1/p-1/2}$, which is nearly optimal \cite{MW2021}, by giving an analysis which bounds the error of the approximation with respect to the optimal rank $k$ approximation, rather than to the optimal rank $k$ approximation given by a column subset selection algorithm. We give our full discussion of this result in Section \ref{sec:entrywise}.

\subsubsection{Euclidean \texorpdfstring{$(k,p)$}{(k,p)}-Clustering}

Another important problem that is often considered together with subspace approximation is the \emph{Euclidean $(k,p)$-clustering problem}, in which one wishes to find a set $C^*\subseteq\mathbb R^d$ of size at most $k$ such that
\[
    \sum_{i=1}^n d(\bfa_i, C^*)^p \leq (1+\eps) \min_{C\subseteq\mathbb R^d, \abs*{C}\leq k} \sum_{i=1}^n d(\bfa_i, C)^p.
\]
Here, $d(\bfx,C) \coloneqq \min_{\bfy\in C}\norm*{\bfx-\bfy}_2$. This includes the special cases of $p = 2$ and $p = 1$, which correspond to $k$-means and $k$-median, respectively. We provide the first results for clustering in the online coreset model:

\begin{Theorem}[Informal Restatement of Theorem \ref{thm:coreset-clustering}]
Let $w^\OL$ be a lower bound on all nonzero costs for $(k,p)$-clustering $\bfA_i$ for $i\in[n]$, and let $W^\OL$ similarly be an upper bound. Then, there is a strong online coreset algorithm which, with probability at least $0.99$, samples at most
\[
    \min\braces*{\tilde O\parens*{\eps^{-4}k^2(\log n)^4\log\frac{W^\OL}{w^\OL}}, \tilde O\parens*{\eps^{-p-3}k(\log n)^3\log\frac{W^\OL}{w^\OL}}}
\]
points $S\subseteq[n]$ with weights $\bfw\in\mathbb R^S$, and satisfies
\[
    \mbox{for all $C\subseteq\mathbb R^d$ with $\abs*{C}\leq k$}\qquad\sum_{i=1}^n d(\bfa_i, C)^p = (1\pm\eps) \min_{C\subseteq\mathbb R^d, \abs*{C}\leq k} \sum_{i\in S}  \bfw_i d(\bfa_i, C)^p.
\]
\end{Theorem}

\begin{Remark}
In the literature of coresets for clustering, a lot of work goes into removing even a logarithmic dependence on $d$ or $n$ from the coreset size (see, e.g., \cite{SW2018, HV2020, CSS2021, CLSS2022}). However, the online setting already introduces $\log n$ factors, and we do not optimize $\log n$ factors in favor of a simpler argument. Note that one can compose these coreset constructions, even in an online fashion, to weaken the dependence on $n$.
\end{Remark}

Our results are based on an online implementation of \cite{FL2011}. While more recent algorithms have a better dependence on $\eps$ \cite{CSS2021, CLSS2022}, we adopt \cite{FL2011} due to the simpler proofs which make it easier to make the adjustments we need. Note that we improve the guarantee of \cite{FL2011}, by giving an analysis with a dependence on $k$ and $\eps$ of $\eps^{-p-3}k$ rather than $\eps^{-2p-2}k$. This is off by only a single $\eps$ factor from the best result we are aware of, which is $\eps^{-p-2}k$ of \cite{CLSS2022}. We prove our results in this setting in Section \ref{sec:online-coreset-clustering}.


While the online coreset model for Euclidean clustering is new to the best of our knowledge, a couple of other works have studied other variants of ``online clustering''. \cite{Mey2001} studied the related online facility location problem, in which incoming points must be irrevocably assigned to a facility location, while \cite{LSS2016} studied a similar version of the $k$-means algorithm, in which points are irrevocably assigned to clusters. We will adapt the algorithm of \cite{LSS2016} to general $(k,p)$-clustering for the purposes of our algorithms. A slightly different approximation guarantee is considered for $k$-means clustering in \cite{CGKR2021}.

\subsubsection{Online Active \texorpdfstring{$\ell_p$}{lp} Linear Regression}

As our final contribution to online algorithms for data analysis, we provide the first online and offline algorithms for \emph{active $\ell_p$ linear regression} with nearly optimal query complexity. In this problem, we are given a design matrix $\bfA\in\mathbb R^{n\times d}$ and query access to a target vector $\bfb\in\mathbb R^n$, and we seek $\tilde\bfx\in\mathbb R^d$ such that
\begin{equation}\label{eq:lp-regression-objective}
    \norm*{\bfA\tilde\bfx-\bfb}_p^p \leq (1+\eps)\min_{\bfx\in\mathbb R^d}\norm*{\bfA\bfx-\bfb}_p^p
\end{equation}
while reading as few entries of $\bfb$ as possible. We note that active regression is intimately related to the subset selection problem for low rank approximation. Indeed, a common approach for active regression is to select a subset of rows of $\bfA$ such that these rows are sufficient to solve regression for \emph{any} target vector $\bfb$. This implies an algorithm for the multiple response regression problem which aims to minimize $\norm*{\bfA\bfX-\bfB}$ over $\bfX$, where $\bfX\in\mathbb R^{d\times m}$ and $\bfB\in\mathbb R^{n\times m}$ are now matrices. In turn, this is useful for subset selection for low rank approximation: if we know one of the factors of the low rank approximation, then computing the other factor is simply a multiple response regression problem. Indeed, this connection is used in prior work on $\ell_p$ subspace approximation \cite{FKW2021} as well as our work on entrywise $\ell_p$ low rank approximation in Section \ref{sec:intro:entrywise-lp}.

Prior work (see Table \ref{tab:active-regression-prior}) resolved the query complexity up to polylogarithmic factors for all $0 < p\leq 2$. However, for $p > 2$, the best known result due to \cite{MMWY2022} is an upper bound of $\tilde O(d^{p/2}/\eps^p)$ and a lower bound of $\Omega(d^{p/2} + 1/\eps^{p-1})$, leaving a gap in the query complexity of active $\ell_p$ regression for $p>2$. Our main result of this section is a resolution to this problem, showing an algorithm which makes $\tilde O(d^{p/2}/\eps^{p-1})$ queries and a matching lower bound of $\Omega(d^{p/2}/\eps^{p-1})$:

\begin{Theorem}[Informal version of Theorems \ref{thm:main-lp-regression} and \ref{thm:opt-lb}]\label{thm:active-lp}
Let $p > 2$. There is an algorithm which, with probability at least $1-\delta$, outputs $\tilde\bfx$ satisfying \eqref{eq:lp-regression-objective}, while reading at most
\[
    \frac{d^{p/2}}{\eps^{p-1}}\cdot\poly\log(d, 1/\eps, 1/\delta)
\]
entries of $\bfb$. Furthermore, this bound is tight, up to polylogarithmic factors.
\end{Theorem}


\begin{table}[ht]
\caption{Prior results for active $\ell_p$ regression}
\label{tab:active-regression-prior}
\centering
\begin{tabular}{ c c c c c c c } 
\toprule 
$p$ & Distortion & Query Bound & Work \\
\midrule
$2$ & $(1+\eps)$ & $\Theta(d/\eps)$ & \cite{CP2019} \\
$1$ & $(1+\eps)$  & $\tilde\Theta(d/\eps^2)$ & \cite{CD2021, PPP2021} \\
$(1, 2)$ & $(1+\eps)$  & $\tilde\Theta(d/\eps)$ & \cite{MMWY2022} \\
$(0, 1)$ & $(1+\eps)$  & $\tilde\Theta(d/\eps^2)$ & \cite{MMWY2022} \\
$(2, \infty)$ & $(1+\eps)$  & $\tilde O(d^{p/2}/\eps^p)$, $\Omega(d^{p/2} + 1/\eps^{p-1})$ & \cite{MMWY2022, CSS2021b} \\
\rowcolor{blue!15}$(2, \infty)$ & $(1+\eps)$  & $\tilde \Theta(d^{p/2}/\eps^{p-1})$ & Our work, Theorems \ref{thm:main-lp-regression}, \ref{thm:opt-lb} \\
\midrule
\rowcolor{blue!15} $\infty$ & $O(\sqrt d)$ & $\tilde O(d)$ & Our work, Theorem \ref{thm:l-inf-active-regression} \\
\rowcolor{blue!15}$(2, \infty)$ & $O(d^{\frac12\parens*{1-\frac{q}{p}}})$  & $\tilde O(d^{q/2})$ & Our work, Theorem \ref{thm:large-lp-active-regression} \\
\rowcolor{blue!15} $\infty$ & $o(\sqrt d)$ & $d^{\omega(1)}$ & Our work, Theorem \ref{thm:l-inf-active-regression-lb} \\
\rowcolor{blue!15} $(2, \infty)$ & $O(d^{\frac12\parens*{1-\frac{q}{p}}})$ & $\Omega(d^{q/2})$ & Our work, Theorem \ref{thm:lp-active-regression-lb} \\
\bottomrule
\end{tabular}
\end{table}

In \cite{MMWY2022}, the optimized dependence on $\eps$ for $1 < p < 2$ is achieved through an iterative size reduction argument based on the strong convexity of the $\ell_p$ norm in this range---a near optimal solution must be close to the true optimum, which means we only need to approximate the objective function in a restricted domain, which then allows for an even more accurate solution. The main obstacle for applying this argument for $p > 2$ is the lack of strong convexity for $\ell_p$ norms in this range. We get around this by using a bound on the Bregman divergence of the $\ell_p$ norm shown in \cite{AKPS2019}.

Our algorithm only uses independent sampling with an $\alpha$-one-sided Lewis weight distribution. Thus, by using the \emph{online} Lewis weights due to \cite{WY2022b}, we also obtain the first nearly optimal \emph{online} active $\ell_p$ regression algorithm for $p > 2$. This answers the main question of \cite{CLS2022}.

\begin{Theorem}[Informal version of Corollary \ref{thm:online-lp-regression}]
Let $p > 2$. There is an algorithm which, with probability at least $1-\delta$, outputs $\tilde\bfx$ satisfying \eqref{eq:lp-regression-objective}, while reading at most
\[
    \frac{d^{p/2}}{\eps^{p-1}}\cdot\poly\log(n, 1/\eps, 1/\delta, \kappa^\OL)
\]
entries of $\bfb$, \emph{in an online manner}.
\end{Theorem}

Our results are given in Section \ref{sec:active-lp-regression}. As presented in Table \ref{tab:active-regression-prior}, we also present some of the first results for nearly optimal active regression with large distortion for $p>2$. While these results are of independent interest on their own to show what can be done when $\tilde O(d^{p/2})$ queries is too expensive, they are also applied in our result for obtaining tighter bounds for bicriteria subset selection for the entrywise $\ell_p$ low rank approximation problem, as discussed in Section \ref{sec:intro:entrywise-lp}. Our full discussion of these results can be found in Section \ref{sec:active-large-dist}.

\subsection{Open Directions}

We highlight several directions left open by our work.

\paragraph{Subset Selection for Entrywise Loss Low Rank Approximation.}

While we have substantially sharpened various upper bounds for entrywise low rank approximation, both for general loss functions and for $\ell_p$ norms, we still leave a few important gaps in our understanding of the possibilities and limitations in this area. The most obvious gap is showing a matching lower bound for entrywise $\ell_p$ low rank approximation for $2 < p < \infty$. We showed an upper bound of $O(k^{1/2-1/p})$ distortion by selecting $O(k\log d)$ columns and used this to obtain the first $(1+\epsilon)$-approximate bicriteria low rank approximations, but is our $O(k^{1/2-1/p})$ bound tight for column subset selection? Our tight lower bound for $p = \infty$ does not seem to extend to $p < \infty$. Another natural question is obtaining optimal bounds for the entrywise Huber loss: here, we have shown an upper bound of $O(k)$ distortion by selecting $O(k(\log\log k)\log d)$ columns, but is it possible to obtain $O(\sqrt k)$ distortion with the same number of columns? If so, this would be optimal by a reduction to entrywise $\ell_1$ low rank approximation \cite{SWZ2017}. 

\paragraph{Online Coresets for $\ell_p$ Subspace Approximation.}

The most important question left open by our work is improving our dependence on $k$ with an efficient algorithm for online coresets. For $p > 2$, a coreset of size $k^{p/2 + O(1)} / \poly(\eps)$ can be achieved efficiently in the offline setting \cite{HV2020}, while we get a result of size $k^{p + O(1)}$, and it is an interesting question to match the offline result up to $k^{O(1)}$ factors, where the $O(1)$ does not depend on $p$. 

More generally, settling the size of coresets for $\ell_p$ subspace approximation is an interesting direction. The dependence on $k$ for strong coresets is already resolved, if we are allowed an inefficient algorithm. Recall that $k^{1\lor(p/2)}$ can be achieved using an inefficient algorithm, by the result of \cite{SW2018}. Furthermore, one can show that this is optimal for strong coresets by a reduction to a subspace embedding; for $d = k+1$, note that for any $\ell_2$ unit vector $\bfx$, we can query for the $k$-dimensional projection given by the orthogonal complement of $\bfx$, which must approximate
\[
    \norm*{\bfA\bfx\bfx^\top}_{p,2} = \parens*{\sum_{i=1}^n \norm*{\bfa_i^\top\bfx\bfx^\top}_2^p}^{1/p} = \parens*{\sum_{i=1}^n \abs*{\bfa_i^\top\bfx}^p\norm*{\bfx}_2^p}^{1/p} = \norm*{\bfA\bfx}_p.
\]
Thus, strong coresets for $\ell_p$ subspace approximation imply strong coresets for $\ell_p$ subspace embeddings, which have a lower bound of $\Omega(k/\eps^2)$ for $p \leq 2$ and $\Omega(k^{p/2} + 1/\eps^2)$ \cite{LWW2021}. However, matching this with an efficient algorithm, even in the offline setting, is open. Settling the $\eps$ dependence is also an interesting direction, as well as related questions for weak coresets, or the size of any subset which spans a $(1+\eps)$-approximately optimal solution. 

\subsection{Roadmap}

We give preliminaries in Section \ref{sec:prelim}.

Sections \ref{sec:well-cond} through \ref{sec:lp-css} are devoted to our results on offline low rank approximation. We first develop our theory of well-conditioned spanning sets in Section \ref{sec:well-cond}. This is first applied to oblivious $\ell_p$ subspace embeddings in Section \ref{sec:ose}. In Section \ref{sec:m-css}, we then apply our well-conditioned spanning sets to $g$-norm low rank approximation. For the special case of the Huber loss, we specialize our technique in Section \ref{sec:huber}. Finally, we discuss our results on entrywise $\ell_p$ low rank approximation in Section \ref{sec:lp-css}.

Sections \ref{sec:online-lp-subspace-apx} through \ref{sec:online-coreset-clustering} are devoted to our online coreset results. Section \ref{sec:online-lp-subspace-apx} develops our main online coreset algorithm for $\ell_p$ subspace approximation. Section \ref{sec:entrywise} then shows how to apply this to online coresets for entrywise $\ell_p$ low rank approximation. Finally, Section \ref{sec:online-coreset-clustering} shows our online coresets for Euclidean $(k,p)$-clustering.

Sections \ref{sec:active-lp-regression} through \ref{sec:active-large-dist} are devoted to our active regression results. Section \ref{sec:active-lp-regression} proves the main active regression algorithm, except for the main technical lemma on the quality of approximation given by Lewis weight sampling, which is given in Section \ref{sec:close-points-distortion}. Section \ref{sec:active-lb} gives our nearly optimal lower bound for the query complexity. Finally, Section \ref{sec:active-large-dist} collects our results on active regression with large distortion.

\section{Preliminaries}
\label{sec:prelim}

For a matrix $\bfA\in\mathbb R^{n\times d}$, we denote its $i$th row by $\bfa_i$ and its $j$th column by $\bfa^j$. If $S$ is a subset of row or column indices, then we denote the restriction of $\bfA$ to these rows by $\bfA\vert_S$ and the restriction of $\bfA$ to these columns by $\bfA\vert^S$.

\subsection{Lewis Weights}

We need a relaxed notion of $\ell_p$ Lewis weights, known as one-sided $\ell_p$ Lewis weights, given in \cite{WY2022a} (see also \cite{JLS2022}):

\begin{Definition}[One-sided $\ell_p$ Lewis weights and bases \cite{WY2022a}]\label{def:one-sided-lewis}
Let $\bfA\in\mathbb R^{n\times d}$ and $p\in(0,\infty)$. Let $\alpha\in(0,1]$. Then, weights $\bfw\in\mathbb R^n$ are \emph{$\alpha$-one-sided $\ell_p$ Lewis weights} if
\[
    \bfw_i \geq \alpha \cdot \bftau_i(\bfW^{1/2-1/p}\bfA),
\]
where $\bfW\coloneqq\diag(\bfw)$, or equivalently,
\[
    \bfw_i \geq \alpha^{p/2} \bracks*{\bfa_i^\top(\bfA^\top\bfW^{1-2/p}\bfA)\bfa_i}^{p/2}.
\]
If $\alpha = 1$, we just say that $\bfw$ are \emph{one-sided $\ell_p$ Lewis weights} Let $\bfR\in\mathbb R^{d\times d}$ be a change of basis matrix such that $\bfW^{1/2-1/p}\bfA\bfR$ has orthonormal columns. Then, $\bfA\bfR$ is a \emph{one-sided $\ell_p$ Lewis basis}.
\end{Definition}

The following lemma collects basic properties of Lewis weights.

\begin{Lemma}[Lemmas 2.8 and 2.10 of \cite{WY2022a}]\label{lem:oslw-sensitivity}
Let $\bfA\in\mathbb R^{n\times d}$ and $p\in(0,\infty)$. Let $\bfw$ be $\alpha$-one-sided $\ell_p$ Lewis weights for $\bfA$ and let $\bfR$ be a one-sided $\ell_p$ Lewis basis. Then,
\begin{itemize}
    \item for every $i\in[n]$,
    \[
        \frac{\bfw_i}{\alpha^{p/2}} \geq \norm*{\bfe_i^\top\bfA\bfR}_2^p
    \]
    \item for all $\bfx\in\mathbb R^d$,
    \[
        \norm*{\bfW^{1/2-1/p}\bfA\bfx}_2 \leq \begin{cases} \norm*{\bfw}_1^{1/2-1/p}\norm*{\bfA\bfx}_p & \text{if $p\geq 2$} \\
        \alpha^{1/2-1/p}\norm*{\bfA\bfx}_p & \text{if $p < 2$}
        \end{cases}
    \]
    \item for every $i\in[n]$, 
    \[
        \sup_{\norm*{\bfA\bfx}_p > 0} \frac{\abs*{\angle*{\bfa_i,\bfx}}^p}{\norm*{\bfA\bfx}_p^p} \leq 
        \begin{cases} \norm*{\bfw}_1^{p/2-1}\bfw_i & \text{if $p\geq 2$} \\
        \alpha^{p/2-1}\bfw_i & \text{if $p < 2$}
        \end{cases}
    \]
\end{itemize}
\end{Lemma}

The main utility of Lewis weights is that they provide $\ell_p$ subspace embeddings, given by the following theorem:

\begin{Theorem}[Theorem 1.3 of \cite{WY2022b}, see also \cite{CP2015}]\label{thm:lewis-weight-sampling}
Let $p>2$ and let $\bfA\in\mathbb R^{n\times d}$. Let $\delta\in(0,1)$ be a failure rate parameter and let $\eps\in(0,1)$ be an accuracy parameter. Let $\bfw\in\mathbb R^n$ be one-sided $\ell_p$ Lewis weights with $\norm*{\bfw}_1 \leq O(d)$, which can be computed in
\[
    \tilde O(\nnz(\bfA) + d^\omega)
\]
time \cite[Theorem 5.3.1]{Lee2016}, \cite[Lemma 2.5]{JLS2022}. Let
\[
    \alpha = O\parens*{\frac{d^{p/2-1}}{\eps^2}\parens*{(\log d)^2(\log n) + \log\frac1\delta}}
\]
be an oversampling parameter. Suppose that weights $\bfs\in\mathbb R^n$ are sampled by independently setting $\bfs_i = 1/\bfp_i^{1/p}$ with probability $\bfp_i \coloneqq \min\{\alpha\bfw_i, 1\}$ and $\bfs_i = 0$ otherwise. Let $\bfS = \diag(\bfs)$. Then, with probability at least $1-\delta$,
\[
    \mbox{for all $\bfx\in\mathbb R^d$, }\norm*{\bfS\bfA}_p = (1\pm\eps)\norm*{\bfA\bfx}_p
\]
and the sample complexity of $\bfS$ is at most
\[
    r = O\parens*{\frac{d^{p/2}}{\eps^2}\parens*{(\log d)^2(\log n) + \log\frac1\delta}}.
\]
By a standard argument, the $\log n$ dependence can be replaced by a $\log(d/\eps)$ dependence (see, e.g., \cite{MMWY2022}).
\end{Theorem}

We will also frequently use the following result of \cite{WY2022a}, which shows that Lewis weights allow one to convert between $\ell_p$ and $\ell_q$ norms for vectors in a $d$-dimensional subspace with a small distortion. 

\begin{Theorem}[Theorem 1.23 of \cite{WY2022a}]\label{thm:lewis-reweighting}
Let $\bfA\in\mathbb R^{n\times d}$. Let $\bfw$ be one-sided $\ell_p$ Lewis weights for $\bfA$ summing to $T$. Then, for an appropriate scaling factor $c_{p,q}>0$,
\[
    \norm*{\bfA\bfx}_p \leq c_{p,q}\norm*{\bfW^{1/q-1/p}\bfA\bfx}_q \leq \kappa_{p,q} \norm*{\bfA\bfx}_p
\]
for all $\bfx\in\mathbb R^d$, where
\[
    \kappa_{p,q} = \begin{cases}
        T^{\abs{\frac1q-\frac1p}} & \text{if $p\land q \leq 2$} \\
        T^{\frac12\parens*{1 - \frac{p\land q}{p\lor q}}} & \text{if $p\land q \geq 2$}
    \end{cases}
\]
\end{Theorem}

For $p>2$, their proof in fact shows that \emph{any} vector does not expand by more than a $\kappa_{p,q}$ factor, which we state below and provide a self-contained proof:

\begin{Lemma}\label{lem:lewis-no-expansion}
Let $p > q \geq 2$. Let $\bfw$ be any nonnegative weights. Then, for any $\bfy\in\mathbb R^n$,
\[
    \norm*{\bfW^{1/q-1/p}\bfy}_q \leq \norm*{\bfw}_1^{1/q-1/p}\norm*{\bfy}_p.
\]
If $q = 2$ and $p = \infty$, then
\[
    \norm*{\bfW^{1/2}\bfy}_2 \leq \norm*{\bfw}_1^{1/2}\norm*{\bfy}_\infty.
\]
\end{Lemma}
\begin{proof}
We have that
\begin{align*}
\norm*{\bfW^{1/q-1/p}\bfy}_q^q &= \sum_{i=1}^n \bfw_i^{1-q/p}\abs*{\bfy(i)}^q \\
&\leq \parens*{\sum_{i=1}^n \bfw_i^{\frac{1-q/p}{1-q/p}}}^{1-q/p} \parens*{\sum_{i=1}^n \abs*{\bfy(i)}^{p}}^{q/p} && \text{H\"older's inequality} \\
&\leq \norm*{\bfw}_1^{1-q/p} \norm*{\bfy}_p^q
\end{align*}
which rearranges to the desired inequality. For $q = 2$ and $p = \infty$, we have that
\[
\norm*{\bfW^{1/2}\bfy}_2^2 = \sum_{i=1}^n \bfw_i\abs*{\bfy(i)}^2 \leq \norm*{\bfw}_1\norm*{\bfy}_\infty^2 
\]
\end{proof}

\section{Well-Conditioned Spanning Sets}
\label{sec:well-cond}

When designing algorithms for matrix and subspace approximations, it is often desirable to select subsets with vectors with a ``well-conditioning'' property, which are roughly properties which are analogous to orthonormal bases, in norms other than the $\ell_2$ norm. We describe several such results in this section.

\subsection{Semi-Optimal Linear Bases}

We first recall \emph{optimal} and \emph{semi-optimal} linear bases, as introduced by \cite{Knu1985}:

\begin{Definition}[Optimal and Semi-Optimal Linear Bases]\label{def:optimal-linear-bases}
    Let $\{\bfa_i\}_{i=1}^n \subseteq\mathbb R^d$ and $\eps\geq 0$. Then, $\{\bfa_i\}_{i\in S}$ for a subset $S\subseteq[n]$ of size $\abs{S} = d$ is a \emph{$(1+\eps)$-semi-optimal linear basis} if for each $i\in[n]$,
    \begin{equation}\label{eq:semi-optimal-basis}
        \norm*{\bfA\vert_S^{-\top}\bfa_i}_\infty \leq 1+\eps.
    \end{equation}
    That is, if $\bfx$ is the unique solution to $\bfA\vert_S^{-\top}\bfx = \bfa_i$, then each entry of $\bfx$ is at most $1+\eps$ in absolute value. If $\eps=0$, we say that $\{\bfa_i\}_{i\in S}$ is an \emph{optimal linear basis}. Note that by Cramer's rule, a subset $S$ which maximizes the determinant is an optimal linear basis.
\end{Definition}

While an optimal linear basis naively requires an exponential time algorithm to compute, \cite{Knu1985} showed that a semi-optimal linear basis can be computed efficiently via an iterative algorithm:

\begin{Theorem}[Semi-Optimal Linear Bases \cite{Knu1985}]
    Let $\bfA\in\mathbb R^{n\times d}$ and $\eps>0$. There is an algorithm which runs in $\poly(n,d,\eps^{-1})$ time and outputs $S\subseteq[n]$ satisfying \eqref{eq:semi-optimal-basis}.
\end{Theorem}

\subsection{John Ellipsoids and \texorpdfstring{$\ell_2$}{l2}-Well-Conditioned Spanning Sets}

Note that Definition \ref{def:optimal-linear-bases} can be thought of as an $\ell_\infty$-well-conditioning, in the sense that the coefficients vector $\bfA\vert_S^{-\top}\bfa_i$ for writing $\bfa_i$ as a linear combination of $\bfA\vert_S$ is bounded in the $\ell_\infty$ norm. This is a rather weak property since $\ell_\infty$ is a very ``small'' norm, since $\ell_\infty$ is bounded above by all $\ell_p$ norms. We will show that by selecting slightly more than $d$ vectors, we can in fact get a subset of vectors such that the coefficient vector satisfies the much stronger guarantee of being bounded in $\ell_2$. We also show that without this relaxation of choosing more than $d$ vectors, such a result is not possible. Our result are based on the theory of coresets for John ellipsoids. 

We give the following definition:

\begin{Definition}[$\ell_p$-Well-Conditioned Spanning Set]
    Let $p > 0$. Let $\{\bfa_i\}_{i=1}^n \subseteq\mathbb R^d$ and $\eps\geq 0$. Then, $\{\bfa_i\}_{i\in S}$ for a subset $S\subseteq[n]$ is a \emph{$(1+\eps)$-approximate $\ell_p$-well-conditioned coreset} if for each $i\in[n]$, there exists $\bfx\in \mathbb R^S$ such that $\bfa_i = \bfA\vert_S^\top\bfx$ and
    \begin{equation}\label{eq:lp-well-conditioned-coreset}
        \norm*{\bfx}_p \leq 1+\eps.
    \end{equation}
\end{Definition}

\subsubsection{Coresets via Coordinate Ascent}

Our first result uses results on coresets for L\"owner--John ellipsoids \cite{Tod2016}, and gives a deterministic algorithm based on coordinate ascent which selects $O(d\log\log d)$ unweighted rows of $\bfA$ that is sufficient to approximate a John ellipsoid for all the rows. 

\begin{Theorem}[Coresets for L\"owner--John Ellipsoids, Proposition 3.17, \cite{Tod2016}]\label{thm:LJ-coreset}
    Let $\bfA\in\mathbb R^{n\times d}$ and $\eps>0$. There exists $S\subseteq[n]$ with $\abs*{S} = O(d\log\log d + d/\eps)$ and nonnegative weights $\bfu\in\mathbb R^n$ supported on $S$ such that $\norm*{\bfu}_\infty \leq 1$, $\norm*{\bfu}_1 = d$, and
    \[
    (1+\eps)\braces*{\bfx \in\mathbb R^d : \bfx^\top(\bfA^\top\diag(\bfu)\bfA)^{-1}\bfx \leq 1} \supseteq \conv(\braces*{\pm\bfa_i}_{i=1}^n)
    \]
    Furthermore, $S$ and $\bfu$ can be computed in $\tilde O((d/\eps)(\nnz(\bfA) + d^2))$ time.
\end{Theorem}

That is, there exists a set $S$ of $O(d\log\log d)$ rows of $\bfA$ and an $O(1)$-approximate John ellipsoid for $\bfA\vert_S$ containing all rows of $\bfA$. This gives the following corollary:

\begin{Corollary}[$\ell_2$-Well-Conditioned Spanning Set via Coordinate Ascent]\label{cor:l2-well-cond}
    Let $\bfA\in\mathbb R^{n\times d}$ and $\eps>0$. Let $S\subseteq[n]$ be the subset given by the algorithm in in Theorem \ref{thm:LJ-coreset}. Then, $\{\bfa_i\}_{i\in S}$ is a $(1+\eps)$-approximate $\ell_2$-well-conditioned spanning set.
\end{Corollary}
\begin{proof}
    We take the coefficients to be $\bfx = (\bfA\vert_S^\top)^-\bfa_i$. Then,
    \begin{align*}
        \norm*{(\bfA\vert_S^\top)^-\bfa_i}_2^2 &= \bfa_i^\top(\bfA\vert_S^\top\bfA\vert_S)^{-1}\bfa_i \\
        &\leq \bfa_i^\top(\bfA\vert_S^\top\diag(\bfu\vert_S)\bfA\vert_S)^{-1}\bfa_i && \norm*{\bfu}_\infty \leq 1 \\
        &= \bfa_i^\top(\bfA^\top\diag(\bfu)\bfA)^{-1}\bfa_i \\
        &\leq 1+\eps && \text{Theorem \ref{thm:LJ-coreset}}
    \end{align*}
    as claimed.
\end{proof}

We also note here that Theorem \ref{thm:LJ-coreset} also yields unweighted coresets for $\ell_\infty$ subspace embeddings.

\begin{Corollary}[$\ell_\infty$ Subspace Embedding]\label{cor:linf-subspace-embedding}
    Let $\bfA\in\mathbb R^{n\times d}$. There exists $S\subseteq[n]$ with $\abs*{S} = O(d\log\log d)$ such that for all $\bfx\in\mathbb R^d$,
    \[
    \norm*{\bfA\vert_S\bfx}_\infty \leq \norm*{\bfA\bfx}_\infty \leq O(\sqrt d)\norm*{\bfA\vert_S\bfx}_\infty.
    \]
    Furthermore, $S$ can be computed in $\tilde O(d\nnz(\bfA) + d^3)$ time.
\end{Corollary}
\begin{proof}
    Let $S\subseteq[n]$ and $\bfu\in\mathbb R^n$ be given by Theorem \ref{thm:LJ-coreset} with $\eps = 1/2$. By scaling, it suffices to prove that
    \[
    \braces*{\bfx\in\mathbb R^d : \norm*{\bfA\bfx}_\infty \leq 1} \subseteq \braces*{\bfx\in\mathbb R^d : \norm*{\bfA\vert_S\bfx}_\infty \leq 1}  \subseteq O(\sqrt d)\braces*{\bfx\in\mathbb R^d : \norm*{\bfA\bfx}_\infty \leq 1} 
    \]
    The first inclusion is immediate, so it suffices to prove the second inclusion. 
    
    Note that by taking polars in the inclusion in the result of Theorem \ref{thm:LJ-coreset}, we have that
    \begin{align*}
        \frac1{1+\eps}\braces*{\bfx \in\mathbb R^d : \bfx^\top\bfA^\top\diag(\bfu)\bfA\bfx \leq 1} &= ((1+\eps)\braces*{\bfx \in\mathbb R^d : \bfx^\top(\bfA^\top\diag(\bfu)\bfA)^{-1}\bfx \leq 1})^\circ \\
        &\subseteq (\conv(\braces*{\pm\bfa_i}_{i=1}^n))^\circ = \braces*{\bfx\in\mathbb R^d : \norm*{\bfA\bfx}_\infty \leq 1} 
    \end{align*}
    Now suppose that $\bfx\in\mathbb R^d$ satisfies $\norm*{\bfA\vert_S\bfx}_\infty \leq 1$. Then, $\abs*{\angle*{\bfa_i,\bfx}} \leq 1$ for every $i\in[n]$, so we have that
    \[
    \bfx^\top\bfA^\top\diag(\bfu)\bfA\bfx \leq \sum_{i=1}^n \bfu_i \angle*{\bfa_i,\bfx}^2 \leq \sum_{i=1}^n \bfu_i \leq d
    \]
    and thus
    \begin{align*}
        \braces*{\bfx \in\mathbb R^d : \norm*{\bfA_S\bfx}_\infty \leq 1} &\subseteq \braces*{\bfx \in\mathbb R^d : \bfx^\top\bfA^\top\diag(\bfu)\bfA\bfx\leq d} \\
        &= \sqrt d\braces*{\bfx \in\mathbb R^d : \bfx^\top\bfA^\top\diag(\bfu)\bfA\bfx\leq 1} \\
        &\subseteq (1+\eps)\sqrt d\braces*{\norm*{\bfA\bfx}_\infty \leq 1}
    \end{align*}
    which was the desired result.
\end{proof}

We note that if only $d$ rows are selected, as opposed to $O(d)$, then a result like Theorem \ref{cor:l2-well-cond} is not possible:

\begin{Theorem}\label{thm:lower-bound-spanning-set}
    There exists a matrix $\bfA\in\mathbb R^{n\times d}$ for $n = d+1$ such that for any subset $S$ of $d$ rows of $\bfA$, we have that
    \[
    \norm*{\bfA\vert_S^{-\top}\bfa_i}_2 = \sqrt d
    \]
    for $[n]\setminus S = \{i\}$. 
\end{Theorem}
\begin{proof}
    Let $\bfA\in\mathbb R^{n\times d}$ be the identity matrix concatenated with the all ones vector. If the all ones vector is not selected as a part of $S$, then we have that
    \[
    \norm*{\bfA\vert_S^{-\top}\bfa_i}_2 = \norm*{\bfa_i}_2 = \sqrt d
    \]
    On the other hand, if one of the standard basis vectors $\bfe_i$ for $i\in[d]$ is not selected as a part of $S$, then $\bfA\vert_S^{-\top}\bfa_i$ is $-1$ on the entry corresponding to a standard basis vector, and $1$ on the entry corresponding to the all ones vector. Thus, we again have
    \[
    \norm*{\bfA\vert_S^{-\top}\bfa_i}_2 = \sqrt d.
    \]
\end{proof}

\subsubsection{Spanning Sets via Leverage Score Sampling}

Our second result uses leverage score sampling to obtain a significantly faster algorithm, at the expense of randomization and a slightly larger coreset. For this result, we use a much faster John ellipsoid algorithm due to \cite{CCLY2019} which does not a priori yield coresets. We show how to use sampling to turn this result into a coreset. 

\begin{Theorem}[Theorem 3.6, \cite{CCLY2019}]\label{thm:ccly}
    Let $\bfA\in\mathbb R^{n\times d}$ and $\eps>0$. There is an algorithm which computes nonnegative weights $\bfu\in\mathbb R^n$ such that $\norm*{\bfu}_\infty \leq 1$, $\norm*{\bfu}_1 = d$, and
    \[
    \bfa_i^\top(\bfA^\top\diag(\bfu)\bfA)^{-1}\bfa_i \leq 1+\eps
    \]
    Furthermore, the algorithm runs in $\tilde O(\eps^{-2}(\nnz(\bfA) + d^\omega))$ time. 
\end{Theorem}

We now give our result for obtaining coresets via sampling:

\begin{Theorem}\label{thm:linf-embedding-LJ}
    Let $\bfu\in\mathbb R^n$ be weights computed in Theorem \ref{thm:ccly}. Let $\beta = O(\eps^{-2}(\log d)(\log\frac1\delta))$ and suppose we sample independently sample each $i\in[n]$ with probability $p_i = \min\{1,(1+\eps)\beta\bfu_i\}$ to form $S\subseteq[n]$. Then, with probability at least $1-\delta$, $\{\bfa_i\}_{i\in S}$ is a $(1+O(\eps))$-approximate $\ell_2$-well-conditioned coreset, and
    \[
    \abs*{S} \leq O(1)\eps^{-2}d(\log d)\log\frac1\delta.
    \]
    Furthermore, with probability at least $1-\delta$, simultaneously for all $\bfx\in\mathbb R^d$, we have that
    \[
    \norm*{\bfA\vert_S\bfx}_\infty \leq \norm*{\bfA\bfx}_\infty \leq \frac{1+\eps}{1-\eps}\sqrt{\abs*{S}}\cdot\norm*{\bfA\vert_S\bfx}_\infty,
    \]
    that is, $S$ is a coreset for an $\ell_\infty$ subspace embedding.
\end{Theorem}
\begin{proof}
    Note that $\bfu$ satisfies
    \begin{equation}\label{eq:LJ-leverage-score}
        \bftau_i(\diag(\bfu)^{1/2}\bfA) = \bfu_i^{1/2}\bfa_i^\top(\bfA^\top\diag(\bfu)\bfA)^{-1}\bfu_i^{1/2}\bfa_i \leq (1+\eps)\bfu_i.
    \end{equation}
    Thus, $(1+\eps)\bfu_i$ are leverage score upper bounds for $\diag(\bfu)^{1/2}\bfA$. Thus, if we sample rows $S\subseteq[n]$ as in the theorem statement and scale each sampled row by $\bfs_i = 1/\sqrt{p_i}$, then we have that
    \begin{equation}\label{eq:LJ-subspace-embedding}
        \norm*{\diag(\bfs)\diag(\bfu)^{1/2}\bfA\bfx}_2 = (1\pm\eps)\norm*{\diag(\bfu)^{1/2}\bfA\bfx}_2
    \end{equation}
    for every $\bfx\in\mathbb R^d$ (see, e.g., \cite[Lemma 4]{CLMMPS2015}). Also note that for each $i\in[n]$, $\bfs_i^2\bfu_i \leq 1$ for every $i\in[n]$. Thus,
    \begin{align*}
        \norm*{\bfA\vert_S^{-\top}\bfa_i}_2^2 &= \bfa_i^\top(\bfA\vert_S^\top\bfA\vert_S)^{-1}\bfa_i \\
        &\leq \bfa_i^\top(\bfA\vert_S^\top\diag(\bfs)^2\diag(\bfu)\bfA\vert_S)^{-1}\bfa_i \\ 
        &\leq (1+\eps)\bfa_i^\top(\bfA^\top\diag(\bfu)\bfA)^{-1}\bfa_i && \text{\eqref{eq:LJ-subspace-embedding}} \\ 
        &\leq (1+\eps)^2 && \text{Theorem \ref{thm:ccly}}
    \end{align*}
    so $S$ is a $(1+O(\eps))$-approximate $\ell_2$-well-conditioned coreset. 
    
    To see that $\bfA\vert_S$ is an $\ell_\infty$ subspace embedding, first note that
    \begin{align*}
        \norm*{\bfA\vert_S\bfx}_\infty &\geq \frac1{\sqrt{\abs*{S}}}\norm*{\bfA\vert_S\bfx}_2 \\
        &\geq \frac1{\sqrt{\abs*{S}}}\norm*{\diag(\bfs)\diag(\bfu)^{1/2}\bfA\bfx}_2 \\
        &\geq (1-\eps)\frac1{\sqrt{\abs*{S}}}\norm*{\diag(\bfu)^{1/2}\bfA\bfx}_2 && \text{\eqref{eq:LJ-subspace-embedding}}
    \end{align*}
    Now note that for any $i\in[n]$,
    \begin{align*}
        \bfu_i\angle*{\bfa_i,\bfx}^2 &\leq \bftau_i(\diag(\bfu)^{1/2}\bfA) \norm*{\diag(\bfu)^{1/2}\bfA\bfx}_2^2 && \text{properties of leverage scores} \\
        &\leq (1+\eps)\bfu_i \norm*{\diag(\bfu)^{1/2}\bfA\bfx}_2^2 && \eqref{eq:LJ-leverage-score}
    \end{align*}
    so $\norm*{\bfA\bfx}_\infty \leq (1+\eps)\norm*{\diag(\bfu)^{1/2}\bfA\bfx}_2$. Furthermore, for every $\bfx\in\mathbb R^d$ we have that
    \[
    \norm*{\bfA\vert_S\bfx}_\infty \leq \norm*{\bfA\bfx}_\infty.
    \]
    Combining these bounds yields that
    \[
    \norm*{\bfA\vert_S\bfx}_\infty \leq \norm*{\bfA\bfx}_\infty \leq \frac{1+\eps}{1-\eps}\sqrt{\abs*{S}}\cdot\norm*{\bfA\vert_S\bfx}_\infty
    \]
    for every $\bfx\in\mathbb R^d$. 
\end{proof}

\subsection{Applications: Subspace Embeddings with Large Distortion}
\label{sec:subspace-embeddings-large-dist}

We now obtain several new results on coresets for subspace embeddings using our new notion of well-conditioned coresets. 

\subsubsection{Average Top \texorpdfstring{$k$}{k} Subspace Embeddings}\label{sec:avg-top-k}

We start with a generalization of the $\ell_\infty$ loss known as the \emph{average top $k$ loss}. The various benefits of considering this loss function is studied in depth by \cite{FLYH2017}.

\begin{Definition}[Average Top $k$ Loss \cite{FLYH2017}]
    Let $k\in[n]$. For $\bfy\in\mathbb R^n$, the \emph{average top $k$ loss} is defined as
    \[
    \norm*{\bfy}_{\AT_k} \coloneqq \frac1k \sum_{i=1}^k \abs*{\bfy_{[i]}},
    \]
    where $\bfy_{[i]}$ denotes the $i$th largest entry in $\bfy$, with ties broken arbitrarily. 
\end{Definition}

We obtain the following subspace embedding results:

\begin{Theorem}[Average Top $k$ Subspace Embedding, Small $k$]\label{thm:at-small-k}
    Let $S\subseteq[n]$ be an $O(1)$-approximate $\ell_2$-well-conditioned coreset. Let $k \leq \abs*{S}$. Then for all $\bfx\in\mathbb R^d$, we have that
    \[
    \norm*{\bfA\vert_S\bfx}_{\AT_k} \leq \norm*{\bfA\bfx}_{\AT_k} \leq O(\sqrt{k\abs*{S}})\cdot\norm*{\bfA\vert_S\bfx}_{\AT_k}.
    \]
    For instance, we can use the result of Corollary \ref{cor:l2-well-cond} and set $\abs*{S} = O(d\log\log d)$, so that we obtain an algorithm which samples $O(d\log\log d)$ rows and achieves a distortion of $O(\sqrt{kd\log\log d})$. 
\end{Theorem}
\begin{proof}
    We assume without loss of generality that $\abs*{S}$ is a multiple of $k$, by reading more entries if needed.
    
    By the guarantee of $S$, for each $i\in[n]$, we have that $\bfa_i = \bfA\vert_S\bfc$ for some $\bfc\in\mathbb R^S$ with $\norm*{\bfc}_2^2 = O(1)$. Then for any $\bfx\in\mathbb R^d$ and any $i\in[n]$, we have that
    \[
    \angle*{\bfa_i,\bfx}^2 = \abs*{\sum_{j\in S}\bfc_j\angle*{\bfa_j,\bfx}}^2 \leq \norm*{\bfc}_2^2 \sum_{j\in S}\angle*{\bfa_j,\bfx}^2 = O(1)\sum_{j\in S}\angle*{\bfa_j,\bfx}^2.
    \]
    Now consider partitioning $S$ into $\abs*{S}/k$ subsets of size $k$, say $S = \bigcup_{l=1}^{\abs*{S}/k} S_l$. Then, 
    \[
    \sum_{j\in S}\angle*{\bfa_j,\bfx}^2 = \sum_{l=1}^{\abs*{S}/k}\norm*{\bfA\vert_{S_l}\bfx}_2^2 \leq \sum_{l=1}^{\abs*{S}/k}\norm*{\bfA\vert_{S_l}\bfx}_1^2 \leq \frac{\abs*{S}}{k}\cdot k^2\norm*{\bfA\vert_S\bfx}_{\AT_k}^2
    \]
    so by combining these two bounds, we have that
    \[
    \frac1k \abs*{\angle*{\bfa_i,\bfx}} \leq O(1)\sqrt{\abs*{S}/k}\norm*{\bfA\vert_S\bfx}_{\AT_k}.
    \]
    Summing over the $k$ indices witnessing $\norm*{\bfA\bfx}_{\AT_k}$ yields
    \[
    \norm*{\bfA\bfx}_{\AT_k} \leq O(\sqrt{k\abs*{S}})\cdot\norm*{\bfA\vert_S\bfx}_{\AT_k}. \qedhere
    \]
\end{proof}

\begin{Theorem}[Average Top $k$ Subspace Embedding, Large $k$]\label{thm:at-large-k}
    Let $k \geq k_0$ for some $k_0 = O(d + \log\frac1\delta)$. Let $N_1, N_2, \dots, N_{k/t}$ be a random partition of $[n]$ into $k/t$ parts for some $t = O(d + \log\frac1\delta)$. Let $S_l$ be $O(1)$-approximate $\ell_2$-well-conditioned coresets for $\bfA\vert_{N_l}$ for each $l\in[k/t]$, each of size at most $s$, and let $S = \bigcup_{l=1}^{k/t} S_l$. Then, with probability at least $1-\delta$, simultaneously for all $\bfx\in\mathbb R^d$,
    \[
    \norm*{\bfA\vert_S\bfx}_{\AT_k} \leq \norm*{\bfA\bfx}_{\AT_k} \leq O(\sqrt{ts})\cdot\norm*{\bfA\vert_S\bfx}_{\AT_k}
    \]
    For instance, we can use the result of Corollary \ref{cor:l2-well-cond} and set $s = O(d\log\log d)$, so that we obtain an algorithm which samples $O(k\log\log d)$ rows and achieves a distortion of $O(d\sqrt{\log\log d})$ with constant probability. 
\end{Theorem}
\begin{proof}
    We first fix a vector $\bfx\in\mathbb R^d$ with $\norm*{\bfA\bfx}_{\AT_k} = 1$. By Chernoff bounds, with probability at least $1-\delta / \exp(10d)$, each of the partitions $N_l$ contains at most $O(t)$ elements of the top $k$ entries of $\bfA\bfx$. Then conditioned on this event, by the guarantee of Theorem \ref{thm:at-small-k}, we have that
    \begin{align*}
        \norm*{\bfA\bfx}_{\AT_{k}} &\leq \frac1{k}\sum_{l=1}^{k/t}O(t)\norm*{\bfA\vert_{N_l}\bfx}_{\AT_{O(t)}} \\
        &\leq \frac{1}{k}\sum_{l=1}^{k/t}\sqrt{t\abs*{S_l}}\cdot O(t)\norm*{\bfA\vert_{S_l}\bfx}_{\AT_{O(t)}} \\
        &\leq O(1)\sqrt{ts}\norm*{\bfA\vert_S\bfx}_{\AT_{O(k)}} \\
        &\leq O(1)\sqrt{ts}\norm*{\bfA\vert_S\bfx}_{\AT_{k}}
    \end{align*}
    A standard net argument over the $\norm*{\cdot}_{\AT_k}$-unit ball in the column space of $\bfA$ then completes the argument.
\end{proof}

\subsubsection{Cascaded Norm Subspace Embeddings}

Next, we show several results for embedding a subspace of \emph{matrices} under various cascaded norms, which are matrix norms formed by taking norms of rows using one norm, and then taking another norm of the vector formed by the row norms \cite{JW2009}. 

Our first result is an embedding for the cascaded $(\infty, \norm*{\cdot})$-norm, which takes an arbitrary norm $\norm*{\cdot}$ and outputs the maximum value over the $n$ rows. 

\begin{Theorem}[$(\infty,\norm*{\cdot})$-Subspace Embedding]
    Let $\bfA\in\mathbb R^{n\times d}$ and let $\norm*{\cdot}$ be any norm on $\mathbb R^m$. Define the \emph{cascaded $(\infty, \norm*{\cdot})$-norm} of an $n\times m$ matrix $\bfB$ as 
    \[
    \norm*{\bfB}_{\infty,\norm*{\cdot}} \coloneqq \max_{i=1}^n \norm{\bfe_i^\top\bfB}
    \]
    Let $S\subseteq[n]$ be an $O(1)$-$\ell_2$-well-conditioned coreset. Then, for every $\bfX\in\mathbb R^{d\times m}$,
    \[
    \norm*{\bfA\vert_S\bfX}_{\infty,\norm*{\cdot}} \leq \norm*{\bfA\bfX}_{\infty,\norm*{\cdot}} \leq O(\sqrt{\abs{S}})\norm*{\bfA\vert_S\bfX}_{\infty,\norm*{\cdot}}.
    \]
\end{Theorem}
\begin{proof}
    For any $i\in[n]$, we write $\bfa_i = \bfA\vert_S^\top\bfx$ for $\bfx\in\mathbb R^S$ with $\norm*{\bfx}_2^2 = O(1)$, as given by the definition of an $O(1)$-$\ell_2$-well-conditioned coreset. Then for any $\bfX\in\mathbb R^{d\times m}$, 
    \begin{align*}
        \norm*{\bfa_i^\top\bfX} &= \norm[\Big]{\sum_{j\in S}\bfx_j\bfa_j^\top\bfX} \\
        &\leq \sum_{j\in S}\abs{\bfx_j} \cdot \norm{\bfa_j^\top\bfX} && \text{triangle inequality and homogeneity} \\
        &\leq \parens*{\sum_{j\in S}\abs{\bfx_j}^2}^{1/2} \parens*{\sum_{j\in S}\norm{\bfa_j^\top\bfX}^2}^{1/2} && \text{Cauchy--Schwarz} \\
        &\leq O(\sqrt{\abs{S}}) \cdot \norm*{\bfA\vert_S\bfX}_{\infty,\norm{\cdot}}.
    \end{align*}
    Taking the max over $i\in[n]$ yields the claim.
\end{proof}

This result is perhaps surprising, in that it \emph{directly} sparsifies the structure of the norm. This is in contrast to what we can prove for the cascaded $(p,\norm{\cdot})$-norm, for which we need to embed the $\norm{\cdot}$ norm into the $\ell_2$ norm first, which causes an $O(d)$ factor loss:

\begin{Theorem}
    Let $p \geq 1$. Let $\bfA\in\mathbb R^{n\times d}$ and let $\norm*{\cdot}$ be any norm on $\mathbb R^m$. Define the \emph{cascaded $(p, \norm*{\cdot})$-norm} of an $n\times m$ matrix $\bfB$ as 
    \[
    \norm*{\bfB}_{\infty,\norm*{\cdot}} \coloneqq \parens*{\sum_{i=1}^n \norm{\bfe_i^\top\bfB}^p}^{1/p}
    \]
    Suppose that $\bfS$ satisfies
    \[
    \norm*{\bfA\bfx}_p \leq \norm*{\bfS\bfA\bfx}_p \leq \kappa \norm*{\bfA\bfx}_p
    \]
    for every $\bfx\in\mathbb R^d$. Then, for every $\bfX\in\mathbb R^{d\times m}$,
    \[
    \norm*{\bfS\bfA\vert\bfX}_{p,\norm*{\cdot}} \leq \norm*{\bfA\bfX}_{p,\norm*{\cdot}} \leq O(\kappa d)\norm*{\bfS\bfA\bfX}_{p,\norm*{\cdot}}.
    \]
\end{Theorem}
\begin{proof}
    We first obtain a result for $m = d$ and $\norm{\cdot} = \norm{\cdot}_2$. In this case, Dvoretzky's theorem \cite{Dvo1961} (see also \cite{SW2018}) states that for a $d^{O(p)}\times d$ Gaussian matrix $\bfG$, with constant probability, we have for all $\bfx\in\mathbb R^{d}$ that
    \[
    \norm{\bfx}_{2} = \Theta(1)\norm{\bfG\bfx}_{p}.
    \]
    Then, by the guarantee of $\bfS$,
    \[
    \norm{\bfA\bfX\bfG^\top}_{p,p} \leq \norm{\bfS\bfA\bfX\bfG^\top}_{p,p} \leq \kappa\norm{\bfS\bfA\bfX\bfG^\top}_{p,p}.
    \]
    Note that Dvoretzky's theorem ensures that
    \[
    \norm{\bfS\bfA\bfX\bfG^\top}_{p,p} = \Theta(1)\norm{\bfS\bfA\bfX}_{p,2}
    \]
    and
    \[
    \norm{\bfA\bfX\bfG^\top}_{p,p} = \Theta(1)\norm{\bfA\bfX}_{p,2}
    \]
    for every $\bfX\in\mathbb R^{d\times d}$, so we have that
    \[
    \Omega(1)\norm*{\bfA\bfX}_{p,2} \leq \norm*{\bfS\bfA\bfX}_{p,2} \leq O(\kappa)\norm*{\bfA\bfX}_{p,2}.
    \]
    
    Next, we use L\"owner--John ellipsoids to show that the previous result in fact implies a result for general cascading $(p,\norm{\cdot})$-norms as well, and in $m$ dimensions. Fix an $\bfX\in\mathbb R^{d\times m}$. We then consider the symmetric convex body given by $K \coloneqq \braces{\bfa\in\mathbb R^d : \norm{\bfa^\top\bfX} \leq 1}$. Then, there exists a $\bfH\in\mathbb R^{d\times d}$ such that the ellipsoid $E \coloneqq \braces{\bfa\in\mathbb R^d : \norm{\bfa^\top\bfH}_2 \leq 1}$ satisfies $E\subseteq K\subseteq \sqrt d E$. Note then that by the above guarantee for $\bfS$, we have that
    \begin{align*}
        \norm*{\bfS\bfA\bfX}_{p,\norm{\cdot}} &\leq \norm*{\bfS\bfA\bfH}_{p,2} && \text{$E\subseteq K$} \\
        &\leq O(\kappa)\norm*{\bfA\bfH}_{p,2} \\
        &\leq O(\kappa\sqrt d)\norm*{\bfA\bfX}_{p,\norm{\cdot}} && \text{$K\subseteq \sqrt d E$}  \\
        &\leq O(\kappa\sqrt d)\norm*{\bfA\bfH}_{p,2} && \text{$E\subseteq K$} \\
        &\leq O(\kappa\sqrt d)\norm*{\bfS\bfA\bfH}_{p,2} \\
        &\leq O(\kappa d)\norm*{\bfS\bfA\bfX}_{p,\norm{\cdot}} && \text{$K\subseteq \sqrt d E$} \qedhere
    \end{align*}
\end{proof}

\section{Nearly Optimal Oblivious \texorpdfstring{$\ell_p$}{lp} Subspace Embeddings}
\label{sec:ose}

We first show the following lemma, which shows how to apply Corollary \ref{cor:l2-well-cond}, even when the set of vectors is a whole subspace of points, rather than a finite set. 

\begin{Lemma}[Well-Conditioned Spanning Sets for Subspaces of $\ell_p$]
\label{lem:well-cond-spanning-lp}
Let $p\in(0,\infty)$ and let $\bfA\in\mathbb R^{n\times d}$. There exists $\bfR\in\mathbb R^{d\times s}$ for $s = O(d\log\log d)$ such that $\norm*{\bfA\bfR\bfe_i}_p = 1$, for every $i\in[s]$, and for any $\bfx\in\mathbb R^d$ with $\norm*{\bfA\bfx}_p = 1$, there exists $\bfy\in\mathbb R^s$ such that $\bfA\bfx = \bfA\bfR\bfy$ and $\norm*{\bfy}_2 \leq O(1)$.
\end{Lemma}
\begin{proof}
Our proof proceeds by handling a net over the $\ell_p$ ball in the column space of $\bfA$ using Corollary \ref{cor:l2-well-cond}, and the difference from the net using Lewis bases. 

Let $\eps>0$ be to be determined. By a standard volume argument, there exists a set $\mathcal N\subseteq\mathbb R^d$ such that for every $\bfx\in\mathbb R^d$ with $\norm*{\bfA\bfx}_p = 1$, there exists $\bfx'\in\mathcal N$ such that $\norm*{\bfA\bfx-\bfA\bfx'}_p \leq \eps$, with $\abs{\mathcal N} \leq (1/\eps)^{O(d)}$. We may then apply Corollary \ref{cor:l2-well-cond} to identify a set $\mathcal S_1 \subseteq\mathcal N$ of size at most $s_1 = \abs{\mathcal S_1} \leq O(d\log\log d)$ such that for any $\bfx'\in\mathcal N$, there exists $\bfy\in\mathbb R^{s_1}$ such that $\bfA\bfx' = \bfA\bfR_1\bfy$ and $\norm*{\bfy}_2 \leq O(1)$, where $\bfR_1\in\mathbb R^{d\times s_1}$ is the matrix which enumerates the vectors of $\mathcal S_1$ in its columns. Note that this proves the result for every $\bfx\in\mathcal N$.

We now let $\bfR_2\in\mathbb R^{d\times d}$ be a $\ell_p$ Lewis change of basis matrix for $\bfA$. We then let $\bfR\in\mathbb R^{d\times(s_1+d)}$ be the horizontal concatenation of $\bfR_1$ and $\bfR_2$. Now let $\bfx\in\mathbb R^d$ be any vector with $\norm*{\bfA\bfx}_p = 1$. Then, we can find $\bfx'\in\mathcal N$ that satisfies $\norm*{\bfA\bfx-\bfA\bfx'}_p \leq \eps$. By the prior paragraph, we may write $\bfA\bfx' = \bfA\bfR_1\bfy_1$ for some $\bfy\in\mathbb R^{s_1}$ with $\norm*{\bfy_1}_2 = O(1)$. On the other hand, if we write $\bfx-\bfx'$ uniquely as a linear combination $\bfx-\bfx' = \bfR_2\bfy_2$ of the columns of $\bfR_2$, then  letting $\bfW$ be the diagonal matrix for $\ell_p$ Lewis weights, we have that
\begin{align*}
    \norm*{\bfy_2}_2 &= \norm{\bfW^{1/2-1/p}\bfA\bfR_2\bfy_2}_2 && \text{Definition \ref{def:one-sided-lewis}} \\
    &\leq d^{0\lor(1/2-1/p)}\norm*{\bfA\bfR_2\bfy_2}_p && \text{Lemma \ref{lem:lewis-no-expansion}} \\
    &\leq d^{0\lor(1/2-1/p)} \norm*{\bfA(\bfx-\bfx')}_p \\
    &\leq d^{0\lor(1/2-1/p)} \eps
\end{align*}
Then by taking $\eps = 1/\sqrt d$, we obtain $\norm*{\bfy_2}_2 = O(1)$. Thus, we can write
\[
    \bfA\bfx = \bfA\bfx' + (\bfA\bfx - \bfA\bfx') = \bfA\bfR_1\bfy_1 + \bfA\bfR_2\bfy_2 = \bfA\bfR\bfy
\]
for some $\bfy\in\mathbb R^{s_1+d}$ with
\[
    \norm*{\bfy}_2 \leq \norm*{\bfy_1}_2 + \norm*{\bfy_2}_2 = O(1)
\]
as claimed.
\end{proof}

Given the above lemma, the proof of Theorem \ref{thm:well-cond-for-ose} is immediate:

\begin{proof}[Proof of Theorem \ref{thm:well-cond-for-ose}]
We simply translate the guarantees of Lemma \ref{lem:well-cond-spanning-lp} into that of Theorem \ref{thm:well-cond-for-ose}. First, we take $\bfU = \bfA\bfR$, where $\bfR$ is given by Lemma \ref{lem:well-cond-spanning-lp}. Then, the entrywise $\ell_p$ norm of $\bfU$ is bounded since
\[
    \norm*{\bfU}_{p,p}^p = \sum_{j=1}^s \norm*{\bfU\bfe_j}_p^p = \sum_{j=1}^s \norm*{\bfA\bfR\bfe_j}_p^p = s.
\]
Next, let $\bfx\in\mathbb R^d$ satisfy $\norm*{\bfA\bfx}_p = 1$. Then, by Lemma \ref{lem:well-cond-spanning-lp}, we may identify a $\bfz\in\mathbb R^s$ such that $\bfA\bfx = \bfU\bfz$ and
\[
    \norm*{\bfz}_2 \leq O(1) \leq O(1)\cdot \norm*{\bfA\bfx}_p = O(1)\norm*{\bfU\bfz}_p.
\]
The result for general $\bfx\in\mathbb R^d$ follows by scaling.
\end{proof}

A related result we can obtain is a well-conditioned factorization of a matrix. This result sharpens \cite[Lemma 9]{BRW2021}, who obtained a similar result using Auerbach bases.

\MatrixDecomp*
\begin{proof}
We first write $\bfL = \bfL' \bfD$ where $\bfD\in\mathbb R^{d\times d}$ is the diagonal matrix with $\norm*{\bfL\bfe_j}_p$ as its $j$th diagonal entry, end $\bfL'$ has columns with unit $\ell_p$ norm. Next, we apply Corollary \ref{cor:l2-well-cond} to obtain a set $S\subseteq [d]$ with $s = O(k\log\log k)$ columns such that for each $j\in[d]$, there exists $\bfz\in\mathbb R^s$ with $\norm*{\bfz}_2 \leq O(1)$ such that $\bfL'\bfe_j = \bfL'\vert^S\bfz$. We may then set $\bfU = \bfL'$ and $\bfV^\top = \bfZ\bfD$, where $\bfZ\in\mathbb R^{s\times d}$ is the matrix with $\bfz$ in its columns. This clearly satisfies the conditions of the theorem.
\end{proof}

\subsection{A Proof of Corollary \ref{cor:lp-ose}}
\label{sec:proof-lp-ose}

We provide a proof of Corollary \ref{cor:lp-ose}. For simplicity, we present a simple proof based on the $\ell_1$ embeddings of \cite{SW2011}, which has suboptimal running time. By using techniques shown in \cite{MM2013, WW2019}, it is possible to use a more sophisticated algorithm running in input sparsity time with similar guarantees, by using our Theorem \ref{thm:well-cond-for-ose} in a similar way.

\begin{proof}[Proof of Corollary \ref{cor:lp-ose}]
We take $\bfS\in\mathbb R^{r\times d}$ to be drawn with i.i.d.\ $p$-stable random variables \cite{Nol2005}, scaled by $C/r^{1/p}$ for some large enough constant $C$. For every $(i,j)\in[r]\times[d]$, $\bfe_i^\top\bfS\bfU\bfe_j$ is distributed as $C\norm*{\bfU\bfe_j}_p/r^{1/p}$ times a $p$-stable variable $X_{i,j}$, by definition of $p$-stable variables. With probability at least $1 - 1/\poly(rd)$, $\abs{X_{i,j}}$ is at most $\poly(rd)$, so by a union bound over all $rd$ choices of $(i,j)$, this is true for every $(i,j)\in[r]\times [d]$. Call this event $\mathcal E$. Conditioned on this event, the expectation of $\abs{X_{i,j}}$ is $O(\log(rd))$, so by linearity of expectation, we have
\[
    \E\bracks*{\norm*{\bfS\bfU}_{p,p}^p\vert \mathcal E} = \sum_{i=1}^r\sum_{j=1}^d \E\bracks*{\abs*{\bfe_i^\top\bfS\bfU\bfe_j}^p\vert \mathcal E} \leq O(1)\sum_{i=1}^r \sum_{j=1}^d \frac{\norm*{\bfU\bfe_j}_p^p}{r} = O(\norm*{\bfU}_{p,p}^p \log(rd)).
\]
By Markov's inequality, this bound holds up to constant factors with probability at least $199/200$. We condition on this event. Then, for any $\bfx\in\mathbb R^{n\times d}$, we write $\bfA\bfx = \bfU\bfz$ for $\bfz$ promised by Theorem \ref{thm:well-cond-for-ose}, so that
\begin{align*}
\norm*{\bfS\bfU\bfz}_p^p &= \sum_{i=1}^n \abs*{\bfe_i^\top\bfS\bfU\bfz}^p \\
&\leq \norm*{\bfz}_q^p \sum_{i=1}^n \abs*{\bfe_i^\top\bfS\bfU}_p^p && \text{H\"older's inequality} \\
&\leq \norm*{\bfz}_2^p \norm*{\bfS\bfU}_{p,p}^p \\
&\leq \norm*{\bfU\bfz}_p^p \norm*{\bfS\bfU}_{p,p}^p \\
&\leq \norm*{\bfU\bfz}_p^p O(\norm*{\bfU}_{p,p}^p\log(rd)) \\
&\leq O(d\log(rd)\log\log d)\norm*{\bfU\bfz}_p^p.
\end{align*}
Taking $p$th roots gives the upper inequality.

For the lower inequality, we use \cite[Lemma 2.12]{WW2019}:

\begin{Lemma}[Lemma 2.12 of \cite{WW2019}]
Let $\{X_i\}_{i=1}^n$ be independent $p$-stable random variables. Then for sufficiently large $n$ and $T$,
\[
    \Pr\braces*{\sum_{i=1}^n \abs{X_i}^p \geq L_p n\log\frac{n}{\log T}}\geq 1 - \frac1T
\]
for some constant $L_p$. 
\end{Lemma}

For every $\bfx\in\mathbb R^d$ with $\norm*{\bfA\bfx}_p = 1$, $\norm*{\bfS\bfA\bfx}_p^p$ is the sum of $r$ independent $p$-stable random variables, raised to the $p$ and scaled by $r$. We then apply the above lemma with $n = r$ and $T = \exp(r)$ to conclude that for every $\bfx\in\mathbb R^d$ with $\norm*{\bfA\bfx}_p = 1$, $\norm*{\bfS\bfA\bfx}_p^p \geq 1$ with probability at least $1 - \exp(-r)$, by choosing our constant $C$ large enough. By a standard net argument (see, e.g., \cite{SW2011}), this is true for every $\bfx\in\mathbb R^d$ with $\norm*{\bfA\bfx}_p = 1$. This in turn implies the lower tail inequality for every $\bfx\in\mathbb R^d$ by scaling.
\end{proof}

\section{Sharp Column Subset Selection for \texorpdfstring{$g$}{g}-Norm Low Rank Approximation}
\label{sec:m-css}

In this section, we show our new results on column subset selection for the $g$-norm, also known as $M$-estimators. 

\begin{algorithm}
    \caption{Column Subset Selection for $M$-Estimators}
    \textbf{input:} Input matrix $\bfA\in\mathbb R^{n\times d}$, rank $k$, loss function $g$. \\
    \textbf{output:} Subset $T\subseteq[d]$ of $O(k\log^2 d)$ columns. 
    \begin{algorithmic}[1]
        \State{$T_0 \gets [d]$}
        \State{$s\gets O(k\log\log k)$}\Comment{Given by Corollary \ref{cor:l2-well-cond}}
        \While{$\abs{T_l} \geq 1000s$}
        \State{$t_l\gets 160s\log_2 d_l$}
        \For{$t = 1, 2, \dots, O(\log\log d)$}
        \State{Sample $H\sim\binom{T_l}{t_l}$}
        \State{Let $\bfx^{j}$ minimize $\min_{\bfx}\norm{\bfA\vert^H\bfx-\bfa^j}_g$ up to a $\reg_{g,t_l}$ factor for each $j\in T_l$}
        \State{Let $F_{l,t}$ be the $d_l / 960 = \abs{T_l} / 960$ columns with smallest regression cost $\norm{\bfA\vert^H\bfx^j-\bfa^j}_g$}
        \State{$C_{l,t} \gets \sum_{j\in F_{l,t}}\norm{\bfA\vert^H\bfx^j-\bfa^j}_g$}
        \EndFor{}
        \State{Let $t^*$ be the $t$ with smallest $C_{l,t}$}
        \State{$T_{l+1} \gets T_l \setminus F_{l,t^*}$}
        \EndWhile{}
    \end{algorithmic}\label{alg:swz2019}
\end{algorithm}

\subsection{An Improved Structural Result on Uniform Sampling}

We first give a slight more useful form of Corollary \ref{cor:l2-well-cond} to our setting.

\begin{Lemma}\label{lem:lra-l2-well-conditioned}
    Let $\bfA_*\in\mathbb R^{n\times d}$ be a rank $k$ matrix. Then, there exists a set $S\subseteq[d]$ of size $s = O(k\log\log k)$ such that for every $j\in[d]$,
    \[
    \norm{(\bfA_*\vert^{S})^- \bfa_*^j}_2^2 \leq O(1).
    \]
\end{Lemma}
\begin{proof}
    Since $\bfA_*$ has rank $k$, we can write $\bfA_* = \bfQ\bfR$ for some orthonormal $\bfQ\in\mathbb R^{n\times k}$ and $\bfR\in\mathbb R^{k\times d}$. Then by Corollary \ref{cor:l2-well-cond}, there exists a set $S\subseteq [d]$ of size $s$ such that for every $j\in H\cup\{i\}$, we have that $\norm{(\bfR\vert^{S})^- \bfr^j}_2^2 \leq O(1)$. The result then follows since
    \begin{align*}
        \norm{(\bfA_*\vert^{S})^- \bfa_*^j}_2^2 &= (\bfa_*^j)^\top(\bfA_*\vert^S)^{-\top}(\bfA_*\vert^S)^{-}\bfa_*^j \\
        &= (\bfr^j)^\top\bfQ^\top \bfQ(\bfR\vert^S)^{-\top} (\bfR\vert^S)^- \bfQ^\top\bfQ\bfr^j \\
        &= (\bfr^j)^\top(\bfR\vert^S)^{-\top} (\bfR\vert^S)^- \bfr^j \\
        &= \norm{(\bfR\vert^{S})^- \bfr^j}_2^2.\qedhere
    \end{align*}
\end{proof}

Using Lemma \ref{lem:lra-l2-well-conditioned}, we now obtain the following lemma, which gives an improved version of Lemmas 2.1 and 2.2 of \cite{SWZ2019}. 

\begin{Lemma}\label{lem:improved-swz2019-2.1}
    Let $\bfA\in\mathbb R^{n\times d}$. Let $\bfA_*\in\mathbb R^{n\times d}$ be any rank $k$ matrix and let $\bfD = \bfA - \bfA_*$. Let $s \geq O(k\log\log k)$ and let $H\sim\binom{[d]}{2s}$ and let $i\sim[d]\setminus H$. Let $R = R(H\cup\{i\})$ be the set of size $s$ given by Lemma \ref{lem:lra-l2-well-conditioned} for $\bfA_*\vert^{H\cup\{i\}}$. The following hold:
    \begin{itemize}
        \item With probability at least $1/2$, $i\notin R$
        \item If $i\notin R$, then there is $\bfx\in\mathbb R^H$ such that
        \begin{equation}\label{eq:avg-sq}
            \min_{\bfx\in\mathbb R^H}\norm*{\bfA\vert^H\bfx - \bfa^i}_g^2 \leq O(1)\frac{\ati_{g,s+1}^2}{\lin_g^2}\sum_{j\in H\cup\{i\}}\norm*{\bfd^j}_g^2
        \end{equation}
        \item With probability at least $1/4$ over $H\sim\binom{[d]}{2s}$,
        \[
        \abs*{\braces*{i\in[d]\setminus H : i\notin R(H\cup\{i\})}} \geq \frac{d}{4}
        \]
    \end{itemize}
\end{Lemma}
\begin{proof}
    By symmetry, $i$ is a uniformly random index of $H\cup\{i\}$, so $\Pr\braces{i\notin R} \geq 1 - s/(2s+1) > 1/2$, which gives the first conclusion. 
    
    Let $\alpha_j$ denote the $j$th entry of $(\bfA_*\vert^R)^-\bfa^i_*$ for each $j\in R$ and $\alpha_j = 0$ for $j\in H\setminus R$. We then have that
    \begin{align*}
        \min_{\bfx\in\mathbb R^H} \norm[\Big]{\bfA\vert^H\bfx - \bfa^i}_g &\leq \norm[\Big]{\sum_{j\in H}\alpha_j \bfa^j - \bfa^i}_g \\
        &\leq \norm[\Big]{\sum_{j\in H}\alpha_j (\bfa_*^j + \bfd^j) - (\bfa_*^i + \bfd^i)}_g \\
        &= \norm[\Big]{\sum_{j\in R}\alpha_j \bfd^j - \bfd^i}_g && \text{since $\bfA_*\vert^R(\bfA_*\vert^R)^-\bfa_*^i = \bfa_*^i$} \\
        &\leq \ati_{g,s+1}\parens[\Big]{\sum_{j\in R}\norm{\alpha_j \bfd^j}_g + \norm{\bfd^i}_g} && \text{approximate triangle inequality} \\
        &\leq \frac{\ati_{g,s+1}}{\lin_g}\parens[\Big]{\sum_{j\in R}\alpha_j\norm{\bfd^j}_g + \norm{\bfd^i}_g} && \text{at least linear growth} \\
        &\leq \frac{\ati_{g,s+1}}{\lin_g}\parens[\Big]{\parens[\Big]{\sum_{j\in R}\alpha_j^2}^{1/2}\parens[\Big]{\sum_{j\in R}\norm{\bfd^j}_g^2}^{1/2} + \norm{\bfd^i}_g} && \text{Cauchy--Schwarz} \\
        &\leq O(1)\frac{\ati_{g,s+1}}{\lin_g}\parens[\Big]{\parens[\Big]{\sum_{j\in R}\norm{\bfd^j}_g^2}^{1/2} + \norm{\bfd^i}_g}.
    \end{align*}
    Squaring both sides yields the second conclusion. 
    
    The third conclusion follows from the same proof as Lemma 2.2 of \cite{SWZ2019}.
\end{proof}

\subsection{Sharper Guarantees for the \texorpdfstring{\cite{SWZ2019}}{[SWZ2019]} Algorithm}

We now use the result of Lemma \ref{lem:improved-swz2019-2.1} to improve the analysis of the \cite{SWZ2019} algorithm. 

\subsubsection{Level Sets}
Let $\bfA = \bfA_* + \bfDelta$, where $\bfA_*$ is the best rank $k$ approximation in the $g$-norm. Let the columns of $\bfDelta$ be $\bfdelta^1, \bfdelta^2, \dots, \bfdelta^d$. To gain fine-grained control over the costs of the columns, we will need to consider a partition of the columns into $O(\log d)$ level sets based on $\norm{\bfdelta^j}_g$. 

\begin{Definition}\label{def:useful}
    Let $l\in\mathbb N$. Then:
    \begin{itemize}
        \item Let $s = O(k\log\log k)$ denote the maximum size of an $\ell_2$-well-conditioned subset given by Corollary \ref{cor:l2-well-cond} in $k$ dimensions.
        \item Let $T_l\subseteq[d]$ denote the subset of columns surviving after the $l$th round of the algorithm. We assume without loss of generality that $T_l = [d_l]$ for some $d_l \leq d$. Furthermore, we assume without loss of generality that $\norm{\bfdelta^1}_g \geq \norm{\bfdelta^2}_g \geq \dots \geq \norm{\bfdelta^{d_l}}_g$.
        \item Let $\Res_l \coloneqq \sum_{j=d_l/4}^{d_l} \norm{\bfdelta^j}_g$ denote the residual cost, after restricting to the surviving columns and after removing the columns with cost in the top quarter.
        \item Let 
        \[
        R_l^i \coloneqq \begin{cases}
            \braces*{j\in[d_l]\setminus[d_l/4] : \norm{\bfdelta^j}_g \leq \frac1{d_l^2}\Res_l} & \text{if $i = \infty$} \\
            \braces*{j\in[d_l]\setminus[d_l/4] : 2^{-i} \cdot \Res_l < \norm{\bfdelta^j}_g \leq 2^{-i+1} \cdot \Res_l} & \text{if $0 < i < 2\log_2 d_l$}
        \end{cases}
        \]
    \end{itemize}
\end{Definition}

Recall that our goal is to show that with constant probability, the $d_l / 80$ columns with the smallest regression cost when fit on $\bfA\vert^H$ each have a cost of at most $O(\sqrt{k\log\log k})\Res_l / d_l$. We first show that we may assume with out loss of generality that $R_l^\infty$ is small in cardinality. 

\begin{Lemma}\label{lem:Rlinf-small}
    If $\abs*{R_l^\infty} > d_l / 4$, then with probability at least $1/6$ over the randomness of $H$,
    \[
    \abs*{\braces*{j\in T_l : \min_{\bfx\in\mathbb R^H}\norm*{\bfA\vert^H\bfx - \bfa^j}_g \leq \frac{1}{d_l}\Res_l}} \geq \frac1{80} d_l
    \]
\end{Lemma}
\begin{proof}
    Note that $\E\abs{R_l^\infty\cap H} \geq 20s$. By Chernoff bounds, with probability at least $99/100$, we have that $\abs*{R_l^\infty \cap H} \geq 4s \geq 2k$. Then by conditioning on the size of $R_l^\infty \cap H$, we can apply the same proof from Lemma 2.5 of \cite{SWZ2019} restricted to $R_l^\infty$ to show that with probability at least $1/5 - 1/100 \geq 1/6$ over the randomness of $H$,
    \[
    \abs*{\braces*{j\in T_l : \min_{\bfx\in\mathbb R^H}\norm*{\bfA\vert^H\bfx - \bfa^j}_g \leq \frac{\abs{H}}{d_l^2}\Res_l}} \geq \frac1{20} \abs*{R_l^\infty} \geq \frac1{20}\cdot \frac{d_l}{4} = \frac1{80} d_l.
    \]
    Note that $\abs{H}\leq d_l$, which gives the claimed result.
\end{proof}

By Lemma \ref{lem:Rlinf-small}, we may assume that $\abs{R_l^\infty} \leq d_l / 4$. In this case, we show that we must have many columns which belong to a large level set.

\begin{Lemma}\label{lem:large-level-sets}
    Suppose that $\abs{R_l^\infty}\leq d_l/4$. Then, at least $d_l / 4$ columns belong to a level set $R_l^i$ such that $\abs{R_l^i} \geq d_l / 8\log_2 d_l$. 
\end{Lemma}
\begin{proof}
    Note that the number of columns which can belong in a level set of size less than $d_l / 8\log_2 d_l$ is less than
    \[
    2(\log_2 d_l) \cdot \frac{d_l}{8\log_2 d_l} = \frac{d_l}{4}
    \]
    since there are only $2\log_2 d_l$ level sets. Since there are at most $d_l/4$ columns in $R_l^\infty$ and at most $d_l/4$ that are excluded for being in the top quarter, we conclude as desired.
\end{proof}

\subsubsection{Fitting a Constant Fraction of Columns}

We will now show that we can fit a constant fraction of columns in a large level set with small cost. We first show the following lemma for a single level set:

\begin{Lemma}\label{lem:swz2019-2.3-2.4-analogue}
    Let $i\in[2\log_2 d_l]$ be such that $\abs{R_l^i}\geq d_l/8\log_2 d_l$. Then, with probability at least $1/6$, there are at least $\abs{R_l^i}/20$ indices $j\in R_l^i$ such that there exists $\bfx$ satisfying
    \[
    \min_{\bfx\in\mathbb R^H}\norm*{\bfA\vert^H\bfx - \bfa^{j'}}_g \leq O(\sqrt s)\frac{\ati_{g,s+1}}{\lin_g}\frac{\Res_l}{2^{i}}
    \]
\end{Lemma}
\begin{proof}
    The proof is based on adapting Lemmas 2.3, 2.4, and 2.5 of \cite{SWZ2019}. 
    
    Note that $\bfE\abs{R_l^i\cap H} \geq 20s$. By Chernoff bounds, with probability at least $99/100$, we have that $\abs{R_l^\infty\cap H} \geq 4s$. We condition on this event. Then, let $H'\subseteq R_l^i\cap H$ be a uniformly random subset of $R_l^i\cap H$ of size $2s$. Then by Markov's inequality, 
    \begin{align*}
        \Pr_{H'}\braces*{\sum_{j\in H'}\norm*{\bfdelta^j}_g^2 \geq 40\frac{s}{\abs{R_l^i}}\sum_{j\in R_l^i} \norm*{\bfdelta^j}_g^2} &\leq \frac{\E\bracks*{\sum_{j\in H'}\norm*{\bfdelta^j}_g^2}}{40\frac{s}{\abs{R_l^i}}\sum_{j\in R_l^i} \norm*{\bfdelta^j}_g^2} \leq \frac{\frac{2s}{\abs{R_l^i}}\sum_{j\in R_l^i}\norm*{\bfdelta^j}_g^2}{40\frac{s}{\abs{R_l^i}}\sum_{j\in R_l^i} \norm*{\bfdelta^j}_g^2} \leq \frac{1}{20}
    \end{align*}
    Furthermore, by an averaging argument, we have that
    \[
    \abs[\Bigg]{\braces[\Bigg]{j'\in R_l^i : \norm*{\bfdelta^{j'}}_g^2 \geq \frac{5}{\abs{R_l^i}}\sum_{j\in R_l^i}\norm*{\bfdelta^j}_g^2}} \leq \frac15 \abs{R_l^i}
    \]
    
    Now note that $H'$ is a uniformly random subset of $R_l^i$ of size $2s$. Then, by Lemma \ref{lem:improved-swz2019-2.1}, we have that with probability at least $1/4$, there are at least $\abs{R_l^i} / 4$ indices $j'\in R_l^i$  for which \eqref{eq:avg-sq} holds. Thus, for at least $\abs{R_l^i}/4 - \abs{R_l^i}/5 = \abs{R_l^i}/20$ indices $j'\in R_l^i$, we have that
    \begin{align*}
        \min_{\bfx\in\mathbb R^{H'}}\norm*{\bfA\vert^{H'}\bfx - \bfa^{j'}}_g^2 &\leq O(1)\frac{\ati_{g,s+1}^2}{\lin_g^2}\sum_{j\in H'\cup\{j'\}}\norm*{\bfdelta^j}_g^2 && \text{Lemma \ref{lem:improved-swz2019-2.1}} \\
        &\leq O(1)\frac{\ati_{g,s+1}^2}{\lin_g^2}\frac{20s + 5}{\abs{R_l^i}}\sum_{j\in R_l^i} \norm*{\bfdelta^j}_g^2 && \text{Lemma \ref{lem:swz2019-2.3-2.4-analogue}} \\
        &\leq O(1)\frac{\ati_{g,s+1}^2}{\lin_g^2}\frac{s}{2^{2i}}\Res_l^2 && \text{Definition \ref{def:useful}}
    \end{align*}
    By padding $\bfx$ with zeros on $H\setminus H'$ and taking square roots, we get the desired result.
\end{proof}

Next, we apply an averaging argument to show that if we sum across all large level sets, we fit a constant fraction of columns all $d_l$ with constant probability. 

\begin{Lemma}\label{lem:fit-constant-frac}
    Suppose that $\abs{R_l^\infty} \leq d_l / 4$. Then with probability at least $1/960$, there is a set of size $F\subseteq[d_l]$ such that $\abs{F} \geq d_l / 960$ and
    \[
    \sum_{j\in F}\min_{\bfx\in\mathbb R^H}\norm{\bfA\vert^H \bfx-\bfa^j}_g \leq O(\sqrt s)\frac{\ati_{g,s+1}}{\lin_g}\cdot\Res_l
    \]
\end{Lemma}
\begin{proof}
    By Lemma \ref{lem:swz2019-2.3-2.4-analogue}, for a fixed level set $i$ with $\abs{R_l^i} \geq d_l / 8\log_2 d_l$, with probability at least $1/6$, we fit at least $\abs{R_l^i}/20$ columns with cost at most
    \[
    O(\sqrt s)\frac{\ati_{g,s+1}}{\lin_g}\frac{\Res_l}{2^{i}}
    \]
    each. Then, let $X_i$ be the random variable that represents the number of such columns in $R_l^i$, and define
    \[
    X\coloneqq \sum_{i : \abs{R_l^i} \geq d_l / 8\log_2 d_l} X_i
    \]
    Note then that
    \[
    \E[X] \geq \sum_{i : \abs{R_l^i} \geq d_l / 8\log_2 d_l}\frac16\cdot\frac1{20}\abs{R_l^i} \geq \frac1{6\cdot 20\cdot 4} d_l = \frac1{480} d_l
    \]
    where the last inequality is by Lemma \ref{lem:large-level-sets}. 
    Then by a standard averaging argument,
    \begin{align*}
        \frac1{480} d_l &\leq d_l \cdot \Pr\{X \geq d_l / 960\} + \frac{d_l}{960}\Pr\{X < d_l / 960\} \\
        &\leq d_l \cdot \Pr\{X \geq d_l / 960\} + \frac{d_l}{960}
    \end{align*}
    so $X$ is at least $d_l / 960$ with probability at least $1/960$. Furthermore, the total cost of all of the columns which are fit well is at most
    \[
    \sum_i O(\sqrt s)\frac{\ati_{g,s+1}}{\lin_g}\frac{\Res_l}{2^{i}} \cdot \abs{R_l^i} \leq O(\sqrt s)\frac{\ati_{g,s+1}}{\lin_g}\cdot\Res_l.\qedhere
    \]
\end{proof}

\subsubsection{Proof of Theorem \ref{thm:improved-swz2019}}

We now give proofs for the various guarantees of our algorithm. 

\begin{proof}[Proof of Theorem \ref{thm:improved-swz2019}]
    Note first that the algorithm decreases the size of $T_l$ by a $(1 - 1/960)$ factor at each iteration. Thus, the algorithm makes at most $L = O(\log d)$ iterations of the outer loop. By Lemma \ref{lem:fit-constant-frac}, we have a constant probability of success of choosing $d_l / 960$ columns such that the total cost is at most
    \[
    O(\sqrt s)\frac{\ati_{g,s+1}}{\lin_g}\cdot\Res_l.
    \]
    Since we repeat $O(\log L) = O(\log\log d)$ times and use an $\reg_{g,t_l}$-approximate regression algorithm, we with probability at least $1 - 1/100L$, we find $d_l / 960$ columns $F_l\subseteq T_l$ and corresponding coefficients $\bfX$ such that
    \[
    \norm*{\bfA\vert^{F_l} - \bfA\vert^{S_l}\bfX}_g \leq O(\sqrt s)\frac{\reg_{g,t_l}\cdot\ati_{g,s+1}}{\lin_g}\cdot\Res_l.
    \]
    Thus, our total cost is
    \[
    \sum_{l=1}^{O(\log d)}O(\sqrt s)\frac{\reg_{g,t_l}\cdot\ati_{g,s+1}}{\lin_g}\cdot\Res_l.
    \]
    Finally, as argued in \cite{SWZ2019, MW2021}, we show that $\sum_l \Res_l = O(\norm{\bfDelta}_g)$. Note that if a column $j$ contributes to $\Res_l$, then it must be in the bottom $3/4$ fraction of the $\norm{\bfdelta^j}_g$ in round $l$. Then since the bottom $1/960$ fraction of $\norm{\bfdelta^j}_g$ is fitted and removed in each round, $\norm{\bfdelta^j}_g$ can only contribute to $\Res_l$ in $O(1)$ rounds. Thus, the sum is bounded by $O(1)\sum_j \norm{\bfdelta^j}_g = O(\norm{\bfDelta}_g)$. 
    
    The total number of columns selected is $O(s\log d)$ in each of the $O(\log d)$ rounds, for a total of $O(s\log^2 d)$. 
\end{proof}

\section{Huber Column Subset Selection}
\label{sec:huber}

For the important case of the Huber loss, the result of Theorem \ref{thm:improved-swz2019} only yields a distortion of $\tilde O(k^{3/2})$, due to a $k$ factor loss from the approximate triangle inequality term. We further optimize our argument specifically for the Huber loss and obtain a distortion of $O(k)$ instead.

\HuberCSS*

Our improvement comes from the following structural result, which yields Theorem \ref{thm:huber-css} when combined with Theorem \ref{thm:generalized-mw2021}:

\begin{Lemma}
    Let $\bfA\in\mathbb R^{n\times d}$ and let $\bfA_*$ denote the optimal rank $k$ approximation to $\bfA$ in the entrywise Huber norm. Then, there exists a set $S\subseteq[d]$ of $O(k\log\log k)$ columns of $\bfA$ and $\bfX\in\mathbb R^{S\times d}$ such that
    \[
    \norm*{\bfA - \bfA\vert^S \bfX}_{H} \leq O(d)\norm*{\bfA - \bfA_*}_{H}.
    \]
\end{Lemma}
\begin{proof}
    Let $S\subseteq[d]$ be an $\ell_2$-well-conditioned coreset for the columns of $\bfA_*$, given by Corollary \ref{cor:l2-well-cond}. For each $j\notin S$, we let the $j$th column of $\bfX$ be the coefficient vector for fitting $\bfa_*^j$ by $\bfA_*\vert^S$. 
    
    Following \cite[Lemma 37]{CW2015b}, we have that for any $\bfx\in\mathbb R^d$,
    \[
    H(\norm*{\bfx}_2) \leq \frac{\norm*{\bfx}_2^2}{\norm*{\bfx}_\infty^2}H(\norm*{\bfx}_\infty) = \sum_{j=1}^d \frac{\bfx_i^2}{\norm*{\bfx}_\infty^2}H(\norm*{\bfx}_\infty) \leq \sum_{j=1}^d H(\bfx_i) = \norm{\bfx}_H.
    \]
    Then,
    \begin{align*}
        \norm*{\bfA - \bfA\vert^S\bfX}_H &= \norm*{(\bfA^* + \bfDelta) - (\bfA^* + \bfDelta)\vert^S\bfX}_H \\
        &= \norm*{\bfDelta - \bfDelta\vert^S\bfX}_H \\
        &\leq O(1)(\norm*{\bfDelta}_H + \norm*{\bfDelta\vert^S\bfX}_H) \\
    \end{align*}
    so it suffices to bound $\norm*{\bfDelta\vert^S\bfX}_H$. We have
    \begin{align*}
        \norm*{\bfDelta\vert^S\bfX}_H &= \sum_{j=1}^d \sum_{i=1}^n H(\bfe_i^\top\bfDelta\vert^S\bfx^j) \\
        &\leq \sum_{j=1}^d \sum_{i=1}^n H(\norm*{\bfe_i^\top\bfDelta\vert^S}_2\norm*{\bfx^j}_2) && \text{Cauchy--Schwarz} \\
        &\leq O(1)\sum_{j=1}^d \sum_{i=1}^n H(\norm*{\bfe_i^\top\bfDelta\vert^S}_2) \\
        &\leq O(1)\sum_{j=1}^d \sum_{i=1}^n \norm*{\bfe_i^\top\bfDelta\vert^S}_H \\
        &\leq O(1)\sum_{j=1}^d \norm*{\bfDelta\vert^S}_H \\
        &\leq O(d) \norm*{\bfDelta}_H
    \end{align*}
    as claimed.
\end{proof}

\section{\texorpdfstring{$\ell_p$}{lp} Column Subset Selection, \texorpdfstring{$p>2$}{p>2}}
\label{sec:lp-css}

We improve the analysis of column subset selection algorithms which select more than $k$ columns, by showing a randomized polynomial time algorithm for selecting $O(k\log d)$ columns with a distortion of $O(k^{1/2-1/p})$. This improves the algorithms of \cite{CGKLPW2017, DWZZR2019} in this regime and circumvents the lower bound of $\Omega(k^{1-1/p})$ distortion for selecting exactly $k$ columns.

\subsection{Improved Existential Result for Bicriteria Column Subset Selection}

Our main improvement comes from the following lemma, which is inspired by the techniques of \cite{MW2021} and our active $\ell_p$ regression techniques with large distortion. Note that the proof techniques of \cite{SWZ2017} and \cite[Theorem 2.4]{MW2021} do not apply for this result, since they use $p$-stable random variables, which do not exist for $p>2$.

\begin{Lemma}\label{lem:lp-columns}
    Let $2\leq p\leq\infty$. Let $\bfA\in\mathbb R^{n\times d}$ and let $\bfA_*$ denote the optimal rank $k$ approximation to $\bfA$ in the entrywise $\ell_p$ norm. Then, there exists a set $S\subseteq[d]$ of $O(k)$ columns of $\bfA$ and $\bfR\in\mathbb R^{k\times d}$ such that
    \begin{equation}\label{eq:lp-css-guarantee-exist}
        \norm*{\bfA - \bfA\vert^S \bfR}_{p,p} \leq O(k^{1/2-1/p})\norm*{\bfA - \bfA_*}_{p,p}.
    \end{equation}
\end{Lemma}
\begin{proof}
    Let $\bfA_* = \bfU\bfV^\top$ for some $\bfU\in\mathbb R^{n\times k}$ and $\bfV^\top\in\mathbb R^{k\times d}$. Now let $\bfw$ be the $\ell_p$ Lewis weights of $\bfV$ and let $\hat\bfX$ minimize
    \[
    \min_{\bfX\in\mathbb R^{n\times k}}\norm*{(\bfA - \bfX\bfV^\top)\bfW^{1/2-1/p}}_{p,2}
    \]
    up to a factor of $2$. We have
    \begin{align*}
        \norm{\bfA - \hat\bfX\bfV^\top}_{p,p} &\leq \norm{\bfA - \bfU\bfV^\top}_{p,p} + \norm{\bfU\bfV^\top - \hat\bfX\bfV^\top}_{p,p} && \text{Theorem \ref{thm:lewis-reweighting}} \\
        &\leq \norm{\bfA - \bfU\bfV^\top}_{p,p} + \norm{(\bfU\bfV^\top - \hat\bfX\bfV^\top)\bfW^{1/2-1/p}}_{p,2} \\
        &\leq \norm{\bfA - \bfU\bfV^\top}_{p,p} + \norm{(\bfU\bfV^\top - \bfA)\bfW^{1/2-1/p}}_{p,2} \\
        &\hspace{5em}+ \norm{(\bfA - \hat\bfX\bfV^\top)\bfW^{1/2-1/p}}_{p,2} \\
        &\leq \norm{\bfA - \bfU\bfV^\top}_{p,p} + 3\norm{(\bfU\bfV^\top - \bfA)\bfW^{1/2-1/p}}_{p,2} && \text{near optimality} \\
        &\leq \norm{\bfA - \bfU\bfV^\top}_{p,p} + 3k^{1/2-1/p}\norm{\bfU\bfV^\top - \bfA}_{p,p} && \text{Lemma \ref{lem:lewis-no-expansion}} \\
        &= O(k^{1/2-1/p})\norm{\bfA - \bfU\bfV^\top}_{p,p}.
    \end{align*}
    Thus, we have reduced the problem to an $\ell_2$ problem, at a cost of $O(k^{1/2-1/p})$ distortion. Lemma 27 of \cite{CW2015b} then shows that if $\bfS^\top$ is an $\ell_2$ sparsifier for $\bfV^\top\bfW^{1/2-1/p}$ which samples $O(k)$ columns (see \cite[Lemma C.25]{SWZ2019b}, based on \cite[Theorem 3.1]{BSS2012}), then a minimizer $\hat\bfU$ of
    \[
    \min_{\bfX\in\mathbb R^{n\times k}}\norm*{(\bfA - \bfX\bfV^\top)\bfW^{1/2-1/p}\bfS^\top}_{p,2}
    \]
    satisfies
    \[
    \norm*{(\bfA - \hat\bfU\bfV^\top)\bfW^{1/2-1/p}}_{p,2} \leq 2\min_{\bfX\in\mathbb R^{n\times k}}\norm*{(\bfA - \bfX\bfV^\top)\bfW^{1/2-1/p}}_{p,2}.
    \]
    It follows that
    \[
    \norm{\bfA - \hat\bfU\bfV^\top}_{p,p} \leq O(k^{1/2-1/p})\norm{\bfA - \bfU\bfV^\top}_{p,p}.
    \]
    Finally, note that $\hat\bfU$ can be written as
    \[
    \hat\bfU = \bfA\bfW^{1/2-1/p}\bfS^\top(\bfV^\top\bfW^{1/2-1/p}\bfS^\top)^-.
    \]
    Thus, there exists an $O(k^{1/2-1/p})$-approximate solution with a left factor formed by $O(k)$ columns of $\bfA$.
\end{proof}

With Lemma \ref{lem:lp-columns} in hand, we can now apply Theorem \ref{thm:generalized-mw2021} to obtain the following:

\begin{Theorem}\label{thm:lp-css}
    Let $2\leq p < \infty$. Let $\bfA\in\mathbb R^{n\times d}$ and let $k\geq 1$. There is an algorithm which outputs a subset $S\subseteq[d]$ of $\abs{S} = O(k\log d)$ columns and $\bfX\in\mathbb R^{S\times d}$ such that
    \begin{align*}
        \norm*{\bfA - \bfA\vert^S\bfX}_{p,p} &\leq O(k^{1/2-1/p})\min_{\rank(\hat\bfA) \leq k}\norm{\bfA-\hat\bfA}_{p,p}.
    \end{align*}
\end{Theorem}

We note that by setting $p = O(\log n)$, we also obtain a result for $p = \infty$.

\begin{Theorem}\label{thm:linf-css}
    Let $\bfA\in\mathbb R^{n\times d}$ and let $k\geq 1$. There is an algorithm which outputs a subset $S\subseteq[d]$ of $\abs{S} = O(k\log d)$ columns and $\bfX\in\mathbb R^{S\times d}$ such that
    \begin{align*}
        \norm*{\bfA - \bfA\vert^S\bfX}_{\infty,\infty} &\leq O(k^{1/2})\min_{\rank(\hat\bfA) \leq k}\norm{\bfA-\hat\bfA}_{\infty,\infty}.
    \end{align*}
\end{Theorem}

\subsection{Lower Bound}

We give an impossibility result for $\ell_\infty$ column subset selection, showing that our new result for $\ell_\infty$ subset selection is approximately tight.

Our result is based on a variation on the ideas of Theorem 1.4 of \cite{SWZ2017}.

\begin{Definition}[Hard distribution]\label{def:hard-linf-css}
    Let $c\geq 1$ be any constant and let $r = k^{c}$. We then define a distribution $\mathcal D$ over $(k + 2^r) \times r$ matrix as follows. We let the first $k$ rows have entries drawn independently from $\mathcal N(0,\bfI_r)$ and scaled by $k$, and we let the last $2^r$ rows be the $2^r$ vectors in $\{\pm1\}^r$. 
\end{Definition}

We will argue that with high probability, no matrix in the column span of $r/2$ columns of $\bfA\sim\mathcal D$ can approximate $\bfA$ by better than a $\sqrt k$ factor. The optimal rank $k$ approximation of any matrix drawn from the distribution in Definition \ref{def:hard-linf-css} has $\ell_\infty$ has cost at most $1$, by setting the rank $k$ approximation to be the first $k$ rows:

\begin{Lemma}\label{lem:opt-cost-linf}
    Let $\bfA\sim\mathcal D$ for $\mathcal D$ defined in Definition \ref{def:hard-linf-css}. Then, with probability $1$,
    \[
    \min_{\rank(\hat\bfA)\leq k} \norm{\bfA - \hat\bfA}_{\infty,\infty} \leq 1.
    \]
\end{Lemma}

Furthermore, the addition of the $2^r$ hypercube vectors to the matrix gives the following property:

\begin{Lemma}\label{lem:linf-align}
    Let $S\subseteq[r]$. Then, for any $\bfX\in\mathbb R^{S\times r}$,
    \[
    \norm*{\bfA - \bfA\vert^S \bfX}_{\infty,\infty} \geq \max_{j=1}^r\norm{\bfX\bfe_j}_1  - 1
    \]
\end{Lemma}
\begin{proof}
    Let $j\in[r]$. Then, there exists a row $i$ of $\bfA\vert^S$ such that for each $j'\in S$, $\bfA_{i,j'} = \sgn(\bfX_{j',j})$, since $\bfA$ contains all sign vectors. Thus,
    \[
    \bfe_i^\top\bfA\vert^S\bfX\bfe_j = \sum_{j'\in S}\bfA_{i,j'}\bfX_{j',j} = \sum_{j'\in S}\sgn(\bfX_{j',j})\bfX_{j',j} = \norm*{\bfX\bfe_j}_1.
    \]
    On the other hand, $\bfA$ has absolute value at most $1$ on this coordinate, thus yielding the claim.
\end{proof}

With these insights in hand, the proof now essentially follows that of \cite[Theorem G.28]{SWZ2017}; it is shown in \cite{SWZ2017} that if $\bfx\in\mathbb R^S$ fits the first $k$ rows well in $\ell_1$ norm, then it must satisfy $\norm{\bfx}_1 = \Omega(k^{0.5 - o(1)})$. Since we scale the first $k$ rows by $k$, this means that we either have a high $\ell_\infty$ cost in the first $k$ rows, or a high $\ell_\infty$ cost in the bottom $2^r$ rows.

\begin{Theorem}\label{thm:linf-css-lb}
    Let $\alpha\in(0,0.5)$, $k\in\mathbb N$, and $r = \poly(k)$. Then, there exists a $(k + r)\times r$ matrix $\bfA$ such that 
    \[
    \min_{\rank(\hat\bfA)\leq k} \norm{\bfA - \hat\bfA}_{\infty,\infty} \leq 1
    \]
    and for any $S\subseteq[r]$ with $\abs{S}\leq r/2$,
    \[
    \min_{\bfX\in\mathbb R^{S\times r}}\norm{\bfA - \bfA\vert^S\bfX}_{\infty,\infty} \geq \Omega(k^{0.5-\alpha}).
    \]
\end{Theorem}
\begin{proof}
    The proof closely follows \cite[Theorem G.28]{SWZ2017}. For $\bfB\sim\mathcal N(0,1)^{k\times s}$ and scalars $\beta,\gamma>0$, we say the event $\mathcal E(\bfB,\beta,\gamma)$ holds if
    \begin{itemize}
        \item $\norm*{\bfB}_2 \leq O(\sqrt s)$
        \item $\bfB\bfx$ has at most $O(k/\log k)$ coordinates with absolute value at least $\Omega(1/\log k)$, whenever $\norm*{\bfx}_1 \leq O(k^\gamma)$ and $\norm*{\bfx}_\infty \leq O(k^{-\beta})$
    \end{itemize}
    (see \cite[Definition G.19]{SWZ2017}). It is shown in \cite[Lemma G.20]{SWZ2017} that if $k\leq s \leq \poly(k)$, $\beta>\gamma>0$, and $\beta+\gamma<1$, then $\Pr\{\mathcal E(\bfB,\beta,\gamma)\} \geq 1- \exp(-\Theta(k))$. We will apply this to the first $k$ rows $\bfA\vert_{[k]}$ of $\bfA$ scaled down by $k$, as well as to restrictions $\bfA\vert_{[k]}^S$ of these rows to columns $S\subseteq[r]$. 
    
    It is shown in \cite[Claim G.29]{SWZ2017} that for any $S\subseteq[r]$,
    \[
    \Pr\braces*{\mathcal E\parens*{\frac1k\bfA\vert_{[k]}^S, 0.5 + \alpha/2, 0.5-\alpha} \Big\vert \mathcal E\parens*{\frac1k\bfA\vert_{[k]}, 0.5 + \alpha/2, 0.5-\alpha}} = 1
    \]
    We thus condition on $\mathcal E(\frac1k\bfA\vert_{[k]}, 0.5 + \alpha/2, 0.5-\alpha)$, which implies $\mathcal E(\frac1k\bfA\vert_{[k]}^S, 0.5 + \alpha/2, 0.5-\alpha)$ for every $S\subseteq[r]$. Then by \cite[Lemma G.22]{SWZ2017}, for any $S\subseteq[r]$ of size at most $r/2$, with probability at least $1 - \exp(-\Theta(rk))$, a constant fraction of the $r/2$ remaining rows $l\in [r]\setminus S$ satisfies that
    \[
    \min_{\bfx\in\mathbb R^S}\norm*{\frac1k\bfA_{[k]}\vert^S\bfx - \bfA\bfe_l}_1 + \norm*{\bfx}_1 = \Omega(k^{0.5-\alpha})
    \]
    By relating the $\ell_1$ and $\ell_\infty$ norms up to a factor of $k$ for the first term and by using Lemma \ref{lem:linf-align} for the second term, this gives a lower bound of $\Omega(k^{0.5-\alpha})$ on some entry of $\bfA - \bfA\vert^S\bfX$ for any $\bfX$, for this fixed $S$. The failure rate of $\exp(-\Theta(rk))$ is small enough for us to union bound over all choices of $S\subseteq[r]$ of size at most $r/2$, thus giving the theorem.
\end{proof}

\subsection{\texorpdfstring{$(1+\eps)$}{(1+eps)}-Approximate Bicriteria Algorithms}

In a recent work of \cite{BRW2021}, additive approximations for low rank approximation in the entrywise $\ell_p$ norm. In this section, we sharpen their argument using our well-conditioned spanning set result from Section \ref{sec:well-cond} and combine it with our results from earlier in this section to obtain relative error $(1+\eps)$ approximations for this problem. While the main focus of the work of \cite{BRW2021} is on weighted low rank approximation, we specialize our discussion to standard low rank approximation. Our improvements apply to the weighted case as well.

While we show the overall idea and a complete proof of the key lemma for our improvement, we refer several lemmas which can be stated verbatim to \cite{BRW2021}.

\subsubsection{Improved Algorithm for Additive Error}

The algorithm of \cite{BRW2021} is based on an iterative process which updates approximations $\bfx^j_{(t)}$ for each column $\bfa^j\in\mathbb R^n$ of $\bfA$. At each iteration $t$, the approximations $\bfx^j_{(t)}$ for $j\in[d]$ are updated along a single direction $\bfz\in\mathbb R^n$, which is obtained by approximately solving for a vector witnessing the $p\mapsto 2$ singular vector of a matrix. Then to update the approximation $\bfx^j_{(t)}$ for each column, the algorithm proceeds as follows:
\begin{itemize}
\item solve for the minimizer of $\eta\mapsto \norm{\bfa^j - (\bfx^j_{(t)} + \eta\bfz)}_p$
\item set $\bfx' \gets \bfx^j_{(t)} + \eta\bfz$
\item solve for the minimizer of $\eta'\mapsto \norm{\bfa^j - \eta'\bfx'}_p$
\item set $\bfx^j_{(t+1)} \gets \eta'\bfx'$
\end{itemize}
Let $\bfX_{(t)}$ denote the matrix with $\bfx^j_{(t)}$ in its columns. The main argument of \cite{BRW2021} is based on a lemma stating that if the cost of the current approximation $\bfX_{(t)}$ is larger than the cost of the optimal rank $k$ solution $\bfL$, then there exists a good direction $\bfz\in\mathbb R^n$ for updating each of the columns of $\bfX_{(t)}$. We will show that we can improve this lemma by using our Theorem \ref{thm:well-cond-decomp} instead of \cite[Lemma 9]{BRW2021}. 

Following \cite{BRW2021}, we first restrict our attention to a set of good columns $\mathcal G\subseteq[d]$, where
\[
     \mathcal G \coloneqq \braces*{j\in[d] : \norm*{\bfL\bfe_j}_p^p \leq \frac1\eps \norm*{\bfa^j}_p^p}.
\]
By combining with the fact that $\norm*{\bfL}_{p,p} \leq \norm*{\bfA}_{p,p}$, it can easily be shown that the total mass on $[d]\setminus\mathcal G$ is at most $\eps\norm{\bfA}_{p,p}^p$, and thus we do not need to fit these columns well (see \cite[Lemma 6]{BRW2021}). 

To analyze the algorithm, we follow \cite{BRW2021} and define the following functions:
\begin{align*}
    f_{j,p}(\bfx) &\coloneqq \norm*{\bfa^j - \bfx}_p^p \\
    Q^{(t)}(\bfz) &\coloneqq \sum_{j\in \mathcal G}\frac{\angle{\nabla f_{j,p}(\bfx^j_{(t)}), \bfz}^2}{(f_{j,p}(\bfx^j_{(t)}))^{1-2/p}} \\
    \delta^* &\coloneqq \frac{\sum_{j\in\mathcal G}f_{j,p}(\bfL\bfe_j)}{\norm*{\bfA\vert^{\mathcal G}}_{p,p}^p} \\
    \delta^{(t)} &\coloneqq \frac{\sum_{j\in\mathcal G}f_{j,p}(\bfx_{(t)}^j)}{\norm*{\bfA\vert^{\mathcal G}}_{p,p}^p}
\end{align*}
The quantity $Q^{(t)}(\bfz)$ is the crucial quantity which represents the improvement that is possible, due to \cite[Lemma 11]{BRW2021}. The quantities $\delta^*$ and $\delta^{(t)}$ represent the costs of the optimal and current approximations, respectively.

\cite[Lemma 13]{BRW2021} states the following:
\begin{Lemma}[Lemma 13, \cite{BRW2021}]
If $\delta^{(t)} > \delta^*$, then there exists $\bfz\in\mathbb R^n$ with $\norm*{\bfz}_p = 1$ such that
\[
    Q^{(t)}(\bfz) \geq \frac{\eps^{2/p}\norm{\bfA\vert^{\mathcal G}}_{p,p}^p(\delta^{(t)} - \delta^*)^2}{k^2}
\]
\end{Lemma}

We improve this to the following:

\begin{Lemma}[Improved Lemma 13, \cite{BRW2021}]\label{lem:improved-brw-lem13}
If $\delta^{(t)} > \delta^*$, then there exists $\bfz\in\mathbb R^n$ with $\norm*{\bfz}_p = 1$ such that
\[
    Q^{(t)}(\bfz) \geq \Omega(1)\frac{\eps^{2/p}\norm{\bfA\vert^{\mathcal G}}_{p,p}^p(\delta^{(t)} - \delta^*)^2}{k\log\log k}
\]
\end{Lemma}
\begin{proof}
We write $\bfL = \bfU\bfV^\top$ using the decomposition given by Theorem \ref{thm:well-cond-decomp}. Then for each $j\in[d]$, we can write $\bfL\bfe_j = \bfU\bfv^j$ with $\norm{\bfv^j}_2 \leq O(1)\norm{\bfL\bfe_j}_p$. Furthermore, since $j\in\mathcal G$, $\norm{\bfL\bfe_j}_p \leq \norm{\bfa^j}_p / \eps$. We then have, for $s = O(k\log\log k)$ and each $j\in[d]$,
\begin{align*}
    \sum_{i=1}^s \angle{\nabla f_{j,p}(\bfx_{(t)}^j),\bfu^i}^2 &\geq \frac{(f_{j,p}(\bfx_{(t)}^j) - f_{j,p}(\bfL\bfe_j))^2}{\norm{\bfv^j}_2^2} && \text{Lemma 5, \cite{BRW2021}} \\
    &\geq \frac{(f_{j,p}(\bfx_{(t)}^j) - f_{j,p}(\bfL\bfe_j))^2}{O(1)\norm{\bfL\bfe_j}_p^2} && \text{Theorem \ref{thm:well-cond-decomp}} \\
    &\geq \eps^{2/p}\frac{(f_{j,p}(\bfx_{(t)}^j) - f_{j,p}(\bfL\bfe_j))^2}{O(1)\norm{\bfa^j}_p^2} && \text{$j\in\mathcal G$}
\end{align*}
Dividing both sides by $f_{j,p}(\bfx_{(t)}^j)^{1-2/p} \leq \norm{\bfa^j}_p^{p-2}$ and summing over $j\in\mathcal G$ gives
\begin{align*}
\sum_{i=1}^s \sum_{j\in\mathcal G}\frac{\angle{\nabla f_{j,p}(\bfx_{(t)}^j),\bfu^i}^2}{f_{j,p}(\bfx_{(t)}^j)^{1-2/p}} &\geq \Omega(\eps^{2/p})\sum_{j\in\mathcal G}\frac{(f_{j,p}(\bfx_{(t)}^j) - f_{j,p}(\bfL\bfe_j))^2}{\norm{\bfa^j}_p^p} \\
&= \Omega(\eps^{2/p})\norm*{\bfA\vert^{\mathcal G}}_{p,p}^p \cdot \sum_{j\in\mathcal G}\frac{(f_{j,p}(\bfx_{(t)}^j) - f_{j,p}(\bfL\bfe_j))^2}{\norm{\bfa^j}_p^{2p}}\cdot \frac{\norm{\bfa^j}_p^p}{\norm*{\bfA\vert^{\mathcal G}}_{p,p}^p} \\
&\geq \Omega(\eps^{2/p})\norm*{\bfA\vert^{\mathcal G}}_{p,p}^p \cdot \bracks*{\sum_{j\in\mathcal G}\frac{(f_{j,p}(\bfx_{(t)}^j) - f_{j,p}(\bfL\bfe_j))}{\norm{\bfa^j}_p^{p}}\cdot \frac{\norm{\bfa^j}_p^p}{\norm*{\bfA\vert^{\mathcal G}}_{p,p}^p}}^2 && \text{Jensen's inequality} \\
&= \Omega(\eps^{2/p})\norm*{\bfA\vert^{\mathcal G}}_{p,p}^p (\delta^{(t)} - \delta^*)^2.
\end{align*}
We conclude by averaging over $i\in[s]$. 
\end{proof}

Given this improvement to the key lemma \cite[Lemma 13]{BRW2021}, we obtain the following improvement to \cite[Theorem 3]{BRW2021}, using the exact same proof:

\begin{Theorem}\label{thm:lp-lra-additive}
Let $\bfA\in\mathbb R^{n\times d}$, let $2 < p < \infty$, and let $k\geq 1$. Then, there exists an efficient algorithm that outputs a matrix $\bfL'$ of rank at most $O(k(\log\log k) / \eps^{1+2/p})$ such that
\[
    \norm*{\bfA - \bfL'}_{p,p}^p \leq \min_{\rank(\hat\bfA)\leq k} \norm{\bfA - \hat\bfA}_{p,p}^p + \eps\norm*{\bfA}_{p,p}^p
\]
\end{Theorem}

\subsubsection{Relative Error Approximation}

We now compose the above additive error algorithm of Theorem \ref{thm:lp-lra-additive} with our relative error algorithm of Theorem \ref{thm:lp-css} to give the first relative error algorithm:

\RelErr*
\begin{proof}
We first apply Theorem \ref{thm:lp-css} to find a matrix $\bfB$ consisting of $r = O(k\log d)$ columns such that
\[
    \norm*{\bfA - \bfB}_{p,p}\leq O(k^{1/2-1/p})\min_{\rank(\hat\bfA)\leq k} \norm{\bfA - \hat\bfA}_{p,p}
\]
Now let $\bfB' = \bfA - \bfB$. Then we apply Theorem \ref{thm:lp-lra-additive} with rank parameter set to $r + k$ and accuracy parameter set to $\eps / k^{p/2-1}$. This produces an approximation $\bfC$ such that
\begin{align*}
    \norm*{\bfB' - \bfC}_{p,p}^p &\leq \min_{\rank(\hat\bfB)\leq r+k} \norm{\bfB' - \hat\bfB}_{p,p}^p + \frac{\eps}{k^{p/2-1}}\norm*{\bfB'}_{p,p}^p \\
    &\leq \min_{\rank(\hat\bfA)\leq k} \norm{\bfA - \hat\bfA}_{p,p}^p + O(\eps) \min_{\rank(\hat\bfA)\leq k} \norm{\bfA - \hat\bfA}_{p,p}^p \\
    &\leq (1+O(\eps))\min_{\rank(\hat\bfA)\leq k} \norm{\bfA - \hat\bfA}_{p,p}^p,
\end{align*}
as desired.
\end{proof}

\section{Online Coresets for \texorpdfstring{$\ell_p$}{lp} Subspace Approximation}
\label{sec:online-lp-subspace-apx}

For this section, we define the optimal value of the $\ell_p$ subspace approximation problem as
\[
    \OPT_{p,k}(\bfA) \coloneqq \min_{F\in\mathcal F_k} \norm*{\bfA(\bfI-\bfP_F)}_{p,2}^p = \min_{\rank(\bfX) \leq k} \norm*{\bfA(\bfI-\bfX)}_{p,2}^p
\]

We show the following theorem in this section:

\begin{Theorem}\label{thm:main-int}
Let $\bfA\in\mathbb Z^{n\times d}$ have entries bounded by $\Delta$, let $\eps,\delta\in(0,1)$, let $p\geq 1$ a constant, and let $k$ be a rank. Suppose that any submatrix of $\bfA$ formed by consecutive rows has online condition number at most $\kappa$. There is an online coreset algorithm, Algorithm \ref{alg:online-lp-subspace}, which stores at most
\[
    s = O\parens*{\mathcal S\parens*{k\cdot\min\{k^2\log k, k^{1\lor(p/2)}\}\log\mathcal S + \eps'^{-2}\log\frac1\delta + \eps^{-2}\eps'^{-1}k^2\log\mathcal S}}
\]
rows, where $\eps' = \eps^{(p+3)\cdot(1\lor(2/p))}$ and
\begin{equation}\label{eq:total-sensitivity-alg}
\begin{aligned}
\mathcal S &= O\left((k\log k + \log^2 n)^2\log^2(n\Delta)\right)^{1\lor(p/2)}(\log^2 k + \log^2\log n)(\log n)\log\frac{n}{\delta} \\
&= k^{2\lor p}\cdot(\log(n\Delta/\delta))^{O(p)}
\end{aligned}
\end{equation}
If $\bfS\in\mathbb R^{n\times n}$ is the resulting sampling matrix, then with probability at least $1-\delta$, simultaneously for all $i\in[n]$ and $F\in\mathcal F_k$, we have that
\[
    \norm*{\bfS_i\bfA_i(\bfI-\bfP_F)}_{p,2}^p = (1\pm\eps)\norm*{\bfA_i(\bfI-\bfP_F)}_{p,2}^p
\]
\end{Theorem}

As we show in Section \ref{sec:int-red}, this immediately gives the following for real valued matrices:

\begin{Theorem}\label{thm:main-real}
Let $\bfA\in\mathbb R^{n\times d}$, let $\eps,\delta\in(0,1)$, let $p\geq 1$ a constant, and let $k$ be a rank. Suppose that any submatrix of $\bfA$ formed by consecutive rows has online condition number at most $\kappa$. There is an online coreset algorithm, Algorithm \ref{alg:online-lp-subspace}, which stores at most
\[
    s = O\parens*{\mathcal S\parens*{k\cdot\min\{k^2\log k, k^{1\lor(p/2)}\}\log\mathcal S + \eps'^{-2}\log\frac1\delta + \eps^{-2}\eps'^{-1}k^2\log\mathcal S}}
\]
rows, where $\eps' = \eps^{(p+3)\cdot(1\lor(2/p))}$ and
\begin{align*}
\mathcal S &= O\left((k\log k + \log^2 n)^2\log^2(n\kappa^\OL)\right)^{1\lor(p/2)}(\log^2 k + \log^2\log n)(\log n)\log\frac{n}{\delta} \\
&= k^{2\lor p}\cdot(\log(n\kappa^\OL/\delta))^{O(p)}
\end{align*}
If $\bfS\in\mathbb R^{n\times n}$ is the resulting sampling matrix, then with probability at least $1-\delta$, simultaneously for all $i\in[n]$ and $F\in\mathcal F_k$, we have that
\[
    \norm*{\bfS_i\bfA_i(\bfI-\bfP_F)}_{p,2}^p = (1\pm\eps)\norm*{\bfA_i(\bfI-\bfP_F)}_{p,2}^p
\]
\end{Theorem}

\begin{proof}[Proof Sketch of Theorem \ref{thm:main-int}.]
The overall approach is based on showing that sensitivity sampling can be made to work online. We show in Section \ref{sec:constant-factor-apx} that a constant factor bicriteria solution can be found in an online manner. In Section \ref{sec:online-sens-apx}, we show how to use this online bicriteria solution to estimate sensitivities. Finally, we show in Section \ref{sec:sens-sample} that sampling by using these weights yields a strong coreset.

By Lemma \ref{lem:online-sensitivities-sum}, Algorithm \ref{alg:online-sensitivity} returns a sensitivity upper bound with constant probability for each $i\in[n]$. We may then repeat $O(\log(n/\delta))$ times so that the sum of the repetitions as taken in Line \ref{line:online-sensitivity-sum} yields a valid sensitivity overestimate simultaneously for every $i\in[n]$ with probability at least $1-\delta$, by a union bound. Furthermore, again by Lemma \ref{lem:online-sensitivities-sum}, the total sensitivity is as claimed in Equation \eqref{eq:total-sensitivity-alg}.

Lemma \ref{lem:hv2020-lem5.7-analogue} shows the number of samples required to reduce the problem of finding strong coresets for $\bfA$ to finding strong coresets for a projection of $\bfA$ onto a lower dimensional space of dimension $O(k/\eps')$. Lemma \ref{lem:low-dim-fl} shows the number of samples required to obtain a strong coreset in the lower dimensional space.
\end{proof}

\begin{algorithm}
\caption{Online Sensitivity Approximation}
\textbf{input:} Stream $\bfA$, rank $k$, total online sensitivity upper bound $\mathcal S$. \\
\textbf{output:} Online coreset for $\ell_p$ subspace approximation.
\begin{algorithmic}[1] 
    \State Obtain online rank $2k$ sensitivity overestimates from $O(\log(n/\delta))$ independent copies of Algorithm \ref{alg:online-sensitivity}
    \State For $i\in[n]$, let $\tilde\bfsigma_i^\OL$ be the sum of the overestimates across the $O(\log(n/\delta))$ copies \label{line:online-sensitivity-sum}
    \State Use $\tilde\bfsigma_i^{\OL}$ and $\mathcal S$ to sample rows as done in Theorem \ref{thm:fl2011-indep}
\end{algorithmic}\label{alg:online-lp-subspace}
\end{algorithm}

\subsection{Reduction to Integer Matrices}\label{sec:int-red}

We first reduce the case of real-valued matrices to the case of integer matrices by rounding the input matrix. This allows us to control the conditioning of the solution in a simple way, since integer matrices have bounded condition number \cite{CW2009}. Let $\tilde\lambda^{p,\OL}$ be a lower bound on 
\[
    \lambda^{p,\OL} \coloneqq \min_{i\in[n] : \rank(\bfA_i) > k} \min_{F\in\mathcal F_k}\norm*{\bfA_i(\bfI-\bfP_F)}_{p,2}^p.
\]
That is, $\lambda^{p,\OL}$ is the smallest nonzero cost of $\bfA_i$ for any $i\in[n]$. Note that by the equivalence of $\ell_p$ norms, this quantity is related up to a factor of $\poly(n)$ with $\lambda^{\OL}\coloneqq\lambda^{2,\OL}$. Then, we may round each entry of $\bfA$ to the nearest integer multiple of $\eps n^{-1/p}d^{-1/2}(\tilde\lambda^{p,\OL})^{1/p}$ to obtain a matrix $\bfA'$, so that
\[
    \norm*{\bfA - \bfA'}_{p,2}^p \leq \sum_{i=1}^n \bracks*{\sum_{j=1}^d \abs*{\bfA[i,j] - \bfA'[i,j]}^2}^{p/2} \leq \sum_{i=1}^n\bracks*{\eps^{2} n^{-2/p}(\tilde\lambda^{p,\OL})^{2/p}}^{p/2} \leq \eps^p \tilde\lambda^{p,\OL}.
\]
Then, for all $F\in\mathcal F_k$
\[
    \norm*{\bfA(\bfI-\bfP_F)}_{p,2} = \norm*{\bfA'(\bfI-\bfP_F)}_{p,2} \pm \eps\norm*{\bfA-\bfA'}_{p,2} \subseteq \norm*{\bfA'(\bfI-\bfP_F)}_{p,2} \pm \eps\norm*{\bfA(\bfI-\bfP_F)}_{p,2}
\]
which implies that
\[
    \norm*{\bfA'(\bfI-\bfP_F)}_{p,2}^p = (1\pm\eps)^p\norm*{\bfA(\bfI-\bfP_F)}_{p,2}^p
\]
for all $F\in\mathcal F_k$, so it suffices to solve our problem on $\bfA'$, which, up to a scaling, is an integer matrix with entries bounded by
\[
    \Delta = \frac{\norm*{\bfA}_\infty}{\eps n^{-1/p}d^{-1/2}(\tilde\lambda^{p,\OL})^{1/p}} \leq \poly(n, \norm*{\bfA}_\infty / \tilde\lambda^{p,\OL}) \leq \poly(n, \kappa^\OL).
\]

\subsection{Constant Factor Approximation}\label{sec:constant-factor-apx}

We first obtain a constant factor bicriteria approximation. For this, our approach is to adapt Theorem 4.1 of \cite{FKW2021}, which shows that a Lewis weight sample from $\bfA\bfG^\top$ for a Gaussian matrix $\bfG$ with $\tilde O(k)$ columns yields rows whose span contains an $O(1)$-approximate solution. Although \cite{FKW2021} only states the result for $p = 1$, we show that the same proof and conclusion holds for all $p\geq 1$. For our online implementation, we will replace Lewis weights with online Lewis weights \cite{WY2022b}. Furthermore, we replace the use of a Gaussian matrix $\bfG$ with an $\ell_2$ subspace embedding with integer entries, so that the sketch also has integer entries. 

\subsubsection{Dimension Reduction}
\label{sec:dim-red-lp-subspace-apx}

We first replace the use of a dense Gaussian matrix in \cite{FKW2021} with an integer subspace embedding, so that the resulting matrix is integer. One possibility\footnote{There are many possible alternatives here, for example a dense sign matrix, but we choose SRHT since the results we need are stated and proven in the existing literature.} is the subsampled randomized Hadamard transform (SRHT), which has the following guarantees:
\begin{Definition}[Subsampled Randomized Hadamard Transform (SRHT), Definition 1.2, \cite{BG2013}]\label{def:srht}
Fix $r$ an integer and $n$ a power of $2$ with $r < n$. An SRHT matrix is an $r\times n$ matrix $\bfS = \sqrt{n/r}\cdot\bfR\bfH\bfD$, where $\bfD\in\mathbb R^{n\times n}$ is a random Rademacher diagonal matrix, $\bfH\in\mathbb R^{n\times n}$ is a Walsh--Hadamard matrix, and $\bfR\in\mathbb R^{r\times n}$ is a sampling matrix which selects $r$ rows uniformly at random without replacement.
\end{Definition}

\begin{Theorem}[SRHT is a subspace embedding, Lemma 4.1 of \cite{BG2013}, \cite{Tro2011b}]\label{thm:srht}
Let $\bfS$ be an $r\times d$ SRHT matrix. Let $\bfA\in\mathbb R^{d\times k}$. If $r = \Theta(\eps^{-2}(k + \log(d/\delta))\log(k/\delta))$, then with probability at least $1-\delta$, for all $\bfx\in\mathbb R^k$,
\[
    \norm*{\bfS\bfA\bfx}_2 = (1\pm\eps)\norm*{\bfA\bfx}_2
\]
\end{Theorem}

Using these properties of the SRHT, we show the following main lemma of this section, Lemma \ref{lem:dim-red-srht}, which allows us to reduce the dimension of the points $\bfa_i$ from $d$ to $t = O(k\log k + \log^2 n)$. This reduced dimensionality will be useful for removing a $d$ dependence from our subsequent discussion. For the rest of Section \ref{sec:dim-red-lp-subspace-apx}, we focus on proving Lemma \ref{lem:dim-red-srht}.

\begin{Lemma}[Dimension Reduction for $\ell_p$ Subspace Approximation]\label{lem:dim-red-srht}
Let $\bfG$ be a $t\times d$ SRHT matrix (Definition \ref{def:srht}). Then, there is $t = O(k\log k + \log^2 n)$ such that, with probability at least $9/10$,
\[
    \min_{\rank(\bfX)\leq k} \norm*{\bfA\bfG^\top\bfX - \bfA}_{p,2}^p \leq \frac32 \OPT_k.
\]
\end{Lemma}

To prove this, we need the notion of lopsided embeddings.

\begin{Definition}[Lopsided Embedding (Definition 26, \cite{CW2015b})]
Consider a constraint set $\mathcal C$ and norm $\norm*{\cdot}$, and matrices $\bfA\in\mathbb R^{k\times d}$ and $\bfB\in\mathbb R^{n\times d}$. Suppose $\bfS\in\mathbb R^{d\times r}$ satisfies:
\begin{itemize}
    \item $\norm*{(\bfY\bfA-\bfB)\bfS^\top} \geq (1-\eps)\norm*{\bfY\bfA-\bfB}$ for all $\bfY\in\mathbb R^{n\times k}$
    \item $\norm*{(\bfY^*\bfA-\bfB)\bfS^\top} \leq (1+\eps)\norm*{\bfY^*\bfA-\bfB}$, where $\bfY^* = \argmin_{\bfY\in\mathcal C}\norm*{\bfY\bfA-\bfB}$
\end{itemize}
Then, $\bfS$ is a $\eps$-lopsided embedding for $(\bfA,\bfB)$ with respect to $\mathcal C$ and $\norm*{\cdot}$.
\end{Definition}

The next lemma from prior work shows the utility of lopsided embeddings for subspace approximation, showing that if we can maintain a sketch $\bfA\bfS^\top$ where $\bfS$ is a lopsided embedding, then solving for the best rank $k$ approximation in the column space of $\bfA\bfS^\top$ is sufficient for obtaining a good subspace approximation solution.

\begin{Lemma}[Lemma B.1, \cite{FKW2021}, \cite{CW2015b}]\label{lem:lop-suff}
Let $\bfU\in\mathbb R^{n\times k}$ and $\bfV\in\mathbb R^{k\times d}$ be matrices such that
\[
    \norm*{\bfU\bfV - \bfA}_{p,2} = \min_{\rank(\bfX) \leq k} \norm*{\bfA(\bfI-\bfX)}_{p,2}.
\]
If $\bfS\in\mathbb R^{t\times d}$ is a lopsided $\eps$-embedding for $(\bfV, \bfA)$ with respect to $\norm*{\cdot}_{p,2}$, then
\[
    \min_{\rank(\bfX) \leq k}\norm*{\bfA\bfS^\top\bfX - \bfA}_{p,2} \leq (1+O(\eps)) \min_{\rank(\bfX) \leq k}\norm*{\bfA(\bfI-\bfX)}_{p,2}
\]
\end{Lemma}

The work of \cite{FKW2021} shows that a random Gaussian matrix $\bfG$ with $\tilde O(k)$ columns gives $O(1)$-lopsided embeddings for the $\norm*{\cdot}_{1,2}$ norm, based on results from \cite{CW2015b}. We show an analogous result for the $\norm*{\cdot}_{p,2}$ norm and for the SRHT. As done in \cite{FKW2021}, we use the sufficient conditions for a lopsided embedding provided by \cite[Lemma 27]{CW2015b}. 

\begin{Lemma}[Lemma 27, \cite{CW2015b}]\label{lem:cw2015-lop}
Let $\bfA\in\mathbb R^{k\times d}$ and $\bfB\in\mathbb R^{n\times d}$. Suppose $\bfS\in\mathbb R^{t\times d}$ satisfies the following:
\begin{itemize}
    \item With probability at least $1-\delta/3$, $\bfS$ is a subspace $\eps$-contraction for $\bfA^\top$, that is, simultaneously for all $\bfx\in\mathbb R^k$,
    \[
        \norm*{\bfS\bfA^\top\bfx}_2 \geq (1-\eps)\norm*{\bfS\bfA^\top\bfx}_2
    \]
    \item For all $i\in[n]$, with probability at least $1-\delta\eps^{p+1} / 3$, $\bfS$ is a subspace $\eps^{p+1}$-contraction for $[\bfA^\top\ \bfB^\top\bfe_i]$, that is, simultaneously for all $\bfx\in\mathbb R^{k+1}$,
    \[
        \norm*{\bfS[\bfA^\top\ \bfB^\top\bfe_i]\bfx}_2 \geq (1-\eps)\norm*{\bfS[\bfA^\top\ \bfB^\top\bfe_i]\bfx}_2
    \]
    \item With probability at least $1-\delta/3$,
    \[
        \norm*{\bfB^*\bfS^\top}_{p,2} \leq (1+\eps^{p+1})\norm*{\bfB^*}_{p,2}
    \]
    where $\bfB^* = \bfY^*\bfA - \bfB$ for $\bfY^* = \argmin_\bfY\norm*{\bfY\bfA - \bfB}_{p,2}$ 
\end{itemize}
Then, $\bfS$ is a lopsided $\eps$-embedding for $(\bfA,\bfB)$ with respect to $\norm*{\cdot}_{p,2}$.
\end{Lemma}

With the above results in hand, we can prove Lemma \ref{lem:dim-red-srht}.

\begin{proof}[Proof of Lemma \ref{lem:dim-red-srht}]
We check the conditions for Lemma \ref{lem:cw2015-lop} to show that $\bfG$ is a lopsided embedding. In turn, we will apply Lemma \ref{lem:lop-suff} to conclude. The first two follow from the subspace embedding guarantees for SRHT in Theorem \ref{thm:srht} with constant $\delta$ and $\eps$. For the last, we apply Theorem \ref{thm:srht} with $\delta = 1/\poly(n)$ and $k = 1$ to show that the norm of each of the $n$ rows of $\bfB^*$ is preserved up to a factor of $(1\pm\eps^{p+1})$, which implies the required condition. 
\end{proof}

\subsubsection{Online Point Reduction}

Next, we use the previous dimension reduction result in combination with online Lewis weights \cite{WY2022b} to obtain a small online coreset for a $O(1)$ approximation. The following is Lemma B.4 of \cite{FKW2021}, whose argument applies directly to all $p\geq 1$, and states that any subspace embedding $\bfL$ for the matrix $\bfA\bfG^\top$ which preserves the $\norm*{\cdot}_{p,2}$-norm of an arbitrary matrix in expectation preserves the subspace approximation cost of $\bfA\bfG^\top$, up to a constant factor. We will apply this lemma with $\bfL$ chosen to be sampled according to the online Lewis weights of \cite{WY2022b}, which indeed satisfy the hypotheses. 

\begin{Lemma}[Lemma B.4 \cite{FKW2021}]\label{lem:olw-point-red}
Let $\bfG\in\mathbb R^{t\times d}$ be an SRHT matrix. Let $\bfL$ be a random matrix such that with probability at least $9/10$, simultaneously for all $\bfy\in\mathbb R^t$,
\[
\alpha\norm*{\bfA\bfG^\top\bfy}_p \leq \norm*{\bfL\bfA\bfG^\top\bfy}_p \leq \beta\norm*{\bfA\bfG^\top\bfy}_p
\]
and
\[
\E_\bfL[\norm*{\bfL\bfM}_{p,2}^p] = \norm*{\bfM}_{p,2}^p
\]
for any matrix $\bfM$. Then, there is $t = O(k\log k + \log^2 n)$ such that with probability at least $3/5$, all matrices $\bfX$ with
\[
\norm*{\bfL\bfA\bfG^\top\bfX - \bfL\bfA}_{p,2}^p \leq 10 \cdot \OPT_k
\]
satisfy
\[
\norm*{\bfA\bfG^\top\bfX - \bfA}_{p,2}^p \leq (2+40/\alpha)^p\OPT_k.
\]
\end{Lemma}

We may now show our online point reduction lemma. While the result only holds for a fixed $i\in[n]$ with constant probability, we may boost the success probability by taking $O(\log(n/\delta))$ independent copies, so that we have at least one good bicriteria approximation for all $i\in[n]$ with probability at least $1-\delta$. This will be enough for our uses.

We use the following online Lewis weight sampling theorem, which provides the subspace embedding $\bfL$ we need in Lemma \ref{lem:olw-point-red}.

\begin{Theorem}[Online Lewis Weight Sampling \cite{WY2022a, WY2022b}]\label{thm:online-lewis-p>2}
    Let $\bfA\in\mathbb Z^{n\times d}$ with entries bounded by $\Delta$ and let $p\in(0,\infty)$. Let $\delta\in(0,1)$ be a failure rate parameter and let $\eps\in(0,1)$ be an accuracy parameter. Then there is an online coreset algorithm $\mathcal A$ such that, with probability at least $1-\delta$, $\mathcal A$ outputs a weighted subset of $m$ rows with sampling matrix $\bfS$ such that
    \begin{equation}\label{eq:lp-subspace-embedding}
        \norm*{\bfS_i\bfA_i\bfx}_p^p = (1\pm\eps)\norm*{\bfA_i\bfx}_p^p
    \end{equation}
    for all $\bfx\in\mathbb R^d$, for every $i\in[n]$, and
    \[
        m = \begin{dcases}
            O\parens*{\frac{d^{p/2}}{\eps^2}}\log(n\Delta)^{p/2+1}\bracks*{(\log d)^2(\log n) + \log\frac1\delta} & p\in(2,\infty) \\
            O\parens*{\frac{d}{\eps^2}}\log(n\Delta)\bracks*{(\log d)^2\log n + \log\frac1\delta} & p \in (1,2) \\
            O\parens*{\frac{d}{\eps^2}}\log(n\Delta)\log\frac{n}{\delta} & p = 1 \\
            O\parens*{\frac{d}{\eps^2}}\log(n\Delta)\bracks*{(\log d)^3 + \log\frac1\delta} & p\in(0,1)
        \end{dcases}
    \]
\end{Theorem}
\begin{proof}
Theorem 3.9 of \cite{WY2022a} shows that one can construct weights $\bfw_i \leq O(1)$ for $i\in[n]$ in an online fashion such that there is a fixed $s = O(n)$ such that $s\cdot \bfw^{1/2-1/p}_i\in\mathbb Z$ for all $i\in[n]$, and satisfies
\[
    \bfw_i \leq O(1)\cdot \bftau_i^\OL(\bfW^{1/2-1/p}\bfA)
\]
for every $i\in[n]$, where $\bfW = \diag(\bfw)$. Since $\bfW^{1/2-1/p}\bfA$ is an $n\times d$ integer matrix with entries bounded by $\Delta$, the proof of \cite[Theorem 1.5]{WY2022a} shows that the sum of the weights is at most
\[
    \sum_{i=1}^n \bfw_i \leq O(1)\sum_{i=1}^n \bftau_i^{\OL}(\bfW^{1/2-1/p}\bfA) \leq O(d\log(n\Delta)).
\]
Furthermore, the weights $\bfw$ constructed in \cite[Theorem 3.9]{WY2022a} satisfy the \emph{one-sided Lewis weight property} (see \cite[Definition 2.2]{WY2022b}), which means that sampling proportionally to these weights gives the guarantee of \eqref{eq:lp-subspace-embedding} by \cite[Theorem 5.2]{WY2022b} for $p>2$ or \cite[Theorem A.2]{WY2022b}, with the sample complexity as stated.
\end{proof}

We then obtain the following lemma, which reduces our original problem of finding a constant factor solution for the matrix $\bfA_i$, to solving a regression problem on a small subsample given by $\bfL_i\bfA_i\bfG^\top$ and $\bfL_i\bfA_i$.

\begin{Lemma}[Online Point Reduction]\label{lem:online-point-red}
Let $\bfG\in\mathbb R^{t\times d}$ be an SRHT matrix with $t = O(k\log k + \log^2 n)$. Let $\bfL$ be an $\ell_p$ online Lewis weight sample generated by applying Theorem \ref{thm:online-lewis-p>2} to $\bfA\bfG^\top$ with $\eps = 1/2$ and $\delta = 9/10$. Let $\bfL_i$ denote the online coreset $\bfL$ at time $i\in[n]$. Let $\tilde\bfY_i$ satisfy
\[
    \norm*{\bfL_i\bfA_i\bfG^\top\tilde\bfY_i - \bfL_i\bfA_i}_{p,2}^p \leq \frac{6}{5} \min_{\bfY\in\mathbb R^{t\times d}}\norm*{\bfL_i\bfA_i\bfG^\top\bfY - \bfL_i\bfA_i}_{p,2}^p.
\]
Note that such a $\tilde\bfY_i$ can be found in polynomial time by converting to an instance of $\ell_p$ regression using Dvoretzky's theorem \cite{SW2018, FKW2021}. Then for each fixed $i\in[n]$, with probability at least $3/10$,
\[
    \norm*{\bfA_i\bfG^\top\tilde\bfY_i - \bfA_i}_{p,2}^p \leq O(1)\min_{\rank(\bfX)\leq k}\norm*{\bfA_i(\bfI-\bfX)}_{p,2}^p
\]
\end{Lemma}
\begin{proof}
Fix $i\in[n]$ and let $\OPT_k$ denote the optimal rank $k$ $\ell_p$ subspace approximation cost for $\bfA_i$. Note that the column space of $\bfA\bfG^\top$ contains a $3/2$-approximate solution by Lemma \ref{lem:dim-red-srht}. Then by Markov's inequality over the draws of $\bfL_i$, with probability at least $1 - 1/5 - 1/10 = 7/10$, there is a rank $k$ projection $\tilde\bfX$ such that
\[
    \norm*{\bfL_i\bfA_i\bfG^\top\tilde\bfX - \bfL_i\bfA_i}_{p,2}^p \leq \frac{15}{2} \cdot \OPT_k.
\]
We may then lower bound this by minimizing over all $t\times d$ matrices $\bfY$ instead of rank $k$ matrices, so
\[
    \norm*{\bfL_i\bfA_i\bfG^\top\tilde\bfY - \bfL_i\bfA_i}_{p,2}^p \leq \frac{6}{5}\min_{\bfY\in\mathbb R^{t\times d}}\norm*{\bfL_i\bfA_i\bfG^\top\bfY - \bfL_i\bfA_i}_{p,2}^p \leq \frac{6}{5}\cdot\frac{15}{2}\cdot\OPT_k < 10\cdot\OPT_k,
\]
where $\tilde\bfY$ is the minimizer over all matrices without the rank constraint. By a union bound with the event from Lemma \ref{lem:olw-point-red}, we have that with probability at least $3/10$, 
\[
    \norm*{\bfA_i\bfG^\top\tilde\bfY - \bfA_i}_{p,2}^p \leq O(1)\OPT_k.\qedhere
\]
\end{proof}

Note that $\bfA_i\bfG^\top\tilde\bfY_i$ found from Lemma \ref{lem:online-point-red} may not necessarily be an integer matrix, even if $\bfA_i\bfG^\top$ is. We thus need to round $\tilde\bfY_i$. If $\rank(\bfA_i) \geq 2t$, then we use the next lemma, Lemma \ref{lem:round-lra}, to carry this out. Otherwise, we just directly store a basis for $\rowspan(\bfA_i)$.

\begin{Lemma}\label{lem:round-lra}
Let $\bfA\in\mathbb Z^{n\times t}$ and $\bfB\in\mathbb Z^{n\times d}$ be integer matrices with entries bounded by $\Delta$, and let $\bfY\in\mathbb R^{t\times d}$. Suppose that $\rank(\bfB)\geq 2t$. Then, rounding each entry of $\bfY$ to the nearest integer multiple of $1/\poly(n\Delta)$ produces a matrix $\tilde\bfY$ such that
\[
    \norm*{\bfA\tilde\bfY - \bfB}_{p,2} \leq (1+\poly(n\Delta)^{-1})\norm*{\bfA\bfY - \bfB}_{p,2}
\]
\end{Lemma}
\begin{proof}
Note that
\[
    \min_{\bfX\in\mathbb R^{t\times d}}\norm*{\bfA\bfX - \bfB}_{p,2} \geq \frac1{\poly(n)}\min_{\bfX\in\mathbb R^{t\times d}}\norm*{\bfA\bfX - \bfB}_{F}.
\]
By Lemma 4.1 of \cite{CW2009}, since the rank of $\bfB$ is at least $2t$, we have that
\[
    \min_{\bfX\in\mathbb R^{t\times d}}\norm*{\bfA\bfX - \bfB}_{F} \geq \frac1{\poly(n\Delta)}.
\]
Then,
\begin{align*}
    \norm*{\bfA\tilde\bfY - \bfB}_{p,2} &\leq \norm*{\bfA\bfY - \bfB}_{p,2} + \norm*{\bfA(\tilde\bfY - \bfY)}_{p,2} \\
    &\leq \norm*{\bfA\bfY - \bfB}_{p,2} + \poly(n)\norm*{\bfA(\tilde\bfY - \bfY)}_{F} \\
    &\leq \norm*{\bfA\bfY - \bfB}_{p,2} + \poly(n)\norm*{\bfA}_2\norm*{\tilde\bfY - \bfY}_{F} \\
    &\leq \norm*{\bfA\bfY - \bfB}_{p,2} + \poly(n\Delta)^{-1} \\
    &\leq (1+\poly(n\Delta)^{-1})\norm*{\bfA\bfY - \bfB}_{p,2} \qedhere
\end{align*}
\end{proof}

\subsection{Online Sensitivity Approximation}\label{sec:online-sens-apx}

In Section \ref{sec:constant-factor-apx}, we have shown how to find a constant factor bicriteria solution in an online fashion. Using our online constant factor bicriteria solution, we now show that we can estimate sensitivities in an online manner, and that they have a small sum. 

\begin{Definition}[Online Sensitivity]
Let $\bfA\in\mathbb R^{n\times d}$. Then, the $i$th online sensitivity for the rank $k$ $\ell_p$ subspace approximation problem is
\[
    \bfsigma_i^\OL(\bfA) \coloneqq \sup_{F\in\mathcal F_k}\frac{\norm*{\bfa_i^\top(\bfI-\bfP_F)}_2^p}{\norm*{\bfA_i(\bfI-\bfP_F)}_{p,2}^p},
\]
where $\mathcal F_k$ is the set of rank $k$ subspaces, and $\bfP_F$ is the orthogonal projection matrix onto the subspace $F$. Equivalently, $\bfsigma_i^\OL(\bfA) = \bfsigma_i(\bfA_i)$, where $\sigma_i$ is the usual sensitivity \cite[Definition 2]{VX2012}
\[
    \bfsigma_i(\bfA) \coloneqq \sup_{F\in\mathcal F_k}\frac{\norm*{\bfa_i^\top(\bfI-\bfP_F)}_2^p}{\norm*{\bfA(\bfI-\bfP_F)}_{p,2}^p}.
\]
\end{Definition}

We first adapt an argument from \cite[Theorem 7]{VX2012}, which shows that it suffices to bound sensitivities over an approximately optimal bicriteria subspace. 

\begin{Lemma}[Theorem 7, \cite{VX2012}]\label{lem:vx12-thm-7}
Let $\tilde F$ be a rank $r$ subspace such that 
\[
    \norm*{\bfA(\bfI-\bfP_{\tilde F})}_{p,2}^p \leq \alpha \cdot \OPT_k = \alpha \min_{\rank(\bfX) \leq k}\norm*{\bfA(\bfI-\bfX)}_{p,2}^p
\]
for some $\alpha\geq 1$. Then,
\[
    \bfsigma_i(\bfA) \leq 2^{2p-1}\alpha\cdot \bfsigma_i(\bfA\bfP_{\tilde F}) + 2^{p-1}\alpha\frac{\norm*{\bfa_i^\top(\bfI-\bfP_{\tilde F})}_2^p}{\norm*{\bfA(\bfI-\bfP_{\tilde F})}_{p,2}^p}.
\]
\end{Lemma}
\begin{proof}
If $\norm*{\bfA(\bfI-\bfP_{\tilde F})}_{p,2}^p = 0$, then $\tilde F$ contains $\rowspan(\bfA)$ so the bound holds, so assume otherwise. For any rank $k$ subspace $F\in\mathcal F$ and $i\in[n]$, we have that
\begin{align*}
    \norm*{\bfa_i^\top(\bfI-\bfP_F)}_2^p &\leq \norm*{\bfa_i^\top - \bfa_i^\top\bfP_{\tilde F}\bfP_F}_2^p && \text{optimality of $\bfa_i^\top\bfP_F$} \\
    &\leq 2^{p-1}\bracks*{\norm*{\bfa_i^\top(\bfI-\bfP_{\tilde F})}_2^p + \norm*{\bfa_i^\top\bfP_{\tilde F}(\bfI-\bfP_F)}_2^p} && \text{triangle inequality} \\
    &\leq 2^{p-1}\bracks*{\norm*{\bfa_i^\top(\bfI-\bfP_{\tilde F})}_2^p + \bfsigma_i(\bfA\bfP_{\tilde F})\norm*{\bfA\bfP_{\tilde F}(\bfI-\bfP_F)}_{p,2}^p}
\end{align*}
Note that
\begin{align*}
    \norm*{\bfA\bfP_{\tilde F}(\bfI-\bfP_F)}_{p,2} &\leq \norm*{\bfA\bfP_{\tilde F}-\bfA\bfP_F}_{p,2} && \text{optimality of $\bfA\bfP_{\tilde F}\bfP_F$} \\
    &\leq \norm*{\bfA\bfP_{\tilde F}-\bfA}_{p,2} + \norm*{\bfA-\bfA\bfP_F}_{p,2} && \text{triangle inequality} \\
    &\leq (\alpha^{1/p}+1) \norm*{\bfA-\bfA\bfP_F}_{p,2} && \text{near optimality of $\tilde F$} \\
    &\leq 2\alpha^{1/p} \norm*{\bfA-\bfA\bfP_F}_{p,2}
\end{align*}
Thus, we continue to bound
\begin{align*}
    \norm*{\bfa_i^\top(\bfI-\bfP_F)}_2^p &\leq 2^{p-1}\bracks*{\frac{\norm*{\bfa_i^\top(\bfI-\bfP_{\tilde F})}_2^p}{\norm*{\bfA-\bfA\bfP_F}_{p,2}^p} + 2^p\alpha \bfsigma_i(\bfA\bfP_{\tilde F})}\norm*{\bfA-\bfA\bfP_F}_{p,2}^p \\
    &\leq 2^{p-1}\bracks*{\alpha\frac{\norm*{\bfa_i^\top(\bfI-\bfP_{\tilde F})}_2^p}{\norm*{\bfA(\bfI-\bfP_{\tilde F})}_{p,2}^p} + 2^p\alpha \bfsigma_i(\bfA\bfP_{\tilde F})}\norm*{\bfA-\bfA\bfP_F}_{p,2}^p
\end{align*}
by near optimality of $\tilde F$. Taking a supremum over $F\in\mathcal F_k$ yields the desired result.
\end{proof}

Consider instantiating the bicriteria solution $\tilde F$ from the above by the row span of $\bfG^\top\tilde\bfY_i\in\mathbb R^{d\times d}$, which is a rank $t$ integer matrix after scaling by $\poly(n\Delta)^t$, and is a good solution by Lemma \ref{lem:online-point-red}. Note that $\bfG^\top\tilde\bfY_i$ only changes when we draw a new row when sampling from the online Lewis weights of $\bfA\bfG^\top$, since $\tilde\bfY_i$ only depends on the rows sampled by the online Lewis weights by the construction in Lemma \ref{lem:online-point-red}. This occurs only $m$ times, where $m$ is the sample complexity of the online Lewis weights (see Theorem \ref{thm:online-lewis-p>2}). Then, we can partition the stream into $m$ segments, and separately bound the online sensitivities of each of these $m$ substreams. The advantage of this is that within each of the substreams, we only need to bound the distances to the fixed subspace $\tilde F = \rowspan(\bfG^\top\tilde\bfY)$ and then online sensitivities within $\bfA\bfP_{\tilde F}$. To bound the sensitivities within $\bfA\bfP_{\tilde F}$, we show that the online Lewis weights of $\bfA\bfP_{\tilde F}$ give a good bound:

\begin{Lemma}[Lewis Weights Bound $(p,2)$-Sensitivities]\label{lem:lewis-p2}
Let $\bfB\in\mathbb R^{n\times r}$, and let $\bfw^{p,\OL}(\bfB)$ be one-sided online Lewis weights for $\bfB$. Then, for every $l\in[n]$,
\[
    \sup_{\bfY\in\mathbb R^{d\times m}} \frac{\norm*{\bfb_l^\top\bfY}_2^p}{\norm*{\bfB_l\bfY}_{p,2}^p} \leq \norm*{\bfw^{p,\OL}(\bfB)}_1^{0 \lor (p/2-1)}\bfw_l^{p,\OL}(\bfB)
\]
\end{Lemma}
\begin{proof}
Let $\bfR\in\mathbb R^{r\times r}$ be a change of basis matrix so that $\bfW^{p,\OL}(\bfB)_l^{1/2-1/p}\bfB_l\bfR$ is orthonormal. By the one-sided Lewis property of online Lewis weights (Lemma 3.6 of \cite{WY2022b}), we have that
\[
    \norm*{\bfb_l^\top \bfR\bfY}_2^p \leq \norm*{\bfb_l^\top\bfR}_2^p \norm*{\bfY}_2^p \leq \bfw_l^{p,\OL}(\bfB)\norm*{\bfY}_2^p.
\]
We then bound
\begin{align*}
    \norm*{\bfY}_2^p &= \sup_{\norm*{\bfx}_2 = 1} \norm*{\bfY\bfx}_2^p \\
    &= \sup_{\norm*{\bfx}_2 = 1} \norm*{\bfW^{p,\OL}(\bfB)_l^{1/2-1/p}\bfB_l\bfR\bfY\bfx}_2^p \\
    &\leq \sup_{\norm*{\bfx}_2 = 1} \norm*{\bfw^{p,\OL}(\bfB)_l}_1^{0\lor(p/2-1)}\norm*{\bfB_l\bfR\bfY\bfx}_p^p && \text{Lemma 2.3 of \cite{WY2022b}} \\
    &= \norm*{\bfw^{p,\OL}(\bfB)_l}_1^{0\lor(p/2-1)} \sup_{\norm*{\bfx}_2 = 1} \sum_{i=1}^l \abs*{\bfe_i^\top\bfB_l\bfR\bfY\bfx}^p \\
    &\leq \norm*{\bfw^{p,\OL}(\bfB)_l}_1^{0\lor(p/2-1)} \sum_{i=1}^l \sup_{\norm*{\bfx}_2 = 1}\abs*{\bfe_i^\top\bfB_l\bfR\bfY\bfx}^p \\
    &= \norm*{\bfw^{p,\OL}(\bfB)_l}_1^{0\lor(p/2-1)} \sum_{i=1}^l \norm*{\bfe_i^\top\bfB_l\bfR\bfY}_2^p \\
    &= \norm*{\bfw^{p,\OL}(\bfB)_l}_1^{0\lor(p/2-1)} \norm*{\bfB_l\bfR\bfY}_{p,2}^p.
\end{align*}
Chaining these bounds together yields the claimed result. 
\end{proof}

We can now apply Lemma \ref{lem:lewis-p2} with $\bfB = \bfA\bfP_{\tilde F} = \bfA(\tilde\bfY^\top\bfG)(\tilde\bfY^\top\bfG)^{-}$. In fact, we can apply Lemma \ref{lem:lewis-p2} with $\bfB = \bfA\tilde\bfY^\top\bfG$, since the online Lewis weights only depend on the column span of the matrix. By Lemma \ref{lem:round-lra}, this is an $n\times t$ integer matrix with entries bounded by $\poly(n\Delta)$ up to scaling, unless $\rank(\bfA) \leq 2t$, in which case we can just use the online Lewis weights of $\bfA$ directly, as we will show. This gives us the following algorithm, Algorithm \ref{alg:online-sensitivity}, for approximating online sensitivities:

\begin{algorithm}
\caption{Online Sensitivity Approximation}
\textbf{input:} $\bfA\in\mathbb Z^{n\times d}$ with entries bounded by $\Delta$, rank $k$. \\
\textbf{output:} Online sensitivity approximations $\tilde\bfsigma_i^\OL$.
\begin{algorithmic}[1] 
    \State Draw a $t\times d$ SRHT matrix $\bfG$ for $t = O(k\log k + \log^2 n)$
    \For{$i\in[n]$}
        \State $\bfL_i = \textsc{OnlineLewis}(\bfA_i\bfG^\top)$ \label{line:olw-AG}
        \If{\text{$\bfL_i$ sampled a new row}}\label{line:new-row}
            \State Solve for $\tilde\bfY_i$ as in Lemma \ref{lem:online-point-red} and rounded as in Lemma \ref{lem:round-lra}
            \State $\tilde F_i\gets \rowspan(\bfA_i\bfG^\top\tilde\bfY_i)$\label{line:proj}
            \State $v_i \gets \norm*{\bfa_i^\top(\bfI - \bfP_{\tilde F_i})}_2^p$
        \Else
            \State $\tilde F_i \gets \tilde F_{i-1}$, $\tilde\bfY_i \gets \tilde\bfY_{i-1}$
            \State $v_i \gets v_i + \norm*{\bfa_i^\top(\bfI - \bfP_{\tilde F_i})}_2^p$
        \EndIf
        \State $\tilde\bfw_i^\OL \gets \textsc{OnlineLewis}(\bfA_i\tilde\bfY_i^\top\bfG)$
        \State $\tilde\bfsigma_i^\OL \gets O(1)\bracks*{\norm*{\bfa_i^\top(\bfI - \bfP_{\tilde F_i})}_2^p / v_i + O(t\log(n\Delta))^{0\lor(p/2-1)}\tilde\bfw_i^\OL}$
    \EndFor
\end{algorithmic}\label{alg:online-sensitivity}
\end{algorithm}

\begin{Lemma}[Sum of Online Sensitivities]\label{lem:online-sensitivities-sum}
Let $\bfA\in\mathbb Z^{n\times d}$ have entries bounded by $\Delta$. Fix $i\in[n]$. Then, with probability at least $3/10$, Algorithm \ref{alg:online-sensitivity} returns an upper bound $\tilde\bfsigma_i^\OL$ on the online sensitivity such that
\[
    \bfsigma_i^\OL(\bfA) \leq \tilde\bfsigma_i^\OL.
\]
Furthermore,
\[
    \sum_{i=1}^n \tilde\bfsigma_i^\OL(\bfA) \leq O(t^2\log^2(n\Delta))^{1\lor(p/2)}(\log^2 t)(\log n)
\]
where $t = O(k\log k + \log^2 n)$.
\end{Lemma}
\begin{proof}
Fix $i\in[n]$. Then, by Lemma \ref{lem:online-point-red}, there is a $3/10$ probability that $\bfL_i\bfA_i$ spans an $O(1)$-approximately optimal solution. Condition on this event and let $\tilde F_i = \rowspan(\bfA_i\bfG^\top\tilde\bfY_i)$. Then by Lemma \ref{lem:vx12-thm-7}, we may bound the $i$th online sensitivity by
\[
    \bfsigma_i^\OL(\bfA) = \bfsigma_i(\bfA_i) \leq O(1)\bracks*{\bfsigma_i(\bfA_i\bfP_{\tilde F}) + \frac{\norm*{\bfa_i^\top(\bfI-\bfP_{\tilde F})}_2^p}{\norm*{\bfA_i(\bfI-\bfP_{\tilde F})}_{p,2}^p}}.
\]
Now let $\bfA_i'$ denote the last $I$ rows of $\bfA_i$, where $I$ is the number of rows that have streamed in since the last time a row was sampled from $\bfA_i\bfG^\top$ in Line \ref{line:olw-AG}. That is, it is the set of rows which used the same subspace $\tilde F_i$ as row $i$. Note then that
\[
    \bfsigma_i^\OL(\bfA) \leq O(1)\bracks*{\bfsigma_i(\bfA_i'\bfP_{\tilde F}) + \frac{\norm*{\bfa_i'^\top(\bfI-\bfP_{\tilde F})}_2^p}{\norm*{\bfA_i'(\bfI-\bfP_{\tilde F})}_{p,2}^p}}.
\]
By Lemma \ref{lem:lewis-p2} and the bit complexity bound from Lemma \ref{lem:round-lra}, the first term can be bounded by the online Lewis weight, i.e., 
\[
    \bfsigma_i(\bfA_i'\bfP_{\tilde F}) \leq  O(t\log(n\Delta))^{0\lor(p/2-1)}\bfw_i^{p,\OL}(\bfA_i'\bfP_{\tilde F}).
\]
We now bound the sum. First fix a segment of the stream which has the same $\tilde F_i$. Note that there are at most $m = O(t\log(n\Delta))^{1\lor(p/2)}(\log^2 t)(\log n)$ such segments by Theorem \ref{thm:online-lewis-p>2}. Similarly, the sum of online Lewis weights in this segment is bounded by $O(t\log(n\Delta))^{1\lor(p/2)}$ by Lemma 3.7 of \cite{WY2022b}. This gives a total contribution of
\[
    O(t\log(n\Delta))^{1\lor(p/2)}\cdot O(t\log(n\Delta))^{1\lor(p/2)}(\log^2 k)(\log n) = O(t^2\log^2(n\Delta))^{1\lor(p/2)}(\log^2 t)(\log n)
\]

To bound the second term, we bound the number of times which $\norm*{\bfA_i'(\bfI-\bfP_{\tilde F})}_{p,2}^p$ can double as $i$ ranges over $[n]$. Note that
\[
    \norm*{\bfA_i'(\bfI-\bfP_{\tilde F})}_{p,2}^p \geq \min_{\rank(\bfX)\leq t}\norm*{\bfA_i'(\bfI-\bfX)}_{p,2}^p \geq \frac1{\poly(n)} \min_{\rank(\bfX)\leq t}\norm*{\bfA_i'(\bfI-\bfX)}_2^p = \sigma_{t+1}(\bfA_i')^p \geq \poly(n\Delta)^{-t}
\]
We also have
\[
    \norm*{\bfA_i'(\bfI-\bfP_{\tilde F})}_{p,2}^p \leq \poly(n\Delta)\norm*{\bfA_i'}_2^p.
\]
Thus, $\norm*{\bfA_i'(\bfI-\bfP_{\tilde F})}_{p,2}^p$ can double at most $O(t\log(n\Delta))$ times. Then in each of these windows in which the the mass $\norm*{\bfA_i'(\bfI-\bfP_{\tilde F})}_{p,2}^p$ does not double, the $\norm*{\bfa_i'^\top(\bfI-\bfP_{\tilde F})}_{2}^p$ must sum to $O(\norm*{\bfA_i'(\bfI-\bfP_{\tilde F})}_{p,2}^p)$ so the second term adds up to at most $O(t\log(n\Delta))$ in each segment, which is dominated by the sum of online Lewis weights.
\end{proof}

\begin{Remark}\label{rem:random-order}
If we do not need to algorithmically approximate the online sensitivities, then we can get better existential bounds. Indeed, the argument in Section \ref{sec:techniques} shows a bound of
\[
    O(k\log(n\Delta))^{1 + (1\lor(p/2))},
\]
for integer inputs, or
\[
    O(k\log(n\kappa^\OL))^{1\lor(p/2)}\log(n\kappa^\OL),
\]
for real inputs. For random order streams, an even simpler argument of \cite{CMP2020} yields a bound of
\[
    O(k^{1\lor(p/2)}\log n).
\]
Indeed, we can view the online sensitivity of row $i$ as the sensitivity of a random row among a subset of $i$ random rows, which is at most $s / i$ in expectation, where $s = O(k^{1\lor(p/2)})$ is an upper bound on the offline total sensitivity, using Lemma \ref{lem:lewis-p2}. By linearity of expectation, this is at most $O(s\log n)$ in expectation.
\end{Remark}

\subsection{Sensitivity Sampling}\label{sec:sens-sample}

With sensitivity estimates in hand, we now use sensitivity sampling results for subspace approximation from \cite{HV2020}.

To show sensitivity sampling bounds that are independent of $d$, \cite{HV2020} use the result of \cite{SW2018} which states that there exists a $O(k/\eps^2)$-dimensional subspace $\Gamma$ which preserves subspace approximation objectives, which then implies that it is sufficient to prove coreset guarantees over the low dimensional subspace $\Gamma$ instead.

We will need the following results of \cite{SW2018}, with the theorem numbering from arXiv version 2. 

\begin{Lemma}[Lemma 6, \cite{SW2018}]\label{lem:sw2018-lem6}
Let $\eps > 0$ and let $\tau = \eps^{1 \lor (2/p)}$. Let $W$ be a subspace of dimension at most $k$. If $V$ is a subspace of any dimension such that
\[
    \norm*{\bfA(\bfI-\bfP_V)}_{p,2}^p \leq (1+\eps)\OPT_k
\]
and
\[
    \norm*{\bfA(\bfI-\bfP_{V})}_{p,2}^p - \norm*{\bfA(\bfI-\bfP_{V\cup W})}_{p,2}^p \leq \tau\OPT_k,
\]
then
\[
    \norm*{\bfA\bfP_V - \bfA\bfP_{V\cup W}}_{p,2}^p \leq O(\eps) \OPT_k.
\]
Such a subspace $V$ can be constructed by Algorithm 1 of \cite{SW2018}.
\end{Lemma}

\begin{Lemma}[Theorem 10, \cite{SW2018}]\label{lem:sw2018-thm10}
Let $V\subseteq\mathbb R^d$ be a subspace such that for all rank $k$ subspaces $W\subseteq\mathbb R^d$,
\[
    \norm*{\bfA\bfP_V - \bfA\bfP_{V\cup W}}_{p,2}^p \leq \eps^{p+3}\cdot \OPT_k.
\]
Let $\bfB\in\mathbb R^{n\times (d+1)}$ be the matrix with $\bfA\bfP_V$ in its first $d$ columns and $\{\norm*{\bfa_i^\top(\bfI-\bfP_V)}_2\}_{i=1}^n$ as its $(d+1)$st column. Then for all rank $k$ subspaces $W\subseteq\mathbb R^d$,
\[
    \norm*{\bfA(\bfI_d-\bfP_W)}_{p,2}^p = (1\pm O(\eps)) \norm*{\bfB(\bfI_{d+1}-\bfP_W')}_{p,2}^p,
\]
where $\bfP_W'\in\mathbb R^{(d+1)\times(d+1)}$ is the projection matrix which applies $\bfP_W$ on the first $d$ coordinates and zeros out the $(d+1)$st coordinate.
\end{Lemma}

We will adapt Lemma 5.7 of \cite{HV2020} to the $\ell_p$ subspace approximation problem. We will need their Theorem 5.10, with a couple of adjustments: their use of results from \cite{FL2011} are replaced by the corresponding ``independent sampling'' version of Theorem \ref{thm:fl2011-indep}, and their use of the existence of a set of $\tilde O(k^2/\eps)$ points \cite{SV2007, SV2012} spanning a $(1+\eps)$-optimal solution is replaced by the strong coresets of \cite{SW2018} of size $O(k^{1\lor(p/2)} / \eps^{(p+3)(1\lor(2/p))})$, which has a better dependence on $k$ for $p < 4$. 

\begin{Lemma}[Weak Coreset for Subspace Approximation, Theorem 5.10, \cite{HV2020}]\label{lem:weak-coreset}
Suppose $\tilde\bfsigma_i(\bfA)$ satisfies 
\[
    \tilde\bfsigma_i(\bfA) \geq \sup_{F\in\mathcal F_k} \frac{\norm*{\bfa_i^\top(\bfI-\bfP_F)}_2^p}{\norm*{\bfA(\bfI-\bfP_F)}_{p,2}^p}
\]
Let $\tilde{\mathfrak{S}} = \sum_{i=1}^n \tilde\bfsigma_i(\bfA)$ denote the total sensitivity and let $\mathcal S \geq \tilde{\mathfrak S}$ be an upper bound on the total sensitivity. Suppose a sampling matrix $\bfS\in\mathbb R^{s\times n}$ is constructed as done in Theorem \ref{thm:fl2011-indep}, with VC-dimension upper bound
\[
    d = O(k)\cdot\min\braces*{\eps^{-1}k^2\log(k/\eps), \eps^{-(p+3)(1\lor(2/p))} k^{1\lor(p/2)}}
\]
Then with probability at least $1-\delta$, we have that
\[
    \min_{F\in\mathcal F_k}\norm*{\bfS\bfA(\bfI-\bfP_F)}_{p,2}^p = (1\pm\eps)\min_{F\in\mathcal F_k}\norm*{\bfA(\bfI-\bfP_F)}_{p,2}^p
\]
and $\bfS$ samples at most
\[
    O\parens*{\frac{\mathcal S}{\eps^2}\parens*{dk\log\mathcal S + \log\frac1\delta}}
\]
rows.
\end{Lemma}

The above results can be used to show a version of Lemmas \ref{lem:sw2018-lem6} and \ref{lem:sw2018-thm10} for the sampled matrix $\bfS\bfA$. This simplifies and sharpens Lemma 5.7 of \cite{HV2020} for $\ell_p$ subspace approximation.

\begin{Lemma}\label{lem:hv2020-lem5.7-analogue}
Let $\eps' = \eps^{(p+3)\cdot(1\lor(2/p))}$. Let $\bfS$ be sampled as in Lemma \ref{lem:weak-coreset}, using rank $2k$ sensitivities, where the expected number of rows sampled is
\[
    s = O\parens*{\mathcal S\parens*{k\cdot\min\{k^2\log k, k^{1\lor(p/2)}\}\log\mathcal S + \eps'^{-2}\log\frac1\delta}}
\]
Suppose $\Gamma\subseteq\mathbb R^d$ is a subspace such that for any subspace $W\subset\mathbb R^d$ of dimension at most $k$,
\begin{equation}\label{eq:rep-subspace-cond}
    \norm*{\bfA(\bfI-\bfP_\Gamma)}_{p,2}^p - \norm*{\bfA(\bfI-\bfP_{\Gamma\cup W})}_{p,2}^p \leq \eps'\cdot\OPT_k(\bfA).
\end{equation}
and also contains $V^*$, where
\[
V^* = \arg\min_{V'\in\mathcal F_k}\norm*{\bfA(\bfI-\bfP_{V'})}_{p,2}^p.
\]
Let $\bfB\in\mathbb R^{n\times (d+1)}$ be the matrix with $\bfA\bfP_\Gamma$ in its first $d$ columns and $\{\norm*{\bfa_i^\top(\bfI-\bfP_\Gamma)}_2\}_{i=1}^n$ as its $(d+1)$st column. Then, with probability at least $1-\delta$, for any rank $k$ subspace $W$,
\[
    \norm*{\bfS\bfA(\bfI_d-\bfP_W)}_{p,2}^p = (1\pm O(\eps)) \norm*{\bfS\bfB(\bfI_{d+1}-\bfP_W')}_{p,2}^p,
\]
where $\bfP_W'\in\mathbb R^{(d+1)\times(d+1)}$ is the projection matrix which applies $\bfP_W$ on the first $d$ coordinates and zeros out the $(d+1)$st coordinate.
\end{Lemma}
\begin{proof}
We will show that \eqref{eq:rep-subspace-cond} implies a similar condition for $\bfS\bfA$, which yields the result by Lemma \ref{lem:sw2018-thm10}. Thus, we will bound
\begin{align*}
    \norm*{\bfS\bfA(\bfI-\bfP_\Gamma)}_{p,2}^p - \norm*{\bfS\bfA(\bfI-\bfP_{\Gamma\cup W})}_{p,2}^p &\leq \norm*{\bfS\bfA(\bfI-\bfP_\Gamma)}_{p,2}^p - \min_{W'\in\mathcal F_k}\norm*{\bfS\bfA(\bfI-\bfP_{\Gamma\cup W'})}_{p,2}^p \\
    &= \norm*{\bfS\bfA(\bfI-\bfP_\Gamma)}_{p,2}^p - \norm*{\bfS\bfA(\bfI-\bfP_{\Gamma\cup W^*})}_{p,2}^p
\end{align*}
where
\[
    W^* = \arg\min_{W'\in\mathcal F_k}\norm*{\bfS\bfA(\bfI-\bfP_{\Gamma\cup W'})}_{p,2}^p.
\]

Recall that $V^*$ is the optimal rank $k$ solution for $\bfA$ achieving the value $\OPT_k(\bfA)$. Note that
\[
    \frac{1}{\tilde\bfsigma_i(\bfA)}\norm*{\bfa_i^\top(\bfI-\bfP_{\Gamma\cup W^*})}_2^p \leq \frac{1}{\tilde\bfsigma_i(\bfA)}\norm*{\bfa_i^\top(\bfI-\bfP_{V^*\cup W^*})}_2^p \leq \norm*{\bfA_i^\top(\bfI-\bfP_{V^*\cup W^*})}_{p,2}^p \leq \OPT_k(\bfA)
\]
since $\tilde\bfsigma_i(\bfA)$ upper bound the rank $2k$ sensitivities. Then by Bernstein's inequality, we have that
\[
    \Pr\braces*{\abs*{\norm*{\bfS\bfA(\bfI-\bfP_{\Gamma\cup W^*})}_{p,2}^p - \norm*{\bfA(\bfI-\bfP_{\Gamma\cup W^*})}_{p,2}^p} \geq \eps' \OPT_k(\bfA)} \leq 2\exp\parens*{-\frac{s}{\mathcal S}\frac{(\eps'\OPT_k(\bfA))^2}{2(1+\eps'/3)\OPT_k(\bfA)^2}} < \delta
\]
since $s \geq C\mathcal{S}\eps'^{-2}\log(1/\delta)$ for a large enough constant $C$. Similarly,
\[
    \Pr\braces*{\abs*{\norm*{\bfS\bfA(\bfI-\bfP_{\Gamma})}_{p,2}^p - \norm*{\bfA(\bfI-\bfP_{\Gamma})}_{p,2}^p} \geq \eps' \OPT_k(\bfA)} \leq 2\exp\parens*{-\frac{s}{\mathcal S}\frac{(\eps'\OPT_k(\bfA))^2}{2(1+\eps'/3)\OPT_k(\bfA)^2}} < \delta.
\]
Thus, conditioned on the success of these events, we have that
\begin{align*}
\norm*{\bfS\bfA(\bfI-\bfP_\Gamma)}_{p,2}^p - \norm*{\bfS\bfA(\bfI-\bfP_{\Gamma\cup W^*})}_{p,2}^p &\leq \norm*{\bfA(\bfI-\bfP_\Gamma)}_{p,2}^p - \norm*{\bfA(\bfI-\bfP_{\Gamma\cup W^*})}_{p,2}^p + 2\eps'\OPT_k(\bfA) \\
&\leq 3\eps'\OPT_k(\bfA).
\end{align*}
Since $\bfS$ preserves the optimal cost up to $O(1)$ factors by Lemma \ref{lem:weak-coreset}, this is bounded by $O(\eps')\OPT_k(\bfS\bfA)$. Finally, we have by Lemma \ref{lem:sw2018-lem6} that
\[
    \norm*{\bfS\bfA\bfP_\Gamma - \bfS\bfA\bfP_{\Gamma\cup W}}_{p,2}^p \leq \eps^{p+3}\cdot\OPT_k(\bfS\bfA)
\]
for all rank $k$ subspaces $W$, and by Lemma \ref{lem:sw2018-thm10} that
\[
    \norm*{\bfS\bfA(\bfI_d-\bfP_W)}_{p,2}^p = (1\pm O(\eps)) \norm*{\bfS\bfB(\bfI_{d+1}-\bfP_W')}_{p,2}^p
\]
for all rank $k$ subspaces $W$.
\end{proof}

By Lemma \ref{lem:hv2020-lem5.7-analogue}, it now suffices to show coreset guarantees for the low dimensional subspace spanned by the rows of $\bfA\bfP_V$, rather than $\bfA$. Indeed, if we show that
\begin{equation}\label{eq:low-dim-coreset}
    \norm*{\bfS\bfB(\bfI_{d+1}-\bfP_W')}_{p,2}^p = (1\pm O(\eps))\norm*{\bfB(\bfI_{d+1}-\bfP_W')}_{p,2}^p
\end{equation}
for all $W\in\mathcal F_k$, then this implies that
\[
    \norm*{\bfS\bfA(\bfI_{d}-\bfP_W)}_{p,2}^p = (1\pm O(\eps))\norm*{\bfA(\bfI_{d}-\bfP_W)}_{p,2}^p
\]
by the chain of approximations from Lemmas \ref{lem:hv2020-lem5.7-analogue} and \ref{lem:sw2018-thm10}, using the subspace $\Gamma$ as constructed in Algorithm 1 of \cite{SW2018}. To show \eqref{eq:low-dim-coreset}, it suffices to use the independent sampling version \cite{FL2011} again (Theorem \ref{thm:fl2011-indep}), this time using the ambient dimension as an upper bound on the VC-dimension, which is just $O(k/\eps^{(p+3)\cdot(1\lor(2/p)}) \times k$. This is analogous to Lemma 5.5 of \cite{HV2020}.

\begin{Lemma}\label{lem:low-dim-fl}
Let $\bfB\in\mathbb R^{n\times d}$ a rank $r$ matrix. Let $\bfS$ be sampled as in Theorem \ref{thm:fl2011-indep} with $d = rk$. Then, with probability at least $1-\delta$, simultaneously for all rank $k$ subspaces $W$,
\[
    \norm*{\bfS\bfB(\bfI_d-\bfP_W)}_{p,2}^p = (1\pm\eps)\norm*{\bfB(\bfI_d - \bfP_W}_{p,2}^p.
\]
\end{Lemma}

\section{Online Coresets for Entrywise \texorpdfstring{$\ell_p$}{lp} Low Rank Approximation}\label{sec:entrywise}

We discuss how an argument of \cite{JLLMW2021}, together with our Theorem \ref{thm:informal-real}, gives the following result on online coresets for entrywise $\ell_p$ low rank approximation.

\Entrywise*

The streaming entrywise low rank approximation algorithm of \cite{JLLMW2021} roughly proceeds as follows. First, an oblivious sketch using \emph{$p$-stable random variables} is used to approximate the low rank approximation cost by the cost of an $n\times t$ matrix for $t = k(\log n)^{O(1)}$. With only $t$ columns, we can afford to approximate the $\ell_p$ norm of the rows by the $\ell_2$ norm of the rows, up to a $t^{1/p-1/2}$ factor, using the equivalence of $\ell_p$ norms. This is now just the subspace approximation problem, for which coreset constructions apply. 

The following two lemmas show that sketching with $p$-stable variables preserves the objective function value. The first shows that the sketch does not contract objective function values for any column subset, while the second shows that the sketch does not expand objective function values by too much with fixed probability. The asymmetry in the sketching guarantees can be attributed to the heavy-tailedness of $p$-stable variables.

\begin{Lemma}[No Contraction, Lemma 2, \cite{JLLMW2021}]\label{lem:jllmw2}
Let $\bfA\in\mathbb R^{n\times d}$ and let $s\in\mathbb N$. Let $\delta\in(0,\delta)$. Let $t = s(\log\frac{n}{\delta})^{O(1)}$, and let $\bfG\in\mathbb R^{t\times d}$ be a matrix whose entries are i.i.d.\ standard $p$-stable random variables (see \cite[Definition 2.4]{JLLMW2021}), rescaled by $\Theta(1/t^{1/p})$. Then with probability at least $1-\delta$, for all subsets $T\subseteq[n]$ with $\abs*{T}\leq s$ and all $\bfV\in\mathbb R^{n\times T}$, 
\[
    \norm*{\bfV\bfS_T\bfA - \bfA}_{p,p} \leq \norm*{\bfV\bfS_T\bfA\bfG^\top - \bfA\bfG^\top}_{p,p}
\]
where $\bfS_T$ is the sampling matrix associated with $T$.
\end{Lemma}


\begin{Lemma}[No Expansion, Lemma E.17, \cite{SWZ2017}]\label{lem:jllmw3}
Let $\bfA\in\mathbb R^{n\times d}$. Let $\delta\in(0,\delta)$. Let $t\in\mathbb N$, and let $\bfG\in\mathbb R^{t\times d}$ be a matrix whose entries are i.i.d.\ standard $p$-stable random variables, rescaled by $\Theta(1/t^{1/p})$. Then with probability at least $1-\delta$,
\[
    \norm*{\bfA\bfG^\top}_{p,p} \leq \frac1\delta\parens*{\log\frac{n}{\delta}}^{O(1)}\norm*{\bfA}_{p,p}.
\]
\end{Lemma}

We now return to the proof of Corollary \ref{cor:entrywise}.

\begin{proof}[Proof of Corollary \ref{cor:entrywise}]
Our online coreset algorithm is to first sketch with a $p$-stable matrix $\bfG\in\mathbb R^{t\times d}$, and then apply our Theorem \ref{thm:informal-real} on $\bfA\bfG^\top$. The $t$ here is chosen as in Lemma \ref{lem:jllmw2}, with $s$ being the sample complexity of our Theorem \ref{thm:informal-real}.

Note that sketching with $\bfG$ preserves the optimal rank $k$ approximation cost in the $\norm*{\cdot}_{p,p}$ norm by Lemmas \ref{lem:jllmw2} and \ref{lem:jllmw3}, which in turn preserves the optimal rank $k$ approximation cost in $\norm*{\cdot}_{p,p}$ norm, both up to a $\poly(n)$ factor. Then, the rounding reduction to integer matrices as in Section \ref{sec:int-red} still works the same way, with sample complexity depending on the online condition number of $\bfA$ rather than $\bfA\bfG^\top$.

Let $\bfS$ be the online coreset sampling matrix given by Theorem \ref{thm:informal-real} with constant $\eps$ and $\delta$. Then, we have that
\[
    \norm*{\bfS\bfA\bfG^\top(\bfI-\bfX)}_{p,2} = \Theta(1)\norm*{\bfA\bfG^\top(\bfI-\bfX)}_{p,2}
\]
for all rank $k$ projections $\bfX$. Now let $\bfX^*$ denote the optimal rank $k$ projection for $\bfS\bfA\bfG^\top$. Note that $\bfX^*$ can be written as $\bfY^*(\bfS\bfA\bfG^\top)$, as it is a rank $k$ projection in the row span of $\bfS\bfA\bfG^\top$. Then,
\[
    \norm*{\bfA\bfG^\top - \bfA\bfG^\top\bfY^*\bfS\bfA\bfG^\top}_{p,2} \leq \Theta(1)\min_{\rank(\bfX) \leq k}\norm*{\bfA\bfG^\top(\bfI-\bfX)}_{p,2} = \Theta(1)\min_{\rank(\bfB) \leq k}\norm*{\bfA\bfG^\top - \bfB}_{p,2}
\]
so
\[
    \min_{\rank(\bfV)\leq k}\norm*{\bfA\bfG^\top - \bfV\bfS\bfA\bfG^\top}_{p,2} \leq \Theta(1)\min_{\rank(\bfB) \leq k}\norm*{\bfA\bfG^\top - \bfB}_{p,2}.
\]
Let $\bfV^*$ witness the minimum on the left hand side, and let $\tilde \bfA_k = \argmin_{\rank(\tilde\bfA)\leq k}\norm{\bfA-\tilde\bfA}_{p,p}$. Then,
\begin{align*}
\min_{\rank(\bfV) \leq k} \norm*{\bfV\bfS\bfA - \bfA}_{p,p} &\leq \norm*{\bfV^*\bfS\bfA - \bfA}_{p,p} \\
&\leq \norm*{\bfV^*\bfS\bfA\bfG^\top - \bfA\bfG^\top}_{p,p} && \text{No contraction (Lemma \ref{lem:jllmw2})} \\
&\leq t^{1/p-1/2}\norm*{\bfV^*\bfS\bfA\bfG^\top - \bfA\bfG^\top}_{p,2} && \text{Equivalence of $\ell_p$ norms in $\mathbb R^t$} \\
&\leq O(t^{1/p-1/2})\min_{\rank(\bfB)\leq k}\norm*{\bfB - \bfA\bfG^\top}_{p,2} \\
&\leq O(t^{1/p-1/2})\min_{\rank(\bfB)\leq k}\norm*{\bfB - \bfA\bfG^\top}_{p,p} && \text{Monotonicity of $\ell_p$ norms} \\
&\leq O(t^{1/p-1/2})\norm*{\tilde\bfA_k\bfG^\top - \bfA\bfG^\top}_{p,p} \\
&\leq O(t^{1/p-1/2})(\log n)^{O(1)}\norm*{\tilde\bfA_k - \bfA}_{p,p} && \text{No expansion (Lemma \ref{lem:jllmw3})} \qedhere
\end{align*}
\end{proof}

\section{Online Coresets for Euclidean \texorpdfstring{$(k,p)$}{(k,p)}-Clustering}\label{sec:online-coreset-clustering}

Our main result of this section is the following:

\begin{Theorem}\label{thm:coreset-clustering}
Let $w^\OL$ be a lower bound on all nonzero costs for $(k,p)$-clustering $\bfA_i$ for $i\in[n]$, and let $W^\OL$ similarly be an upper bound. Then, there is a strong online coreset algorithm which, with probability at least $1-\delta$, samples at most
\[
    \min\braces*{\tilde O\parens*{\eps^{-4}k^2(\log n)^3\parens*{\log\frac{n}{\delta}}\log\frac{W^\OL}{w^\OL}}, \tilde O\parens*{\eps^{-p-3}k(\log n)\parens*{\log\frac{n}{\delta}}^2\log\frac{W^\OL}{w^\OL}}}
\]
points.
\end{Theorem}
\begin{proof}
The bulk of the work is in showing that we can get online sensitivity estimates $\tilde\bfsigma_i^\OL$ which upper bound the true sensitivities and sum to at most
\[
    \sum_{i=1}^n \tilde\bfsigma_i(\bfA) \leq O\parens*{k(\log n)^2\parens*{\log\frac{n}{\delta}}\log\frac{W^\OL}{w^\OL}}
\]
with probability at least $1-\delta$. We show this in Section \ref{sec:online-sensitivity-approximation-clustering}. The result then follows from a standard terminal embedding \cite{NN2019} and sensitivity sampling \cite{FL2011} argument (see Section 3 of \cite{HV2020}), done in an online manner (Theorem \ref{thm:fl2011-indep}). We defer the standard details to similar arguments in \cite{HV2020}.

We also adapt and improve another argument of \cite{FL2011}, which gives the latter sample complexity. The analysis is similar and is described in Section \ref{sec:improved-fl}. 
\end{proof}

\subsection{Online Sensitivity Approximation}\label{sec:online-sensitivity-approximation-clustering}

The sensitivity approximation approach we take is to first compute a bicriteria solution, move the points to the bicriteria solution, and then compute the sensitivities of the resulting points, which is just the reciprocal of the cluster size. 

We use the following result on online $k$ clustering due to \cite{LSS2016} in order to obtain a bicriteria solution in an online manner. While \cite{LSS2016} only state their result for $k$-means clustering corresponding to $p = 2$, we show that their algorithm generalizes straightforwardly to any $p\geq 1$ in Appendix \ref{sec:lss}. 

\begin{restatable}[\cite{LSS2016}]{Theorem}{LSS}
\label{thm:lss2016}
Let $\bfA\in\mathbb R^{n\times d}$ and $p\geq 1$. There is an online algorithm for Euclidean $(k,p)$-clustering, Algorithm \ref{alg:online-k-clustering}, which takes as input a cost lower bound $w^*$ and immediately assigns each incoming point to at most
\[
    O\parens*{k(\log n)\log\frac{W^*}{w^*}}
\]
clusters $\tilde C$ and has cost at most
\[
    \sum_{i=1}^n d(\bfa_i, \tilde C)^p = O(W^*),
\]
where
\[
    W^* = \min_{C\subseteq \mathbb R^d, \abs*{C} \leq k} \sum_{i=1}^n d(\bfa_i, C)^p.
\]
\end{restatable}

\begin{algorithm}
\caption{Online Sensitivity Approximation for Euclidean $(k,p)$-Clustering}
\textbf{input:} $\bfA\in\mathbb R^{n\times d}$, number of clusters $k$, cost lower bound $w^*$. \\
\textbf{output:} Online sensitivity estimates $\tilde\bfsigma_i^\OL(\bfA)$.
\begin{algorithmic}[1] 
    \State $C\gets\varnothing$ \Comment{Bicriteria clustering using Algorithm \ref{alg:online-k-clustering}} 
    \State $v \gets 0$
    \For{$i\in[n]$}
        \State Update clustering $C$ with $\bfa_i$ using Algorithm \ref{alg:online-k-clustering}
        \State $\bfc_i \gets \argmin_{\bfc\in C}\norm*{\bfa_i - \bfc}_2^p$
        \State $v \gets v + \norm*{\bfa_i - \bfc}_2^p$\label{line:bicriteria-cost}
        \State $S_i \gets \braces*{j\in[i] : \bfc_j = \bfc_i}$ \Comment{The set of points in the same cluster as $i$}
        \State $\tilde\bfsigma_i^\OL(\bfA) \gets O(1)\parens*{\frac{d(\bfa_i, C)^p}{v} + \frac1{\abs*{S_i}}}$\label{line:sensitivity-output}
    \EndFor
\end{algorithmic}\label{alg:online-k-clustering-sensitivity}
\end{algorithm}

The next lemma follows \cite[Lemma 5.5]{HV2020}. While our result only works with constant probability for a fixed prefix stream $\bfA_i$, this can be boosted to $\poly(\delta/n)$ probability by repetition and summing over the sensitivities, with only a $\log\frac{n}{\delta}$ factor loss in the total sensitivity.

\begin{Lemma}
Let $\tilde\bfsigma_i(\bfA)$ be output by Algorithm \ref{alg:online-k-clustering-sensitivity} (specifically Line \ref{line:sensitivity-output}). Then, for each $i\in[n]$, with constant probability,
\[
    \tilde\bfsigma_i(\bfA) \geq \bfsigma_i(\bfA) = \sup_{C\subseteq\mathbb R^d, \abs*{C}\leq k}\frac{d(\bfa_i, C)^p}{\sum_{j=1}^n d(\bfa_j, C)^p}
\]
\end{Lemma}
\begin{proof}
We condition on the success of the bicriteria solution at time $i$, which occurs with constant probability.

Let $C$ be a set of $k$ points. Then, 
\begin{align*}
d(\bfa_i, C)^p &\leq 2^{p-1}\parens*{d(\bfa_i, \bfc_i)^p + d(\bfc_i, C)^p} \\
&= 2^{p-1}d(\bfa_i, \bfc_i)^p + 2^{p-1}\frac1{\abs*{S_i}}\sum_{j\in S_i}d(\bfc_j, C)^p \\
&\leq 2^{p-1}d(\bfa_i, \bfc_i)^p + 2^{p-1}\frac1{\abs*{S_i}}\sum_{j\in [n]}d(\bfc_j, C)^p \\
&\leq 2^{p-1}d(\bfa_i, \bfc_i)^p + 2^{p-1}\cdot 2^{p-1}\frac1{\abs*{S_i}}\sum_{j\in [n]}d(\bfc_j, \bfa_j)^p + d(\bfa_j,C)^p \\
&\leq O(1)\parens*{d(\bfa_i, \bfc_i)^p + \frac1{\abs*{S_i}}\sum_{j\in [n]}d(\bfa_j,C)^p}
\end{align*}
Factoring out $\sum_{j\in [n]}d(\bfa_j,C)^p$ yields the desired conclusion.
\end{proof}

Next, we bound the total sensitivity that is output by Algorithm \ref{alg:online-k-clustering-sensitivity}. Note that we can pass this guarantee to having good total sensitivity with probability $1$, but failing to upper bound the true sensitivities, by outputting zeros after the total sensitivity exceeds the desired threshold total.

\begin{Lemma}
Let $w^\OL$ be a lower bound on all nonzero costs for $(k,p)$-clustering $\bfA_i$ for $i\in[n]$, and let $W^\OL$ similarly be an upper bound. Then for each $i\in[n]$, with constant probability,
\[
    \sum_{i=1}^n \tilde\bfsigma_i^\OL(\bfA) \leq O\parens*{k(\log n)^2\log\frac{W^\OL}{w^\OL}}
\]
\end{Lemma}
\begin{proof}
We condition on the success of the bicriteria solution at time $i$, which occurs with constant probability.

Note that the quantity $v$ in Line \ref{line:bicriteria-cost} can double at most $O\parens*{\log\parens*{\frac{W^\OL}{w^\OL}}}$ times. While $v$ has not doubled, the quantity $d(\bfa_i,C)^p/v$ sums to at most $O(1)$. Thus, the total sensitivity contribution from this term is at most $O\parens*{\log\parens*{\frac{W^\OL}{w^\OL}}}$. To analyze the second term $1/\abs*{S_i}$, note that for a given one of the bicriteria clusters, these terms sum to at most $1 + 1/2 + 1/3 + \dots + 1/n = O(\log n)$. Then across all $O(k(\log n)\log\frac{W^\OL}{w^\OL})$ clusters formed by the bicriteria solution, the total sensitivity contribution is $O(k(\log n)^2\log\frac{W^\OL}{w^\OL})$. 
\end{proof}

\subsection{Improved Feldman--Langberg Algorithm}\label{sec:improved-fl}

Next, we show how to adapt an improved importance sampling algorithm of Feldman--Langberg \cite[Theorem 15.5]{FL2011}. We show how to implement this algorithm in the online setting by using independent sampling without replacement in Theorem \ref{thm:fl-coreset}, and along the way, streamline their argument, improve their dependence on $\eps$ by a $\eps^{p-1}$ factor, and generalize to the case when the initial bicriteria clustering algorithm outputs more than $k$ centers. 

Given Theorem \ref{thm:fl-coreset}, our argument proceeds similarly to our first sensitivity sampling algorithm. For a constant probability of success of estimating the sampling score, we use the bicriteria algorithm of Theorem \ref{thm:lss2016}, which outputs $O(k(\log n)\log\frac{W^\OL}{w^\OL})$ centers. Then, the scores contributed by the reciprocal of the cluster size contributes $O(\frac{k}{\eps^2}(\log\frac{k}\delta)(\log n)^2\log\frac{W^\OL}{w^\OL})$. On the other hand, the scores contributed by the residual cost term can be analyzed by a similar cost doubling argument as before, which shows that they sum to $O(\log\frac{W^\OL}{w^\OL})$, so the sample complexity contribution from this term is $O((dk\log n + \log\frac1\delta)\frac{1}{\eps^{p+1}}\log\frac{W^\OL}{w^\OL})$. We can further use a terminal embedding to reduce $d$ to $O(\eps^{-2}\log n)$. Finally, we repeat the bicriteria algorithm $\log\frac{n}{\delta}$ times to union bound over all $n$ times $i\in[n]$, so that our final sample complexity is
\[
    O\parens*{\frac{k}{\eps^2}\parens*{\log\frac{k}\delta}(\log n)^2 + \frac1{\eps^{p+3}}k(\log n)\parens*{\log\frac{n}{\delta}}}\parens*{\log\frac{n}{\delta}}\log\frac{W^\OL}{w^\OL}.
\]

\section{Lewis Weight Sampling for Active \texorpdfstring{$\ell_p$}{lp} Regression}
\label{sec:active-lp-regression}

We present our results for active $\ell_p$ regression in the following sections.

We denote the optimal value of the $\ell_p$ regression problem as
\[
    \OPT(\bfA,\bfb) \coloneqq \min_{\bfx\in\mathbb R^d}\norm*{\bfA\bfx-\bfb}_p^p
\]
and the optimal solution as
\[
    \bfx^*(\bfA,\bfb) \coloneqq \arg\min_{\bfx\in\mathbb R^d}\norm*{\bfA\bfx-\bfb}_p^p
\]
We will often drop the dependence on $\bfA$ and $\bfb$ and simply write $\OPT$ and $\bfx^*$ in many cases, if the design matrix and target vector are $\bfA$ and $\bfb$. However, when we need to consider different design matrices and target vectors, we will explicitly write the dependence on these objects.

\subsection{Overview}

Our algorithm will be to sample a Lewis weight sampling matrix $\bfS$ (Definition \ref{def:lewis-weight-sampling}), sample the corresponding rows of $\bfb$, and the solve an $\ell_p$ regression problem with design matrix $\bfS\bfA$ and target vector $\bfS\bfb$. More formally, consider the following:

\begin{Definition}[Active $\ell_p$ Regression Lewis Weight Sampling Matrix]\label{def:lewis-weight-sampling}
Let $\bfA\in\mathbb R^{n\times d}$ and let $\bfw\in\mathbb R^n$ be $\alpha$-one-sided Lewis weights for $\bfA$. Let $\eps\in(0,1)$, $\gamma>0$, and $\delta\in(0,1)$. Then, we define the \emph{active $\ell_p$ regression Lewis weight sampling matrix} $\bfS\in\mathbb R^{n\times n}$ to be a diagonal matrix with $\bfS_{i,i}$ equal to $1/p_i$ with probability $p_i$ and $0$ otherwise, for
\begin{align*}
    p_i &= \min\braces*{\Theta(1)\frac{(p/2)^{\frac{p/2}{1-2/p}}}{\alpha^{p/2}}\frac{\bfw_i}{d\beta}, 1} \\
    \beta &= \frac{\alpha\eps^p}{\gamma\norm*{\bfw}_1^{p/2}\bracks*{(\log(d\norm*{\bfw}_1))^2(\log n) + \log\frac1\delta}}
\end{align*}
\end{Definition}

\begin{Definition}
Let $\bfS\in\mathbb R^{n\times n}$ be a sampling matrix. Then, let
\[
    \bfx_\bfS^* \coloneqq \bfx^*(\bfS\bfA,\bfS\bfb) = \arg\min_{\bfx\in\mathbb R^d}\norm*{\bfS\bfA\bfx-\bfS\bfb}_p^p.
\]
\end{Definition}

Our main result is the following theorem, which shows that solving the Lewis weight-sampled system yields $(1+\eps)$-approximate solutions, whenever solving $\bfS\bfA$ and $\bfS\bfb$ gives a constant factor solution.

\begin{Theorem}\label{thm:iterative-improvement}
Let $\bfS$ be a sampling matrix as defined in Definition \ref{def:lewis-weight-sampling} with $\gamma = O(\eps/\poly\log(1/\eps))$. Then, with probability at least $1-O(\log\log(1/\eps))\cdot\delta$, the following implication is true:
\begin{align*}
    \OPT(\bfS\bfA,\bfS\bfb) = O(\OPT(\bfA,\bfb)) \implies \norm*{\bfA\bfx_\bfS^* - \bfb}_p^p\leq (1+\eps)\OPT.
\end{align*}
\end{Theorem}

We will first use Theorem \ref{thm:iterative-improvement} to design a high-probability algorithm for solving active $\ell_p$ linear regression in Section \ref{sec:alg}, and then prove Theorem \ref{thm:iterative-improvement} in the rest of the section.

\subsection{Algorithm for Active \texorpdfstring{$\ell_p$}{lp} Regression}
\label{sec:alg}

Given Theorem \ref{thm:iterative-improvement}, all we need to do is to identify a constant factor solution, and this will be a $(1+\eps)$-approximate solution with high probability. \cite[Algorithm 5]{MMWY2022} provides such an algorithm, in which a ``median''-like procedure is used to find a good candidate among $O(\log\frac1\delta)$ constant-probability candidates, reproduced in Algorithm \ref{alg:median} below. Our approach is to combine this algorithm with the analysis of Theorem \ref{thm:iterative-improvement}.

\begin{algorithm}
	\caption{Probability Boosting for Constant Factor Active $\ell_p$ Regression}
	\textbf{input:} $\ell$ candidate solutions $\bfx_1,\ldots,\bfx_\ell$ with at least $9/10\cdot \ell$ satisfying $\norm{\bfA\bfx_i-\bfb}_p \le \alpha \min_{\bfx} \norm{\bfA\bfx-\bfb}_p$. \\
	\textbf{output:} Approximate solution $\tilde \bfx \in \mathbb R^d$ to $\min_{\bfx} \norm{\bfA\bfx-\bfb}_p$. 
	\begin{algorithmic}[1]
		\State{Let $\bfd \in \mathbb R^{\ell^2}$ contain all pairwise distances $\norm{\bfA\bfx_{i}-\bfA\bfx_{j}}_p$ (over ordered pairs $(i,j)$) sorted in increasing order. Let $\tau = \bfd(\floor*{\ell^2 \cdot 8/10})$ be the $80^{th}$ percentile distance.}\label{line:thresh}
		\State{Return any $\bfx_{i}$ such that $\norm{\bfA\bfx _{i}-\bfA\bfx _{j}}_p \le \tau$ for at least $1/2 \cdot \ell$ vectors $\bfx_{j}$. }\label{line:return}
	\end{algorithmic}\label{alg:median}
\end{algorithm}

\begin{algorithm}
	\caption{High Probability Active $\ell_p$ Regression}
	\textbf{input:} $\bfA\in\mathbb R^{n\times d}$, $\bfb\in\mathbb R^n$, $\eps>0$, $\delta\in(0,1)$, $p>2$. \\
	\textbf{output:} Approximate solution $\tilde \bfx \in \mathbb R^d$ to $\min_{\bfx} \norm{\bfA\bfx-\bfb}_p$. 
	\begin{algorithmic}[1] 
		\State Compute one-sided lewis weights $\bfw\in\mathbb R^n$ such that $\norm*{\bfw}_1 = O(d)$ (see, e.g., \cite{JLS2022})\label{line:lewis-weights}
        \State For each $i\in[\ell]$ for $\ell = O(\log\frac1\delta)$, let $\bfS^i$ be a Lewis weight sample generated as in Definition \ref{def:lewis-weight-sampling} \label{line:sample-sketch}
        \State For each $i\in[\ell]$, let $\bfx_i = \arg\min_{\bfx}\norm*{\bfS^i\bfA\bfx-\bfS^i\bfb}_p$ \label{line:solve-sketch}
        \State Run Algorithm \ref{alg:median} on the $\ell$ candidates and output the solution\label{line:return-median}
	\end{algorithmic}\label{alg:lp-regression}
\end{algorithm}

\begin{Theorem}[Active $\ell_p$ Regression]\label{thm:main-lp-regression}
Let $\tilde\bfx$ be the output of Algorithm \ref{alg:lp-regression} with failure rate in the sampling process of Definition \ref{def:lewis-weight-sampling}  set to $\delta/\ell\log\log\frac1\eps$ and $\gamma$ set to $\eps\poly\log(1/\eps)$. Then, with probability at least $1-\delta$, $\tilde\bfx$ satisfies
\[
    \norm*{\bfA\tilde\bfx-\bfb}_p \leq (1+\eps)\OPT.
\]
Furthermore, Algorithm \ref{alg:lp-regression} reads at most
\[
    O\parens*{\frac{d^{p/2}}{\eps^{p-1}}\bracks*{(\log d)^2(\log n) + \log\frac1\delta}\poly\log\frac1\eps\log\frac1\delta}
\]
entries of $\bfb$.
\end{Theorem}
\begin{proof}
We first apply Theorem \ref{thm:iterative-improvement} with failure probability $\delta/\ell\log\log\frac1\eps$ and union bound over all $\ell$ trials in Line \ref{line:sample-sketch}, so that with probability at least $1-\delta$, the conclusion of Theorem \ref{thm:iterative-improvement} holds for all $\ell$ trials simultaneously.

By \cite[Theorem 3.2]{MMWY2022}, Line \ref{line:solve-sketch} of Algorithm \ref{alg:lp-regression} yields a constant factor solution with probability at least $0.99$. Then with probability at least $1-\delta$, at least a $0.9$ fraction of the $\ell$ solutions obtained in Line \ref{line:solve-sketch} are constant factor solutions. Then by the proof of \cite[Theorem 3.3]{MMWY2022}, Line \ref{line:return-median} returns a constant factor solution, and thus a $(1+\eps)$-approximate solution.
\end{proof}

By using \emph{online} Lewis weights \cite{WY2022b} rather than the usual Lewis weights in Theorem \ref{thm:main-lp-regression}, we obtain the first $\ell_p$ online active regression algorithm for $p>2$:

\begin{Corollary}[Online Active $\ell_p$ Regression]\label{thm:online-lp-regression}
Let $\tilde\bfx$ be the output of Algorithm \ref{alg:lp-regression} with failure rate in the sampling process of Definition \ref{def:lewis-weight-sampling} set to $\delta/\ell\log\log\frac1\eps$ and $\gamma$ set to $\eps\poly\log(1/\eps)$, and Lewis weights in Line \ref{line:lewis-weights} replaced by online Lewis weights. Then, with probability at least $1-\delta$, $\tilde\bfx$ satisfies
\[
    \norm*{\bfA\tilde\bfx-\bfb}_p \leq (1+\eps)\OPT.
\]
Furthermore, Algorithm \ref{alg:lp-regression} reads at most
\[
    O\parens*{\frac{d^{p/2}}{\eps^{p-1}}(\log(n\kappa^\OL))^{p/2+1}\bracks*{(\log d)^2(\log n) + \log\frac1\delta}\poly\log\frac1\eps\log\frac1\delta}
\]
entries of $\bfb$, in an online manner, where $\kappa^\OL$ is the \emph{online condition number} of $\bfA$.
\end{Corollary}

\subsection{Closeness of Near-Optimal Solutions}\label{sec:closeness}

We start by showing that a $(1+\gamma)$-approximate solution must be $O(\gamma^{1/p})\OPT$-close to the optimal solution, similar to Theorem 3.19 of \cite{MMWY2022}. In \cite{MMWY2022}, a similar result is shown for $1 < p < 2$ using the strong convexity of $\norm*{\cdot}_p^2$. For $p > 2$, strong convexity unfortunately does not hold. Nonetheless, we still obtain a similar statement by using a bound on the Bregman divergence of $\ell_p$ norms shown by \cite{AKPS2019}.

\begin{Lemma}[Bregman Divergence of $\ell_p$ Norms \cite{AKPS2019}]\label{lem:lp-bregman}
Let $p\geq2$. Then, for $\bfy,\bfy'\in\mathbb R^n$, we have
\[
    \norm*{\bfy'}_p^p \geq \norm*{\bfy}_p^p - p\angle*{\bfy^{p-1}, \bfy-\bfy'} + \frac{p-1}{p2^p}\norm*{\bfy-\bfy'}_p^p,
\]
where $\bfy^{p-1}$ denotes the signed entrywise $(p-1)$th power, i.e., $\bfy_i^{p-1} = \sgn(\bfy_i)\abs*{\bfy_i}^{p-1}$.
\end{Lemma}
\begin{proof}
We set $\Delta = \bfy-\bfy'$ and $\bfx = \bfy$ and apply \cite[Lemma 4.6]{AKPS2019} to bound the Bregman divergence by the $\gamma_p$ function, which is turn lower bounded by the $\ell_p$ norm by \cite[Lemma 3.2]{AKPS2019} for $p\geq 2$. 
\end{proof}

The following is an immediate consequence:

\begin{Lemma}\label{lem:closeness}
Let $p\geq 2$. Let $\bfA\in\mathbb R^{n\times d}$ and $\bfb\in\mathbb R^n$. Then, for any $\bfx\in\mathbb R^d$ such that
\[
    \norm*{\bfA\bfx-\bfb}_p \leq (1+\gamma) \OPT
\]
with $\gamma\in(0,1/p)$, we have that
\[
    \norm*{\bfA\bfx-\bfA\bfx^*}_p \leq O(\gamma^{1/p}) \OPT,
\]
where $\bfx^* \coloneqq \arg\min_{\bfx\in\mathbb R^d}\norm*{\bfA\bfx-\bfb}_p$.
\end{Lemma}
\begin{proof}
The KKT conditions require that
\[
    \angle*{(\bfA\bfx^*-\bfb)^{p-1}, \bfA\bfx} = 0
\]
for all $\bfx\in\mathbb R^d$. We then apply Lemma \ref{lem:lp-bregman} with $\bfy' = \bfA\bfx-\bfb$ and $\bfy = \bfA\bfx^*-\bfb$ to conclude that
\begin{align*}
\norm*{\bfA\bfx-\bfb}_p^p &\geq \norm*{\bfA\bfx^*-\bfb}_p^p - \angle*{(\bfA\bfx^*-\bfb)^{p-1}, \bfA\bfx-\bfA\bfx^*} + \frac{p-1}{p2^p}\norm*{\bfA\bfx-\bfA\bfx^*}_p^p \\
&= \norm*{\bfA\bfx^*-\bfb}_p^p + \frac{p-1}{p2^p}\norm*{\bfA\bfx-\bfA\bfx^*}_p^p.
\end{align*}
Then,
\[
    \norm*{\bfA\bfx-\bfA\bfx^*}_p \leq \parens*{\frac{\norm*{\bfA\bfx-\bfb}_p^p - \norm*{\bfA\bfx^*-\bfb}_p^p}{(p-1) / p2^p}}^{1/p} \leq O(\gamma^{1/p})\OPT.\qedhere
\]
\end{proof}

Lemma \ref{lem:closeness} implies that a $(1+\gamma)$-approximate solution can never output a solution that is more than $O(\gamma^{1/p})\OPT$ from the optimum, in the column space of $\bfA$ equipped with $\ell_p$. In turn, this means that it suffices to bound the distortion of the Lewis weight sampling process over a ball of radius $O(\gamma^{1/p})\OPT$. 

\subsection{Lewis Weight Sampling for Near-Optimal Solutions}

As done in \cite{MMWY2022}, we seek to bound the distortion in the difference between the cost of $\bfx$ and $\bfx^*$, since this suffices to find a near-optimal $\bfx$. That is, we first define the cost difference
\[
    \Delta_i(\bfx) \coloneqq \abs*{[\bfA\bfx-\bfb](i)}^p - \abs*{[\bfA\bfx^*-\bfb](i)}^p
\]
for each $i\in[n]$ and $\bfx\in\mathbb R^d$. We then seek a bound on
\[
    \abs*{\bracks*{\norm*{\bfS\bfA\bfx-\bfS\bfb}_p^p - \norm*{\bfS\bfA\bfx^*-\bfS\bfb}_p^p} - \bracks*{\norm*{\bfA\bfx-\bfb}_p^p - \norm*{\bfA\bfx^*-\bfb}_p^p}} = \abs*{\sum_{i=1}^n (\bfS_{i,i}-1) \Delta_i(\bfx)}
\]
for $\bfx = \bfx_\bfS^*$. The following lemma shows why:

\begin{Lemma}\label{lem:diff-dist}
Suppose that
\[
    \abs*{\bracks*{\norm*{\bfS\bfA\bfx_\bfS^*-\bfS\bfb}_p^p - \norm*{\bfS\bfA\bfx^*-\bfS\bfb}_p^p} - \bracks*{\norm*{\bfA\bfx_\bfS^*-\bfb}_p^p - \norm*{\bfA\bfx^*-\bfb}_p^p}} \leq \eps\OPT^p.
\]
Then,
\[
    \norm*{\bfA\bfx_\bfS^*-\bfb}_p^p \leq \norm*{\bfA\bfx^*-\bfb}_p^p + \eps\OPT^p.
\]
\end{Lemma}
\begin{proof}
We have that
\begin{align*}
    \norm*{\bfA\bfx_\bfS^*-\bfb}_p^p &= \norm*{\bfS\bfA\bfx_\bfS^*-\bfS\bfb}_p^p + \parens*{\norm*{\bfA\bfx_\bfS^*-\bfb}_p^p - \norm*{\bfS\bfA\bfx_\bfS^*-\bfS\bfb}_p^p} \\
    &\leq \norm*{\bfS\bfA\bfx_\bfS^*-\bfS\bfb}_p^p + \parens*{\norm*{\bfA\bfx_\bfS^*-\bfb}_p^p - \norm*{\bfS\bfA\bfx_\bfS^*-\bfS\bfb}_p^p} - \parens*{\norm*{\bfA\bfx^*-\bfb}_p^p - \norm*{\bfS\bfA\bfx^*-\bfS\bfb}_p^p} \\
    &\hspace{5em} + \parens*{\norm*{\bfA\bfx^*-\bfb}_p^p - \norm*{\bfS\bfA\bfx^*-\bfS\bfb}_p^p} \\
    &\leq \norm*{\bfS\bfA\bfx_\bfS^*-\bfS\bfb}_p^p + \abs*{\parens*{\norm*{\bfA\bfx_\bfS^*-\bfb}_p^p - \norm*{\bfA\bfx^*-\bfb}_p^p} - \parens*{\norm*{\bfS\bfA\bfx_\bfS^*-\bfS\bfb}_p^p - \norm*{\bfS\bfA\bfx^*-\bfS\bfb}_p^p}} \\
    &\hspace{5em} + \parens*{\norm*{\bfA\bfx^*-\bfb}_p^p - \norm*{\bfS\bfA\bfx^*-\bfS\bfb}_p^p} \\
    &\leq \norm*{\bfS\bfA\bfx_\bfS^*-\bfS\bfb}_p^p + \eps\OPT^p + \parens*{\norm*{\bfA\bfx^*-\bfb}_p^p - \norm*{\bfS\bfA\bfx^*-\bfS\bfb}_p^p} \\
    &\leq \norm*{\bfA\bfx^*-\bfb}_p^p + \eps\OPT^p
\end{align*}
as claimed.
\end{proof}

In Section \ref{sec:close-points-distortion}, we show the following moment bound on the quality of uniform sampling on a matrix with uniformly bounded Lewis weights.

\begin{restatable}{Theorem}{ClosePoints}\label{thm:close-points}
Let $\bfA\in\mathbb R^{n\times d}$ and let $\bfw\in\mathbb R^n$ be $\alpha$-one-sided Lewis weights for $\bfA$ such that $\bfw_i \leq Wd/n$ for each $i\in[n]$. Let $\bfS\in\mathbb R^{n\times n}$ be a diagonal matrix with $\bfS_{i,i} = 1 + \sigma_i$ for independent Rademacher variables $\sigma_i\in\{\pm1\}$. Let $\gamma>0$. Then, for all $l\geq 1$,
\begin{equation}\label{eq:diff-dist}
    \E\sup_{\norm*{\bfA\bfx-\bfA\bfx^*}_p \leq \gamma^{1/p}\OPT}\abs*{\sum_{i=1}^n \sigma_i\Delta_i(\bfx)}^l \leq \parens*{O(\eps)\OPT^{p}}^l
\end{equation}
where
\[
    \eps = O(1)\bracks*{\gamma\frac{W\norm*{\bfw}_1^{p/2}}{\alpha n}\bracks*{((\log(d\norm*{\bfw}_1)^2(\log n))^{1+1/l} + l}}^{1/p}.
\]
\end{restatable}

Following a symmetrization argument of \cite[Theorem 5.2]{WY2022b}, this gives the following guarantee on Lewis weight sampling:

\begin{Theorem}\label{thm:one-step-reduction}
Let $\bfS$ be a sampling matrix as defined in Definition \ref{def:lewis-weight-sampling}, where $\gamma < \gamma_0$ for a sufficiently small constant $\gamma_0$. Then, with probability at least $1-\delta$, the following implication is true:
\begin{align*}
    &\norm*{\bfA\bfx_\bfS^* - \bfb}_p \leq (1+\gamma)\OPT(\bfA,\bfb) \implies \\
    &\hspace{5em}\sup_{\norm*{\bfA\bfx-\bfA\bfx^*}_p \leq \gamma^{1/p}\OPT}\abs*{\sum_{i=1}^n (\bfS_{i,i}-1)\Delta_i} \leq O(\eps)(\OPT(\bfA,\bfb) + \OPT(\bfS\bfA,\bfS\bfb)).
\end{align*}
\end{Theorem}
\begin{proof}
The proof closely follows \cite[Theorem 5.2]{WY2022b}, and we defer many of the details to their proof. In order to apply the result of Theorem \ref{thm:close-points} to this setting, we will bound
\begin{equation}\label{eq:lewis-moment}
    \E_\bfS\sup_{\norm*{\bfA\bfx-\bfA\bfx^*}_p \leq \gamma^{1/p}\OPT}\abs*{\sum_{i=1}^n (\bfS_{i,i}-1)\Delta_i(\bfx)}^l
\end{equation}
for $l = O(\log\frac1\delta + \log\log n)$. To bound this moment, \cite[Theorem 5.2]{WY2022b} shows a symmetrization argument which allows one to bound \eqref{eq:lewis-moment} by 
\[
2^l \cdot \E_\sigma\sup_{\norm*{\bfA\bfx-\bfA\bfx^*}_p \leq \gamma^{1/p}\OPT}\abs*{\sum_{i=1}^n \sigma_i\Delta_i'(\bfx)}^l,
\]
where $\Delta_i'$ is the analogue of $\Delta_i$ defined with the matrix $\bfA$ and label vector $\bfb$ replaced by a different matrix and vector $\bfA'$ and $\bfb'$, and weights $\bfw$ replaced by a different set of $O(\alpha)$-one-sided weights $\bfw'$. Here, $\bfA'$ is obtained by concatenating a ``flattened'' version of $\bfA$ together with $\bfS\bfA$, and the weights $\bfw'$ are obtained as a ``batch online'' extension of $\bfw$ to $\bfS\bfA$. The label vector $\bfb$ can similarly be flattened and sampled to obtain $\bfb'$. Note then that $\OPT(\bfA',\bfb') \leq O(\OPT(\bfA,\bfb) + \OPT(\bfS\bfA,\bfS\bfb))$. Furthermore, $\bfS\bfA$ is an $O(1)$ $\ell_p$ subspace embedding for $\bfA$ with probability at least $1-\delta$ by \cite[Theorem 5.2]{WY2022b}, so the new matrix and weights have the property that $\norm*{\bfA\bfx}_p = \Theta(1)\norm*{\bfA'\bfx}_p$ for all $\bfx\in\mathbb R^d$. It also follows from the proof of \cite[Theorem 5.2]{WY2022b} that $\norm*{\bfw'}_\infty \leq d\beta$. 

Furthermore, suppose that $\norm*{\bfA\bfx_\bfS^*-\bfb}_p \leq (1+\gamma)\OPT(\bfA,\bfb)$. First note that $\OPT(\bfA',\bfb') \geq \OPT(\bfA,\bfb) + \OPT(\bfS\bfA,\bfS\bfb)$ since separately optimizing the two parts of the concatenation can only decrease the cost. Then, $\bfx^*_\bfS$ is a $(1+\gamma)$-approximate solution for $(\bfA',\bfb')$, since the cost on $(\bfA,\bfb)$ is at most $(1+\gamma)\OPT(\bfA,\bfb)$ and the cost on $(\bfS\bfA,\bfS\bfb)$ is $\OPT(\bfS\bfA,\bfS\bfb)$. Then if $\bfx^*_{\mathrm{concat}}$ is the optimal solution for $(\bfA',\bfb')$, then by Lemma \ref{lem:closeness} and using that $\gamma < \gamma_0$ for a small enough $\gamma_0$, we have that $\norm*{\bfA'\bfx_\bfS^*-\bfA'\bfx_{\mathrm{concat}}^*}_p \leq O(\gamma^{1/p})\OPT(\bfA',\bfb')$. Similarly, $\norm*{\bfA\bfx_\bfS^*-\bfA\bfx^*}_p \leq O(\gamma^{1/p})\OPT(\bfA,\bfb)$. Then for any $\bfx$ such that $\norm*{\bfA\bfx-\bfA\bfx^*}_p \leq \gamma^{1/p}\OPT$,
\begin{align*}
    \norm*{\bfA\bfx-\bfA\bfx^*_{\mathrm{concat}}}_p &\leq \norm*{\bfA\bfx-\bfA\bfx^*}_p + \norm*{\bfA\bfx^*-\bfA\bfx^*_\bfS}_p + \norm*{\bfA\bfx_\bfS^*-\bfA\bfx_{\mathrm{concat}}^*}_p && \text{triangle inequality} \\
    &\leq \gamma^{1/p}\OPT + O(\gamma^{1/p})\OPT + O(\gamma^{1/p})\OPT(\bfA',\bfb') \\
    &= O(\gamma^{1/p})\OPT(\bfA',\bfb')
\end{align*}
and similarly $\norm*{\bfS\bfA\bfx-\bfS\bfA\bfx^*_{\mathrm{concat}}}_p \leq O(\gamma^{1/p})\OPT$ by using that $\bfS$ is a subspace embedding. We may thus bound
\[
\E_\sigma\sup_{\norm*{\bfA\bfx-\bfA\bfx^*}_p \leq \gamma^{1/p}\OPT}\abs*{\sum_{i=1}^n \sigma_i\Delta_i'(\bfx)}^l \leq \E_\sigma\sup_{\norm*{\bfA'\bfx-\bfA'\bfx_{\mathrm{concat}}^*}_p \leq O(\gamma^{1/p})\OPT(\bfA',\bfb')}\abs*{\sum_{i=1}^n \sigma_i\Delta_i'(\bfx)}^l.
\]
Theorem \ref{thm:close-points} now applies on this new matrix and weights, and we bound \eqref{eq:lewis-moment} by $(O(\eps)\OPT(\bfA',\bfb'))^l$. An application of Markov's inequality and taking $l$th roots yields our claim.
\end{proof}

\subsection{Proof of Theorem \ref{thm:iterative-improvement}}

We are now in a position to prove Theorem \ref{thm:iterative-improvement}. 

\begin{proof}[Proof of Theorem \ref{thm:iterative-improvement}]
Let $C$ be a sufficiently large constant to be chosen later.

Define $\beta_i$ recursively by $\beta_1 = (p-1)/p$ and $\beta_{i+1} = (\beta_i + p - 1) / p$, which has the closed form solution
\[
    \beta_i = 1 - p^{-i}.
\]
Then for $i\in[I]$ for $I = O(\log\log\frac1\eps)$, we apply Theorem \ref{thm:one-step-reduction} with $\eps$ and $\gamma$ in the theorem set to $\eps' = \eps^{\beta_{i+1}}$ and $\gamma$ in the theorem set to $\gamma' = C^i\eps^{\beta_i}$. Note that for every $i$, we have
\[
    \frac{\gamma'}{\eps'^p} = O(C^i)\frac{\eps^{\beta_i}}{\eps^{\beta_{i+1}p}} = O(C^i)\eps^{\beta_i - \beta_i - (p-1)} = O(C^i)\frac1{\eps^{p-1}}.
\]
Then, with probability at least $1-\delta$, we have
\begin{align*}
    &\norm*{\bfA\bfx_\bfS^* - \bfb}_p^p \leq (1+C^i\eps^{\beta_i})\OPT(\bfA,\bfb)^p \implies \\
    &\hspace{5em}\sup_{\norm*{\bfA\bfx-\bfA\bfx^*}_p \leq C^{i/p}\eps^{\beta_i/p}\OPT}\abs*{\sum_{i=1}^n (\bfS_{i,i}-1)\Delta_i} \leq O(\eps^{\beta_{i+1}})(\OPT(\bfA,\bfb)^p + \OPT(\bfS\bfA,\bfS\bfb)^p).
\end{align*}
By a union bound, all $I$ of these events occurs with probability $1-I\delta$. Now assume that $\OPT(\bfS\bfA,\bfS\bfb) = O(\OPT(\bfA,\bfb))$. Then, we have that
\[
    \norm*{\bfA\bfx_\bfS^* - \bfb}_p^p \leq (1+C^i\eps^{\beta_i})\OPT^p \implies \sup_{\norm*{\bfA\bfx-\bfA\bfx^*}_p \leq C^{i/p}\eps^{\beta_i/p}\OPT}\abs*{\sum_{i=1}^n (\bfS_{i,i}-1)\Delta_i} \leq O(\eps^{\beta_{i+1}})\OPT^p
\]
Furthermore, by Lemma \ref{lem:closeness}, we have that 
\[
    \norm*{\bfA\bfx_\bfS^* - \bfb}_p^p \leq (1+C^i\eps^{\beta_i})\OPT^p \implies \norm*{\bfA\bfx_\bfS^* - \bfA\bfx^*}_p \leq O(C^{i/p}\eps^{\beta_i/p})\OPT
\]
so by Lemma \ref{lem:diff-dist}, we then have that
\[
    \norm*{\bfA\bfx_\bfS^* - \bfb}_p^p \leq (1+C^i\eps^{\beta_i})\OPT^p \implies \norm*{\bfA\bfx_\bfS^* - \bfb}_p^p \leq (1+C^{i+1}\eps^{\beta_{i+1}})\OPT^p
\]
for large enough $C$. We may now follow the chain of implications to conclude that
\begin{align*}
    &\norm*{\bfA\bfx_\bfS^* - \bfb}_p^p \leq (1+C\eps^{\beta_1})\OPT^p = (1+C\eps^{(p-1)/p})\OPT^p \\
    \implies~&\norm*{\bfA\bfx_\bfS^* - \bfb}_p^p \leq (1+C^{I}\eps^{\beta_{I}})\OPT^p = (1+\eps\poly\log(1/\eps))\OPT^p.
\end{align*}
The hypothesis of the implication is shown by \cite[Theorem 3.4]{MMWY2022}. 
\end{proof}

\section{Improved Distortion Bounds for Close Points}\label{sec:close-points-distortion}

In this section, we prove Theorem \ref{thm:close-points} (restated below), which gives an improved bound on the distortion of points $\bfA\bfx$ that are near the optimal regression solution $\bfA\bfx^*$.

\ClosePoints*

\subsection{Set Up}\label{sec:setup}

We split the sum in \eqref{eq:diff-dist} into two parts: the part that is bounded by the Lewis weights of $\bfA$, and the part that is not. To this end, define a threshold
\[
    \tau_i \coloneqq \gamma\frac{\norm*{\bfw}_1^{p/2-1}}{\eps^{p}}\bfw_i\OPT^p
\]
and define the set of ``good'' entries $G\subseteq[n]$ as
\[
    G = \braces*{i\in[n] : \abs*{[\bfA\bfx^* - \bfb](i)} \leq \tau_i}
\]
We then bound
\begin{align*}
    \sup_{\norm*{\bfA\bfx-\bfA\bfx^*}_p \leq \gamma^{1/p}\OPT}\abs*{\sum_{i=1}^n \sigma_i\Delta_i(\bfx)}^l &\leq 2^{l-1}\sup_{\norm*{\bfA\bfx-\bfA\bfx^*}_p \leq \gamma^{1/p}\OPT}\abs*{\sum_{i\in G} \sigma_i\Delta_i(\bfx)}^l \\
    &+ 2^{l-1}\sup_{\norm*{\bfA\bfx-\bfA\bfx^*}_p \leq \gamma^{1/p}\OPT}\abs*{\sum_{i\in[n]\setminus G} \sigma_i\Delta_i(\bfx)}^l
\end{align*}
using the relaxed triangle inequality, and separately estimate each term. We can think of the first term as the ``sensitivity'' term, where each term in the sum is bounded by the Lewis weights of $\bfA$, and the latter term as the ``outlier'' term, where each term in the sum is much larger than the corresponding Lewis weights.

\subsection{Estimates on the Outlier Term}

We first bound the outlier terms, which is much easier. 

\begin{Lemma}\label{lem:outlier}
With probability $1$, we have that
\[
    \sup_{\norm*{\bfA\bfx-\bfA\bfx^*}_p \leq \gamma^{1/p}\OPT}\sum_{i\in[n]\setminus G}\abs*{\Delta_i(\bfx)} \leq O(\eps)\OPT^p.
\]
\end{Lemma}
\begin{proof}
For each $i\in[n]\setminus G$, we have that
\begin{align*}
    \abs*{[\bfA\bfx-\bfb](i)} &\in \abs*{[\bfA\bfx^*-\bfb](i)} \pm \abs*{[\bfA\bfx^*-\bfA\bfx](i)} \\
    &\in \abs*{[\bfA\bfx^*-\bfb](i)} \pm \norm*{\bfw}_1^{1/2-1/p}\bfw_i^{1/p}\norm*{\bfA\bfx^*-\bfA\bfx}_p \\
    &\in \abs*{[\bfA\bfx^*-\bfb](i)} \pm \gamma^{1/p}\norm*{\bfw}_1^{1/2-1/p}\bfw_i^{1/p}\OPT && \text{Lemma \ref{lem:oslw-sensitivity}} \\
    &\in \abs*{[\bfA\bfx^*-\bfb](i)} \pm \eps \abs*{[\bfA\bfx^*-\bfb](i)} && i\in[n]\setminus G
\end{align*}
Thus,
\[
    \abs*{\Delta_i(\bfx)} \leq O(\eps)\abs*{[\bfA\bfx^*-\bfb](i)}^p
\]
so
\[
    \sum_{i\in[n]\setminus G}\abs*{\Delta_i(\bfx)} \leq \sum_{i=1}^n O(\eps)\abs*{[\bfA\bfx^*-\bfb](i)}^p = O(\eps)\OPT^p.\qedhere
\]
\end{proof}

\subsection{Estimates on the Sensitivity Term}

Next, we estimate the sensitivity term,
\[
    \E\sup_{\norm*{\bfA\bfx-\bfA\bfx^*}_p \leq \gamma^{1/p}\OPT}\abs*{\sum_{i\in G} \sigma_i\Delta_i(\bfx)}^l.
\]
To estimate this moment, we obtain a subgaussian tail bound via the tail form of Dudley's entropy integral, and then integrate it.

\subsubsection{Change of Density}
We follow the Lewis weight chaining argument of \cite{MMWY2022} and \cite{LT1991}. We start with a change of density using the Lewis weights so that
\[
    \sup_{\norm*{\bfA\bfx-\bfA\bfx^*}_p \leq \gamma^{1/p}\OPT}\abs*{\sum_{i\in G} \sigma_i\Delta_i(\bfx)} = \sup_{\norm*{\bfA\bfx-\bfA\bfx^*}_p \leq \gamma^{1/p}\OPT}\abs*{\sum_{i\in G} \bfw_i\sigma_i\bar\Delta_i(\bfx)}
\]
where we define $\bar\Delta_i$ as the corresponding versions of $\Delta_i$ reweighted by $\bfw_i$, i.e., 
\begin{align*}
\bar\Delta_i(\bfx) &\coloneqq \abs*{[\bfW^{-1/p}(\bfA\bfx-\bfb)](i)}^p - \abs*{[\bfW^{-1/p}(\bfA\bfx^*-\bfb)](i)}^p
\end{align*}
Note that $\abs*{\bar\Delta_i(\bfx)}$ is bounded over all $\norm*{\bfA\bfx-\bfA\bfx^*}_p \leq \gamma^{1/p}\OPT$, since
\begin{equation}\label{eq:sensitivity-bound}
\begin{aligned}
    \abs*{\bar\Delta_i(\bfx)} &\leq \abs*{[\bfW^{-1/p}(\bfA\bfx-\bfb)](i)}^p + \abs*{[\bfW^{-1/p}(\bfA\bfx^*-\bfb)](i)}^p \\
    &\leq 2^{p-1}\parens*{\abs*{[\bfW^{-1/p}(\bfA\bfx^*-\bfb)](i)}^p + \abs*{[\bfW^{-1/p}(\bfA\bfx-\bfA\bfx^*)](i)}^p} + \abs*{[\bfW^{-1/p}(\bfA\bfx^*-\bfb)](i)}^p \\
    &\leq (2^{p-1}+1)\tau + \gamma\norm*{\bfw}_1^{p/2-1}\OPT^p \\
    &\leq (2^{p-1}+1+\eps^p)\tau = O(\tau)
\end{aligned}
\end{equation}
using Lemma \ref{lem:oslw-sensitivity}, where
\[
    \tau \coloneqq \gamma\frac{\norm*{\bfw}_1^{p/2-1}}{\eps^{p}}\OPT^p.
\]

\subsubsection{Bounding Low-Sensitivity Entries}

We now separately handle entries $i\in G$ with small Lewis weight. To do this end, define
\[
    J \coloneqq \braces*{i\in G : \bfw_i \geq \frac{\eps^{p+1}}{\gamma n\norm*{\bfw}_1^{p/2-1}}}.
\]
We then bound the mass on the complement of $J$:

\begin{Lemma}\label{lem:not-J}
For all $\norm*{\bfA\bfx-\bfA\bfx^*}_p \leq \gamma^{1/p}\cdot\OPT$, we have that
\[
    \sum_{i\in [n]\setminus J} \bfw_i \abs*{\bar\Delta_i(\bfx)} \leq O(\eps)\OPT^p
\]
\end{Lemma}
\begin{proof}
Note that for each $i\in[n]\setminus G$,
\[
    \bfw_i \leq \frac{\eps}{\tau n}\OPT^p
\]
For each $i\in G\setminus J$, we use \eqref{eq:sensitivity-bound} to bound $\bar\Delta_i(\bfx)$ by $O(\tau)$. Summing the bounds yields the result.
\end{proof}

\subsubsection{Bounding High-Sensitivity Entries: Gaussian Processes}

In order to obtain tail bounds, we first use Panchenko's lemma to bound a Gaussian process instead of a Rademacher process.

\begin{Lemma}[Lemma 1, \cite{Pan2003}]\label{lem:panchenko}
Let $X, Y$ be random variables such that
\[
    \E[\Phi(X)] \leq \E[\Phi(Y)]
\]
for every increasing convex function $\Phi$. If
\[
    \Pr\braces*{Y\geq t} \leq c_1 \exp(-c_2 t^\alpha)\qquad\mbox{for all $t\geq 0$},
\]
for some $c_1,\alpha\geq 1$ and $c_2>0$, then
\[
    \Pr\braces*{X\geq t} \leq c_1 \exp(1-c_2 t^\alpha)\qquad\mbox{for all $t\geq 0$}.
\]
\end{Lemma}

Let $\Phi$ be any increasing convex function. Since $\bfw_i \leq Wd / n$, we bound
\[
    \E\sup_{\norm*{\bfA\bfx-\bfA\bfx^*}_p \leq \gamma^{1/p}\OPT}\Phi\parens*{\abs*{\sum_{i\in J} \sigma_i\bfw_i\bar\Delta_i(\bfx)}} \leq O(1)\sqrt{\frac{W}{n}}\cdot\E\sup_{\norm*{\bfA\bfx-\bfA\bfx^*}_p \leq \gamma^{1/p}\OPT}\Phi\parens*{\abs*{\sum_{i\in J} \sigma_i \sqrt{d\bfw_i} \bar\Delta_i(\bfx)}}
\]
by the Rademacher contraction theorem \cite[Theorem 4.12]{LT1991}. Then by a Gaussian comparison theorem \cite[Equation 4.8]{LT1991}, we may bound the above by
\[
    O(1)\sqrt{\frac{W}{n}}\cdot\E\sup_{\norm*{\bfA\bfx-\bfA\bfx^*}_p \leq \gamma^{1/p}\OPT}\Phi\parens*{\abs*{\sum_{i\in J} g_i \sqrt{d\bfw_i} \bar\Delta_i(\bfx)}},
\]
where the Rademacher variables $\sigma_i$ have been replaced by Gaussian variables $g_i$. Thus by Lemma \ref{lem:panchenko}, it suffices to obtain tail bounds for
\begin{equation}\label{eq:gp}
    O(1)\sqrt{\frac{W}{n}}\cdot\sup_{\norm*{\bfA\bfx-\bfA\bfx^*}_p \leq \gamma^{1/p}\OPT}\abs*{\sum_{i\in J} g_i \sqrt{d\bfw_i} \bar\Delta_i(\bfx)},
\end{equation}
We can now appeal to the theory of Gaussian processes to bound this quantity. 

Let
\[
    Z \coloneqq O(1)\frac{\gamma^{1/2}}{\eps^{p/2-1}}\OPT^p
\]
be a normalizing quantity and define a Gaussian process by
\[
    G_\bfx \coloneqq \frac1{Z}\sum_{i\in J} g_i\sqrt{d\bfw_i} \bar\Delta_i(\bfx)
\]
with pseudo-metric
\[
    d_G(\bfx,\bfx') \coloneqq \parens*{\E_g\abs*{G_\bfx-G_\bfx'}^2}^{1/2} = \Theta(1)\frac1Z\parens*{\sum_{i\in J}d\bfw_i(\bar\Delta_i(\bfx)-\bar\Delta_i(\bfx'))^2}^{1/2}
\]
As we will see later, $Z$ is chosen to scale the Gaussian process to the scale of the ball $\braces*{\norm*{\bfA\bfx-\bfA\bfx^*}_p \leq \gamma^{1/p}\OPT}$.

\subsubsection{Estimates on the Gaussian Process Geometry}

Using the sensitivity bound of \eqref{eq:sensitivity-bound}, we obtain a bound on the pseudo-metric $d_G$, which improves over the bound of \cite{MMWY2022} by introducing a dependence on $\gamma$.

\begin{Lemma}\label{lem:dG-bound}
For $\bfx,\bfx'\in\braces*{\norm*{\bfA\bfx-\bfA\bfx^*}_p\leq \gamma^{1/p}\OPT}$ , we have that
\[
    d_G(\bfx,\bfx') \leq \frac1{\gamma^{1/p}\OPT}\cdot O(\sqrt{d})(\norm*{\bfw}_1^{p/2-1})^{1/2-1/p} \norm*{[\bfW^{-1/p}\bfA(\bfx-\bfx')]\mid_J}_\infty 
\]
\end{Lemma}
\begin{proof}
For $a,b\in\mathbb R$, we have by convexity for $p>1$ that
\[
    \abs*{a}^p - \abs*{b}^p \leq p(\abs*{a}^{p-1}+\abs*{b}^{p-1})\abs*{\abs*{a}-\abs*{b}} \leq p(\abs*{a}^{p-1} + \abs*{b}^{p-1})\abs*{a-b}
\]
Then by applying the above to $a = [\bfW^{-1/p}(\bfA\bfx-\bfb)](i)$ and $b = [\bfW^{-1/p}(\bfA\bfx'-\bfb)](i)$, we have
\begin{align*}
    Z^2 d_G(\bfx,\bfx')^2 &\leq O(\sqrt d)\sum_{i\in J}\bfw_i(\bar\Delta_i(\bfx)-\bar\Delta_i(\bfx'))^2 \\
    &\leq O(\sqrt d)\sum_{i\in J}\bfw_i\parens*{\abs{[\bfW^{-1/p}(\bfA\bfx-\bfb)](i)}^p-\abs{[\bfW^{-1/p}(\bfA\bfx'-\bfb)](i)}^p}^2 \\
    &\leq O(\sqrt d)\norm*{[\bfW^{-1/p}\bfA(\bfx-\bfx')]\mid_J}_\infty^2\sum_{i\in J}\bfw_i\parens*{\abs{[\bfW^{-1/p}(\bfA\bfx-\bfb)](i)}\lor \abs{[\bfW^{-1/p}(\bfA\bfx'-\bfb)](i)}}^{2p-2}.
\end{align*}
Now using \eqref{eq:sensitivity-bound}, we have
\[
    \parens*{\abs{[\bfW^{-1/p}(\bfA\bfx-\bfb)](i)}^p\lor\abs{[\bfW^{-1/p}(\bfA\bfx'-\bfb)](i)}^p} \leq O(\gamma)\frac{\norm*{\bfw}_1^{p/2-1}}{\eps^p}\OPT^p
\]
so
\begin{align*}
    &\sum_{i\in J}\bfw_i\parens*{\abs{[\bfW^{-1/p}(\bfA\bfx-\bfb)](i)}^{2p-2} + \abs{[\bfW^{-1/p}(\bfA\bfx'-\bfb)](i)}^{2p-2}} \\
    \leq~& O(1)\parens*{\gamma\frac{\norm*{\bfw}_1^{p/2-1}}{\eps^p}\OPT^p}^{1-2/p}\sum_{i\in J}\bfw_i\parens*{\abs{[\bfW^{-1/p}(\bfA\bfx-\bfb)](i)}^p + \abs{[\bfW^{-1/p}(\bfA\bfx'-\bfb)](i)}^p} \\
    =~& O(1)\parens*{\gamma\frac{\norm*{\bfw}_1^{p/2-1}}{\eps^p}\OPT^p}^{1-2/p}(\norm*{\bfA\bfx-\bfb}_p^p + \norm*{\bfA\bfx'-\bfb}_p^p) \\
    \leq~& O(1)\parens*{\gamma\frac{\norm*{\bfw}_1^{p/2-1}}{\eps^p}\OPT^p}^{1-2/p}(\norm*{\bfA\bfx-\bfx^*}_p^p + \norm*{\bfA\bfx'-\bfx^*}_p^p + \norm*{\bfA\bfx^*-\bfb}_p^p) \\
    \leq~& O(1)\parens*{\gamma\frac{\norm*{\bfw}_1^{p/2-1}}{\eps^p}\OPT^p}^{1-2/p}\OPT^p.
\end{align*}
Altogether,
\begin{align*}
    d_G(\bfx,\bfx') &\leq O(\sqrt d)\norm*{[\bfW^{-1/p}\bfA(\bfx-\bfx')]\mid_J}_\infty\frac1Z \parens*{\gamma\frac{\norm*{\bfw}_1^{p/2-1}}{\eps^p}}^{1/2-1/p}\OPT^{p-1} \\
    &\leq O(\sqrt d)\norm*{[\bfW^{-1/p}\bfA(\bfx-\bfx')]\mid_J}_\infty \parens*{\norm*{\bfw}_1^{p/2-1}}^{1/2-1/p}\cdot\frac1{\gamma^{1/p}\OPT}
\end{align*}
as claimed.
\end{proof}

Similarly, we obtain bounds on the $d_G$-diameter of the ball $\braces*{\norm*{\bfA\bfx-\bfA\bfx^*}_p \leq \gamma^{1/p}\OPT}$.

\begin{Lemma}\label{lem:diam}
Let $T = \braces*{\bfx\in\mathbb R^d : \norm*{\bfA\bfx-\bfA\bfx^*}_p \leq \gamma^{1/p}\OPT}$. Then,
\[
    \diam(T) = \sup_{\bfx,\bfx'\in T}d_G(\bfx,\bfx') \leq O(\sqrt{d})(\norm*{\bfw}_1^{p/2-1})^{1/2-1/p} 
\]
\end{Lemma}
\begin{proof}
This follows from applying Lemma \ref{lem:dG-bound} and then a sensitivity bound (Lemma \ref{lem:oslw-sensitivity}) to bound 
\[
    \norm*{[\bfW^{-1/p}\bfA(\bfx-\bfx')]\mid_J}_\infty \leq \gamma^{1/p}\OPT.\qedhere
\]
\end{proof}

\subsubsection{Dudley's Entropy Integral}

We will obtain tail bounds on \eqref{eq:gp} via the following tail bound version of Dudley's inequality:

\begin{Theorem}[Theorem 8.1.6, \cite{Ver2018}]\label{thm:dudley-tail}
Let $(X_t)_{t\in T}$ be a Gaussian process with pseudo-metric $d_X(s,t)\coloneqq \norm*{X_s - X_t}_2$. Let $E(T, d_X, u)$ denote the minimal number of $d_X$-balls of radius $u$ required to cover $T$. Then, for every $z\geq 0$, we have that
\[
	\Pr\braces*{\sup_{t\in T}X_t \geq C\bracks*{\int_0^\infty \sqrt{\log E(T, d_X, u)}~du + z\cdot \diam(T)}} \leq 2\exp(-z^2)
\]
\end{Theorem}

We now calculate Dudley's entropy integral, using metric entropy bounds from \cite{WY2022b}.

\begin{Lemma}\label{lem:entropy-integral}
Let $T = \braces*{\bfx\in\mathbb R^d : \norm*{\bfA\bfx-\bfA\bfx^*}_p \leq \gamma^{1/p}\OPT}$. Then,
\[
    \int_0^\infty \sqrt{\log E(T, d_G, u)}~du \leq O(1)\bracks*{\norm*{\bfw}_1^{p/2}(\log(d\norm*{\bfw}_1))^2(\log n)}^{1/2}
\]
\end{Lemma}
\begin{proof}
Let $B_p(\bfA) \coloneqq \colspan(\bfA)\cap B_p$ and note that the supremum we wish to bound is over the set $T \coloneqq O(\gamma^{1/p})\OPT \cdot B_p(\bfA)$. 

Let $\bar\bfw = \bfw / \norm*{\bfw}_1$ be the normalized one-sided Lewis weights and define the norm
\[
    \norm*{\bfy}_{\bar\bfw,q} \coloneqq \parens*{\sum_{i=1}^n \bar\bfw_i \abs*{\bfy_i}^q}^{1/q} = \frac1{\norm*{\bfw}_1^{1/q}}\norm*{\bfW^{1/q}\bfy}_q
\]
As reasoned in \cite{WY2022b}, we have that for $q = O(\log n)$,
\[
    \norm*{[\bfW^{-1/p}\bfA(\bfx-\bfx')]_J}_\infty \leq O(1)\norm*{\bfW^{-1/p}\bfA(\bfx-\bfx')}_{\bar\bfw,q}
\]
since $\bfw_i \geq 1/\poly(n)$ for $i\in J$. Then, we can bound the metric entropy of $T$ with respect to $d_G$-balls of radius $t$ using the above by
\begin{align*}
    &\log E(O(\gamma^{1/p})\OPT \cdot B_p(\bfA), d_G, t) \\
    \leq~& \log E(O(\gamma^{1/p})\OPT \cdot B_p(\bfA), \norm{\bfW^{-1/p}(\cdot)}_{\bar\bfw,q}, (\gamma^{1/p}\OPT)t / O(\sqrt{d})(\norm*{\bfw}_1^{p/2-1})^{1/2-1/p}) && \text{Lemma \ref{lem:dG-bound}} \\
    \leq~& \log E(B_p(\bfA), \norm{\bfW^{-1/p}(\cdot)}_{\bar\bfw,q}, t / O(\sqrt{d})(\norm*{\bfw}_1^{p/2-1})^{1/2-1/p}) && \text{scaling} \\
    \leq~& \log E(\norm*{\bfw}_1^{1/p} B_p(\bfA), \norm{\bfW^{-1/p}(\cdot)}_{\bar\bfw,q},\norm*{\bfw}_1^{1/p}  t / O(\sqrt{d})(\norm*{\bfw}_1^{p/2-1})^{1/2-1/p}) && \text{scaling}
\end{align*}
Now since $\norm*{\bfy}_{\bar\bfw,p} = \norm*{\bfW^{1/p}\bfy}_p / \norm*{\bfw}_1^{1/p}$, we have that $B_p(\bfA) \norm*{\bfw}_1^{1/p} = B_{\bar\bfw,p}(\bfW^{-1/p}\bfA)$, where
\[
    B_{\bar\bfw,p}(\bfW^{-1/p}\bfA) = \braces*{\bfy : \norm*{\bfy}_{\bar\bfw,p}\leq 1} \cap \colspan(\bfW^{-1/p}\bfA)
\]
so the above metric entropy is equal to
\[
    \log E(B_{\bar\bfw,p}(\bfW^{-1/p}\bfA), \norm{\bfW^{-1/p}(\cdot)}_{\bar\bfw,q}, t / O(\sqrt{d}/\sqrt{\norm{\bfw}_1})(\norm*{\bfw}_1^{p/2})^{1/2-1/p})
\]
Then, the net bounds in Corollary B.9 of \cite{WY2022b} show that this is at most
\[
    O\parens*{\frac{\norm*{\bfw}_1^{p/2}}{\alpha t^2}q} = O\parens*{\frac{\norm*{\bfw}_1^{p/2}}{\alpha t^2}\log n}.
\]
Dudley's entropy integral then gives a bound of
\begin{align*}
    &\int_0^\infty \sqrt{\log E(O(\gamma^{1/p})\OPT \cdot B_p(\bfA), d_G, t)} \\
    =~&\int_0^{\diam(T)} \sqrt{\log E(O(\gamma^{1/p})\OPT \cdot B_p(\bfA), d_G, t)} \\
    \leq~& O(1)\int_0^1 \sqrt{d\log\frac{n}{t}}~dt + O(\alpha^{-1/2})\int_1^{\diam(T)} \frac{\sqrt{\norm*{\bfw}_1^{p/2}\log n}}{t}~dt \\
    \leq~&O(\alpha^{-1/2})\bracks*{\norm*{\bfw}_1^{p/2}(\log\diam(T))^2(\log n)}^{1/2} \\
    \leq~&O(\alpha^{-1/2})\bracks*{\norm*{\bfw}_1^{p/2}(\log(d\norm*{\bfw}_1))^2(\log n)}^{1/2} && \text{Lemma \ref{lem:diam}} \qedhere
\end{align*}
\end{proof}

As a result of the above calculations, we have the following tail bound:

\begin{Corollary}\label{cor:tail-bound}
There is $C = \Theta(1)$ such that for every $z\geq 0$, we have that
\[
    \Pr\braces*{\sup_{\norm*{\bfA\bfx-\bfA\bfx^*}_p \leq \gamma^{1/p}\OPT}\abs*{\sum_{i\in J} \frac1Z g_i \sqrt{d\bfw_i} \bar\Delta_i(\bfx)} \geq C\alpha^{-1/2}\norm*{\bfw}_1^{p/4}\parens*{\bracks*{(\log(d\norm*{\bfw}_1))^2(\log n)}^{1/2} + z}} \leq 2\exp(-z^2)
\]
\end{Corollary}
\begin{proof}
This follows from Dudley's tail bound in Theorem \ref{thm:dudley-tail}, the entropy calculation in Lemma \ref{lem:entropy-integral}, and the diameter calculations in Lemma \ref{lem:diam}.
\end{proof}

\subsubsection{Moment Bounds}

With tail bounds in place, we estimate the moments of \eqref{eq:gp}. 

\begin{Lemma}\label{lem:moment}
\[
    \E\bracks*{\sup_{\norm*{\bfA\bfx-\bfA\bfx^*}_p \leq \gamma^{1/p}\OPT}\abs*{\sum_{i\in J} \frac1Z g_i \sqrt{d\bfw_i} \bar\Delta_i(\bfx)}^l} \leq O(\norm*{\bfw}_1^{p/4}\alpha^{-1/2})^l\cdot \bracks*{((\log(d\norm*{\bfw}_1))(\log n)^{1/2})^{l+1} + O(l)^{l/2}}
\]
\end{Lemma}
\begin{proof}
Let
\[
    \Lambda = \frac{1}{C\norm*{\bfw}_1^{p/4}}\sup_{\norm*{\bfA\bfx-\bfA\bfx^*}_p \leq \gamma^{1/p}\OPT}\abs*{\sum_{i\in J}\frac1Z  g_i \sqrt{d\bfw_i} \bar\Delta_i(\bfx)}
\]
and $z_0 = (\log(d\norm*{\bfw}_1))(\log n)^{1/2}$. We have that
\begin{align*}
    \E[\Lambda^l] &= \int_0^\infty l\cdot z^l \Pr(\Lambda\geq z)~dz \\
    &= \int_0^{2z_0} l\cdot z^l \Pr(\Lambda\geq z)~dz + \int_{2z_0}^\infty l\cdot z^l \Pr(\Lambda\geq z)~dz \\
    &\leq (2z_0)^{l+1} + \int_{2z_0}^\infty l\cdot z^l \Pr(\Lambda\geq z/2 + z_0)~dz \\
    &\leq (2z_0)^{l+1} + 2l\int_{2z_0}^\infty z^l \exp(-z^2/4)~dz && \text{Corollary \ref{cor:tail-bound}} \\
    &\leq (2z_0)^{l+1} + l\cdot O(l)^{l/2}.\qedhere
\end{align*}
\end{proof}

We are now ready to prove Theorem \ref{thm:close-points}.

\begin{proof}[Proof of Theorem \ref{thm:close-points}]
We split the sum into the sensitivity term and outlier term, as discussed in Section \ref{sec:setup}. To bound the outlier term, we use the triangle inequality and Lemma \ref{lem:outlier} to bound by $\parens*{O(\eps)\OPT}^l$. To bound the sensitivity term, we split the sum into the the indices in $J$ and those outside of $J$. For those outside of $J$, we use Lemma \ref{lem:not-J} to get a bound of $\parens*{O(\eps)\OPT}^l$. For those in $J$, we have
\begin{align*}
    &O\parens*{\sqrt{\frac{W}{n}}}^l Z^l \E\bracks*{\sup_{\norm*{\bfA\bfx-\bfA\bfx^*}_p \leq \gamma^{1/p}\OPT}\abs*{\sum_{i\in J} \frac1Z g_i \sqrt{d\bfw_i} \bar\Delta_i(\bfx)}^l} \\
    \leq~&O\parens*{\sqrt{\frac{W}{n}}}^l Z^l O(\norm*{\bfw}_1^{p/4}\alpha^{-1/2})^l\cdot \bracks*{((\log(d\norm*{\bfw}_1))(\log n)^{1/2})^{l+1} + O(l)^{l/2}} && \text{Lemma \ref{lem:moment}} \\
    \leq~&\bracks*{O(1)\frac{W}{n}\frac{\gamma}{\alpha\eps^{p-2}} \norm*{\bfw}_1^{p/2} \parens*{[(\log(d\norm*{\bfw}_1))^2(\log n)]^{1+1/l} + l}}^{l/2} (\OPT^p)^l.
\end{align*}
Now note that
\[
    \eps = \bracks*{O(1)\frac{W}{n}\frac{\gamma}{\alpha\eps^{p-2}} \norm*{\bfw}_1^{p/2} \parens*{(\log(d\norm*{\bfw}_1))^2(\log n)]^{1+1/l} + l}}^{1/2}
\]
since this rearranges to
\[
    \eps^p = O(1)\frac{W}{n} \frac{\gamma}{\alpha}\norm*{\bfw}_1^{p/2} \parens*{[(\log(d\norm*{\bfw}_1))^2(\log n)]^{1+1/l} + l}.
\]
This shows the desired result.
\end{proof}

\section{Nearly Optimal Lower Bound for Active \texorpdfstring{$\ell_p$}{lp} Regression}
\label{sec:active-lb}

In this section, we obtain a tight lower bound for active $\ell_p$ regression for $p > 2$, up to polylogarithmic factors.

We will need the following coding theory theorem, which was also used in \cite{LWY2021, MMWY2022} to construct hard instances for linear algebraic problems.
\begin{Theorem}[\cite{PTB2013}]\label{thm:ptb}
For any $q\geq 1$ and $d = 2^k - 1$ for some integer $k$, there exists a set $S\subset\{\pm1\}^d$ and a constant $C_q$ depending on $q$ such that $\abs*{S} = d^q$ and for every $\bfx,\bfy\in S$ with $\bfx\neq\bfy$, we have that $\abs*{\angle*{\bfx,\bfy}} \leq C_q\sqrt d$. 
\end{Theorem}

We present our main lower bound, which generalizes a lower bound of \cite{CSS2021b} to $d$ dimensions for linear regression.

\begin{Theorem}\label{thm:opt-lb}
Let $p>2$. Suppose that a randomized algorithm solves the $\ell_p$ regression up to a relative error of $(1+\eps/3)$ and queries $m$ entries in expectation and is correct with probability at least $0.99$. Then, $m = \Omega(d^{p/2}/\eps^{p-1})$.
\end{Theorem}
\begin{proof}
By Yao's minimax principle, we may assume that the algorithm is deterministic with correctness probability at least $0.99$ over a distributional hard instance. Let $S$ be the set given by Theorem \ref{thm:ptb} with $q = p/2$. Set $n = s\cdot d^{p/2}$ for $s = c / \eps^{p-1}$ with $c$ a sufficiently small constant to be determined. Then, we take our matrix to be the $n \times d$ matrix formed by taking $s$ copies of each of the $d^{p/2}$ vectors in $S$. Furthermore, we take our target vector $\bfb$ to be the zero vector with probability $1/2$ and $d\cdot \bfe_I$ with probability $1/2$, where $I\sim[n]$ is a uniformly random index and $\bfe_i$ is the $i$th standard basis vector for $i\in[n]$.

Call the deterministic algorithm $\mathcal A$. Suppose for contradiction that $m \leq n / 100$. Consider the sequence of entries of $\bfb$ read by $\mathcal A$ when $\bfb = 0$. Note that this sequence is of length at most $2m$, since otherwise $\mathcal A$ already reads more than $m$ entries in expectation. Furthermore, $\mathcal A$ must output $\bfx = 0$ as the solution if it reads a sequence of $2m$ entries of zeros, since otherwise $\mathcal A$ cannot achieve any relative error. Then since $\mathcal A$ is deterministic, $\mathcal A$ will always output $\bfx = 0$ if it reads $2m$ entries of zeros. 

On the other hand, suppose that $\bfb = d\cdot\bfe_I$ for $I\sim[n]$. We first upper bound the optimal cost. If we choose $\bfx = \eps\cdot \bfa_I$, then for the nonzero row of $\bfb$, we pay a cost of
\[
    (d - \eps\cdot\angle*{\bfa_I, \bfa_I})^p = (1-\eps)^p d^p \leq (1-\eps) d^p.
\]
For the other rows of $\bfA$ corresponding to copies of $\bfa_I$, we pay a cost of
\[
    s \cdot (\eps\cdot\angle*{\bfa_I, \bfa_I})^p = \frac{c}{\eps^{p-1}} \cdot \eps^p \cdot d^p = c\eps d^p.
\]
For all other rows of $\bfA$ for $\bfa_j \neq \bfa_I$, we pay a cost of
\[
    s\cdot d^{p/2}\cdot (\eps\cdot\angle*{\bfa_I, \bfa_j})^p = \frac{c}{\eps^{p-1}}\cdot \eps^p \cdot d^{p/2}\cdot C_q^p d^{p/2} = cC_q^p \eps d^p.
\]
Thus, if we choose $c \leq \min\{C_q^p, 1\} / 3$, then the total cost is at most
\[
    (1-\eps) d^p + c\eps d^p + cC_q^p \eps d^p \leq (1-\eps/3) d^p.
\]

Now note that if $\bfb = d\cdot\bfe_I$, then the probability that $I$ lands on one of the $2m$ entries read by $\mathcal A$ when $\bfb = 0$ is at most $2m/n \leq 1/50$. Thus, with probability at least $1 - 1/50$, $\mathcal A$ outputs $\bfx = 0$ on this instance, which has a cost of $d^p$. By the above calculation, this fails to be a $(1+\eps/3)$-approximate solution, which contradicts the guarantee of $\mathcal A$. We thus conclude that $m \geq n/100 = \Omega(d^{p/2}/\eps^{p-1})$. 
\end{proof}

\section{Active Regression with Large Distortion}
\label{sec:active-large-dist}

Our results for active $\ell_p$ regression have assumed that $p$ is a fixed constant to obtain algorithms reading only $\poly(d)$ entries. However, for $p$ very large, an exponential dependence on $p$ is intractable, and our argument gives nothing for the important case of $p = \infty$. In this section, we show how to handle $\ell_p$ norms even for large $p$ by reading only $\poly(d)$ entries, where the degree of the polynomial does not depend on $p$, by trading off for a $\poly(d)$ factor distortion. Our techniques yield many other results on dimension reduction in the large distortion regime. 

\subsection{Reduction to Subspace Embeddings}

Our algorithms for this section are a generalization of the observation that subspace embeddings which satisfy a ``no expansion'' condition in expectation for any vector yields an active regression algorithm. This is used in \cite{MMWY2022} to obtain an initial constant factor approximation for active $\ell_p$ regression. We generalize this to the following:

\begin{Lemma}\label{lem:active-reduction}
Let $\bfA\in\mathbb R^{n\times d}$ and $\bfb\in\mathbb R^n$. Let $\norm*{\cdot}_X$ and $\norm*{\cdot}_Y$ be two norms. Suppose that $\bfS\in\mathbb R^{s\times n}$ random sampling matrix such that
\[
    \Pr\braces*{\forall\bfx\in\mathbb R^d, \norm*{\bfA\bfx}_X \leq \alpha\norm*{\bfS\bfA\bfx}_Y} \geq 1 - \delta_1
\]
and for any fixed $\bfy\in\mathbb R^n$,
\[
    \Pr\braces*{\norm*{\bfS\bfy}_Y \leq \beta \norm*{\bfy}_X} \geq 1 - \delta_2.
\]
Let $\tilde\bfx$ satisfy
\[
    \norm*{\bfS\bfA\tilde\bfx-\bfS\bfb}_Y \leq \gamma \min_{\bfx\in\mathbb R^d}\norm*{\bfS\bfA\bfx-\bfS\bfb}_Y.
\]
Then, with probability at least $1 - (\delta_1 + \delta_2)$,
\[
    \norm*{\bfA\tilde\bfx - \bfb}_X \leq ((\gamma+1)\alpha\beta+1)\min_{\bfx\in\mathbb R^d}\norm*{\bfA\bfx - \bfb}_X
\]
\end{Lemma}
\begin{proof}
We condition on the two guarantees of $\bfS$. Let
\[
    \norm*{\bfA\bfx^*-\bfb}_X = \OPT = \min_{\bfx\in\mathbb R^d}\norm*{\bfA\bfx-\bfb}_X.
\]
We then have that
\begin{align*}
    \norm*{\bfA\tilde\bfx-\bfb}_X &\leq \norm*{\bfA\tilde\bfx-\bfA\bfx^*}_X + \norm*{\bfA\bfx^*-\bfb}_X && \text{triangle inequality} \\
    &= \norm*{\bfA\tilde\bfx-\bfA\bfx^*}_X  + \OPT \\
    &\leq \alpha\norm*{\bfS\bfA\tilde\bfx-\bfS\bfA\bfx^*}_Y  + \OPT \\
    &\leq \alpha\parens*{\norm*{\bfS\bfA\tilde\bfx-\bfS\bfb}_Y + \norm*{\bfS\bfA\bfx^*-\bfS\bfb}_Y}  + \OPT && \text{triangle inequality} \\
    &\leq (\gamma+1)\alpha\norm*{\bfS\bfA\bfx^*-\bfS\bfb}_Y  + \OPT && \text{optimality} \\
    &\leq (\gamma+1)\alpha\beta\norm*{\bfA\bfx^*-\bfb}_X  + \OPT \\
    &\leq ((\gamma+1)\alpha\beta+1)\OPT.\qedhere
\end{align*}
\end{proof}

Note that this can easily be boosted to a $1-\delta$ probability at a loss of a $\log(1/\delta)$ factor in the query complexity by using a boosting procedure described in \cite{MMWY2022}.

\subsection{Upper Bounds}

By combining Lemma \ref{lem:active-reduction} with subspace embedding results, we immediately obtain many results for active regression. The first is a nearly optimal deterministic algorithm for $\ell_\infty$ active regression, using Theorem \ref{cor:linf-subspace-embedding}:

\begin{Theorem}\label{thm:l-inf-active-regression}
There is a deterministic algorithm which, given $\bfA\in\mathbb R^{n\times d}$, reads $O(d\log\log d)$ entries of $\bfb$ and outputs $\tilde\bfx$ such that
\[
    \norm*{\bfA\tilde\bfx - \bfb}_\infty \leq O(\sqrt d)\OPT.
\]
\end{Theorem}

A faster algorithm based running in input sparsity by sampling is also available, via Theorem \ref{thm:linf-embedding-LJ}. This loses a $\log d$ factor in the sample complexity. We can also generalize this result to the average top $k$ loss, using our results in Section \ref{sec:avg-top-k}.

For finite $p$, inspired by \cite{WY2022a}, we obtain a whole set of trade-offs by using Theorem \ref{thm:lewis-reweighting} to switch to the $\ell_q$ norm and then using $\ell_q$ Lewis weight sampling as in Theorem \ref{thm:lewis-weight-sampling} to sample rows. 

\begin{Theorem}\label{thm:large-lp-active-regression}
Let $2\leq q < p < \infty$. There is an algorithm which, given $\bfA\in\mathbb R^{n\times d}$, reads $O(d^{q/2}\log^3 d)$ entries of $\bfb$ and outputs $\tilde\bfx$ such that
\[
    \Pr\braces*{\norm*{\bfA\tilde\bfx - \bfb}_\infty \leq O(d^{\frac12\parens*{1-\frac{q}{p}}})\OPT} \geq \frac{99}{100}.
\]
\end{Theorem}
\begin{proof}
We take $\bfS$ to be the composition of the reweighting matrix $\bfW^{1/q-1/p}$ where $\bfw$ are the $\ell_p$ Lewis weights of $\bfA$ with an appropriate scaling, and the $\ell_q$ Lewis weight sampling matrix which samples $O(d^{q/2}\log^3 d)$ rows of $\bfW^{1/q-1/p}\bfA$. By Theorems \ref{thm:lewis-reweighting} and \ref{thm:lewis-weight-sampling}, $\bfS$ satisfies
\[
    \norm*{\bfA\bfx}_p \leq \norm*{\bfS\bfA\bfx}_q \leq O(d^{\frac12\parens*{1-\frac{q}{p}}})\norm*{\bfA\bfx}_p
\]
with probability at least $1/200$. Furthermore, by Lemma \ref{lem:lewis-no-expansion} and Markov's inequality, we have that $\norm*{\bfS\bfy}_q \leq O(d^{\frac12\parens*{1-\frac{q}{p}}})\norm*{\bfy}_p$ with probability at least $1/200$. Lemma \ref{lem:active-reduction} then gives the claimed result.
\end{proof}

\subsection{Lower Bounds}

We give two lower bounds, one for active regression in $\ell_\infty$ and one for $\ell_p$ with large distortion.

\begin{Theorem}[$\ell_\infty$ Active Regression Lower Bound]\label{thm:l-inf-active-regression-lb}
Let $C > 1$ be a constant. There is a constant $c$ such that any randomized algorithm solves the $\ell_\infty$ regression up to a relative error of $c\sqrt d$ for some sufficiently small constant $c$ and queries $m$ entries in expectation and is correct with probability at least $0.99$. Then, $m = \Omega(d^C)$.
\end{Theorem}
\begin{proof}
We use Theorem \ref{thm:ptb} to construct a set of $m = \Omega(d^C)$ vectors $S\subseteq\{\pm1\}^d$ such that for every distinct $\bfx,\bfy\in S$, we have $\abs*{\angle*{\bfx,\bfy}} \leq c^{-1}\sqrt d$ for some $c < 1$. Now let $\bfA$ be the $m\times d$ matrix with the vectors of $S$ in its rows, and let $\bfb$ be a zero vector with probability $1/2$ and a uniformly random standard basis vector scaled by $d$ with probability $1/2$. As reasoned in Theorem \ref{thm:opt-lb}, we may assume that our algorithm is deterministic and has success probability at least $0.99$ over our hard random instance. Since $\bfb$ is the zero vector with probability $1/2$, the algorithm must output $\bfx = 0$ if it reads $2m$ zeros. However, for any $\bfb = d\cdot \bfe_i$ for $i\in[n]$, the cost is $d$, whereas a cost of $c^{-1}\sqrt d$ can be obtained if we choose $\bfx$ to be the row of $\bfA$ corresponding to this $i$. Thus, in order to obtain a distortion smaller than $c\sqrt d$, the algorithm must read at least $\Omega(m)$ entries. 
\end{proof}

\begin{Theorem}[$\ell_p$ Active Regression Lower Bound]\label{thm:lp-active-regression-lb}
Let $2\leq q < p$. There is a constant $c$ such that any randomized algorithm solves the $\ell_p$ regression up to a relative error of $c d^{\frac12\parens*{1-\frac{q}{p}}}$ for some sufficiently small constant $c$ and queries $m$ entries in expectation and is correct with probability at least $0.99$. Then, $m = \Omega(d^{q/2})$.
\end{Theorem}
\begin{proof}
We use Theorem \ref{thm:ptb} to construct a set of $m = \Omega(d^{q/2})$ vectors $S\subseteq\{\pm1\}^d$ such that for every distinct $\bfx,\bfy\in S$, we have $\abs*{\angle*{\bfx,\bfy}} \leq c^{-1}\sqrt d$ for some $c < 1$. Now let $\bfA$ be the $m\times d$ matrix with the vectors of $S$ in its rows, and let $\bfb$ be a zero vector with probability $1/2$ and a uniformly random standard basis vector scaled by $d$ with probability $1/2$. As reasoned in Theorem \ref{thm:opt-lb}, we may assume that our algorithm is deterministic and has success probability at least $0.99$ over our hard random instance. Since $\bfb$ is the zero vector with probability $1/2$, the algorithm must output $\bfx = 0$ if it reads $2m$ zeros. However, for any $\bfb = d\cdot \bfe_i$ for $i\in[n]$, the cost is $d$, whereas a cost of
\[
    m^{1/p}\cdot c^{-1}\sqrt d = c^{-1}d^{\frac12\parens*{\frac{q}{p} + 1}}
\]
can be obtained if we choose $\bfx$ to be the row of $\bfA$ corresponding to this $i$. Thus, in order to obtain a distortion smaller than $c\sqrt d$, the algorithm must read at least $\Omega(m)$ entries. 
\end{proof}

\section*{Acknowledgements}

We thank the anonymous reviewers for useful feedback on improving the presentation of this work. David P.\ Woodruff and Taisuke Yasuda were supported by a Simons Investigator Award.

\bibliographystyle{alpha}
\bibliography{citations}

\appendix

\section{Reduction from Existential to Algorithmic Column Subset Selection}

We show an improvement and generalization of the argument of \cite{MW2021}, which shows that an existential result showing the existence of $s = s(k)$ columns with a distortion of $\kappa(d)$ on any $n\times d$ instance for rank $k$ approximation implies an algorithmic version which selects $O(s\log d)$ columns with a distortion of $O(\kappa(2s+1))$. Note that the number of columns can only depend on $k$, whereas the distortion can depend on $d$.

\begin{Definition}
    Let $\bfA = \bfA_* + \bfDelta$, where $\bfA_*$ is the best rank $k$ approximation in the entrywise $g$ norm, that is,
    \[
    \norm*{\bfDelta}_g = \min_{\rank(\hat\bfA)\leq k} \norm*{\bfA - \hat\bfA}_g.
    \]
    Let the columns of $\bfDelta$ be $\bfdelta^1, \bfdelta^2, \dots, \bfdelta^d$. 
\end{Definition}

\begin{Definition}\label{def:useful-g}
    Let $l\in\mathbb N$. Then:
    \begin{itemize}
        \item Let $s(k)$ denote the maximum size of a set of columns $S$ for any $n\times d$ instance $\bfB$ for rank $k$ approximation in the entrywise $g$-norm that can achieve a $\kappa(d)$ approximation, that is, there exists a set $S\subseteq[d]$ such that
        \begin{equation}\label{eq:general-css-guarantee}
            \min_{\bfX\in\mathbb R^{S\times d}}\norm*{\bfB - \bfB\vert^S\bfX}_g \leq \kappa(d)\norm*{\bfDelta}_g
        \end{equation}
        \item Let $T_l\subseteq[d]$ denote the subset of columns surviving after the $l$th round of the algorithm. We assume without loss of generality that $T_l = [d_l]$ for some $d_l \leq d$. Furthermore, we assume without loss of generality that $\norm{\bfdelta^1}_g \geq \norm{\bfdelta^2}_g \geq \dots \geq \norm{\bfdelta^{d_l}}_g$.
        \item Let $\Res_l \coloneqq \sum_{j=d_l/4}^{d_l} \norm{\bfdelta^j}_g$ denote the residual cost, after restricting to the surviving columns and after removing the columns with cost in the top quarter.
    \end{itemize}
\end{Definition}

\begin{algorithm}
    \caption{Column Subset Selection for $M$-Estimators}
    \textbf{input:} Input matrix $\bfA\in\mathbb R^{n\times d}$, rank $k$, loss function $g$, parameter $s$. \\
    \textbf{output:} Subset $T\subseteq[d]$ of $O(s\log d)$ columns. 
    \begin{algorithmic}[1]
        \State{$T_0 \gets [d]$}
        \While{$\abs{T_l} \geq 1000s$}
        \State{$t_l\gets 30s$}
        \For{$t = 1, 2, \dots, O(\log\log d)$}
        \State{Sample $H\sim\binom{T_l}{t_l}$}
        \State{Let $\bfx^{j}$ minimize $\min_{\bfx}\norm{\bfA\vert^H\bfx-\bfa^j}_g$ up to a $\reg_{g,t_l}$ factor for each $j\in T_l$}
        \State{Let $F_{l,t}$ be the $d_l / 20 = \abs{T_l} / 20$ columns with smallest regression cost $\norm{\bfA\vert^H\bfx^j-\bfa^j}_g$}
        \State{$C_{l,t} \gets \sum_{j\in F_{l,t}}\norm{\bfA\vert^H\bfx^j-\bfa^j}_g$}
        \EndFor{}
        \State{Let $t^*$ be the $t$ with smallest $C_{l,t}$}
        \State{$T_{l+1} \gets T_l \setminus F_{l,t^*}$}
        \EndWhile{}
    \end{algorithmic}\label{alg:mw2021}
\end{algorithm}

\begin{Theorem}[Generalization and Improvement of \cite{MW2021}]\label{thm:generalized-mw2021}
    Consider the definitions in Definition \ref{def:useful-g}. Suppose that there is an algorithm outputting $\tilde\bfx$ such that
    \[
    \norm*{\bfB\tilde\bfx - \bfb}_g \leq \reg_{g,s} \cdot \min_{\bfx\in\mathbb R^s}\norm*{\bfB\tilde\bfx - \bfb}_g
    \]
    for any $\bfB\in\mathbb R^{n\times s}$ and $\bfb\in\mathbb R^n$. Then, Algorithm \ref{alg:mw2021} outputs a subset $S\subseteq[d]$ of $\abs{S} = O(s\log d)$ columns and $\bfX\in\mathbb R^{S\times d}$ such that
    \begin{align*}
        \norm*{\bfA - \bfA\vert^S\bfX}_g &\leq O(\kappa)\reg_{g,O(s)}\min_{\rank(\hat\bfA) \leq k}\norm{\bfA-\hat\bfA}_g
    \end{align*}
\end{Theorem}

We present the following main lemma, which follows \cite[Claim 2.6]{MW2021} but also makes some additional improvements to remove a log factor:

\begin{Lemma}\label{lem:mw2021-claim2.6}
    Let $\bfA\in\mathbb R^{n\times d}$. Let $s = s(k)$ and $\kappa = \kappa(2s+1)$. Let $H\sim\binom{[d]}{2s}$ and let $i\sim[d]\setminus H$. Then,
    \[
    \Pr\braces*{\min_{\bfx\in\mathbb R^H}\norm*{\bfa^i - \bfA\vert^H\bfx}_g \leq \frac{600\kappa}{d_l}\Res_l}\geq \frac1{10}
    \]
\end{Lemma}
\begin{proof}
    Let $G \coloneqq [d_l]\setminus[d_l/4]$. Note that $\E\abs{G\cap H} \geq 20s$. By Chernoff bounds, with probability at least $99/100$, we have that $\abs*{G \cap H} \geq 4s$. We conditioned on this event.
    
    Let $H'$ be a uniformly random subset of $G\cap H$ of size $2s$. Let $R = R(H'\cup\{i\})$ be the set of $s(k)$ columns satisfying \eqref{eq:general-css-guarantee}. Then by Markov's inequality,
    \begin{align*}
        \Pr_{H'}\braces*{\sum_{j\in H'}\norm*{\bfdelta^j}_g \geq 20\frac{s}{\abs{G}}\sum_{j\in G}\norm*{\bfdelta^j}_g} \leq \frac{\E_{H'}\bracks*{\sum_{j\in H'}\norm*{\bfdelta^j}_g}}{20\frac{s}{\abs{G}}\sum_{j\in G}\norm*{\bfdelta^j}_g} \leq \frac1{10}
    \end{align*}
    and similarly,
    \[
    \Pr_{i}\braces*{\norm*{\bfdelta^i}_g \geq \frac{10}{\abs{G}}\sum_{j\in G}\norm*{\bfdelta^j}_g} \leq \frac{\E_i\bracks*{\norm*{\bfdelta^i}_g}}{\frac{5}{\abs{G}}\sum_{j\in G}\norm*{\bfdelta^j}_g} \leq \frac1{10}
    \]
    
    Now note that conditioned on the choice of $H'\cup\{i\}$, $i$ is a uniformly random element of $H'\cup\{i\}$, so $\Pr\{i\notin R\}\geq 1/2$. Furthermore,
    \[
    \min_{\bfX\in\mathbb R^{R\times (2s+1)}}\norm*{\bfA\vert^{H'\cup\{i\}} - \bfA\vert^R\bfX}_g \leq \kappa \min_{\rank(\hat\bfA)\leq k}\norm*{\bfA\vert^{H'\cup\{i\}} - \hat\bfA}_g \leq \kappa\cdot\norm*{\Delta\vert^{H'\cup\{i\}}}_g
    \]
    so by Markov's inequality,
    \[
    \min_{\bfx\in\mathbb R^R}\norm*{\bfa^i - \bfA\vert^R\bfx} \leq \frac{10\kappa}{s}\norm*{\Delta\vert^{H'\cup\{i\}}}_g
    \]
    with probability at least $9/10$. By a union bound, we have that with probability at least
    \[
    1 - \frac1{100} - \frac1{10} - \frac1{10} - \frac1{10} \geq \frac1{10},
    \]
    we have
    \[
    \min_{\bfx\in\mathbb R^R}\norm*{\bfa^i - \bfA\vert^R\bfx}_g \leq \frac{10\kappa}{s}\parens*{\frac{10}{\abs{G}}\sum_{j\in G}\norm*{\bfdelta^j}_g + 20\frac{s}{\abs{G}}\sum_{j\in G}\norm*{\bfdelta^j}_g} \leq \frac{400\kappa}{\abs{G}}\sum_{j\in G}\norm*{\bfdelta^j}_g.
    \]
    To conclude, note that $\abs{G} = d_l - d_l/4 = 3d_l / 4$ and that we can pad $\bfx$ with zeros on coordinates in $H\setminus R$.
\end{proof}

We then just mimic the proof of Theorem \ref{thm:improved-swz2019} to complete the proof.

\begin{proof}[Proof of Theorem \ref{thm:generalized-mw2021}]
    Note first that the algorithm decreases the size of $T_l$ by a $(1 - 1/20)$ factor at each iteration. Thus, the algorithm makes at most $L = O(\log d)$ iterations of the outer loop. By averaging Lemma \ref{lem:mw2021-claim2.6} over the $3d_l/4$ bottom columns, we have a probability of at least $1/20$ of choosing $d_l / 20$ columns such that the total cost is at most
    \[
    O(\kappa)\cdot\Res_l.
    \]
    Since we repeat $O(\log L) = O(\log\log d)$ times and use an $\reg_{g,t_l}$-approximate regression algorithm, we with probability at least $1 - 1/100L$, we find $d_l / 20$ columns $F_l\subseteq T_l$ and corresponding coefficients $\bfX$ such that
    \[
    \norm*{\bfA\vert^{F_l} - \bfA\vert^{S_l}\bfX}_g \leq O(\kappa)\reg_{g,t_l}\Res_l.
    \]
    Thus, our total cost is
    \[
    \sum_{l=1}^{O(\log d)}O(\kappa)\reg_{g,t_l}\Res_l.
    \]
    Finally, as argued in \cite{SWZ2019, MW2021}, we show that $\sum_l \Res_l = O(\norm{\bfDelta}_g)$. Note that if a column $j$ contributes to $\Res_l$, then it must be in the bottom $3/4$ fraction of the $\norm{\bfdelta^j}_g$ in round $l$. Then since the bottom $1/20$ fraction of $\norm{\bfdelta^j}_g$ is fitted and removed in each round, $\norm{\bfdelta^j}_g$ can only contribute to $\Res_l$ in $O(1)$ rounds. Thus, the sum is bounded by $O(1)\sum_j \norm{\bfdelta^j}_g = O(\norm{\bfDelta}_g)$. 
    
    The total number of columns selected is $O(s)$ in each of the $O(\log d)$ rounds, for a total of $O(s\log d)$. 
\end{proof}

\section{Feldman--Langberg Framework with Independent Sampling}

We adapt arguments in \cite{FL2011, BFL2016, FSS2020}. Recall the following definitions from the preliminaries section of \cite{FSS2020}.

\begin{Definition}[Range space]
A \emph{range space} is a pair $\mathfrak{R} = (F,\ranges)$ where $F$ is a set called a \emph{ground set} and $\ranges$ is a family of subsets of $F$.
\end{Definition}

\begin{Definition}[VC-dimension]
The \emph{VC-dimension} of a range space $\mathfrak{R} = (F,\ranges)$ is the size $\abs*{G}$ of the largest subset $G\subseteq F$ such that
\[
    \abs*{\braces*{G\cap \range} : \range\in \ranges} = 2^{\abs*{G}}
\]
\end{Definition}

\begin{Definition}[$(\eta,\eps)$-approximation]
Let $\eta,\eps>0$ and $\mathfrak R = (F,\ranges)$ be a range space with finite $F\neq\varnothing$. An $(\eta,\eps)$-approximation of $\mathfrak R$ is a set $S\subseteq F$ such that for all $\range\in\ranges$, we have
\[
    \abs*{\frac{\abs*{\range\cap F}}{\abs*{F}} - \frac{\abs*{\range\cap S}}{\abs*{S}}} \leq \begin{dcases}
    \eps\cdot\frac{\abs*{\range\cap F}}{\abs*{F}} & \text{if $\abs*{\range\cap F}\geq\eta\abs*{F}$} \\
    \eps\cdot \eta & \text{if $\abs*{\range\cap F} < \eta\abs*{F}$} \\
    \end{dcases}
\]
\end{Definition}

\begin{Definition}
Let $F$ be a finite set of functions from a set $\mathcal Q$ to $[0,\infty)$. For every $Q\in\mathcal Q$ and $r\geq 0$, let
\begin{align*}
    \range(F, Q, r) &\coloneqq \braces*{f\in F : f(Q) \geq r} \\
    \ranges(F) &\coloneqq \braces*{\range(F, Q, r) : Q\in\mathcal Q, r\geq 0}
\end{align*}
Then, $\mathfrak{R}_{\mathcal Q, F} \coloneqq (F, \ranges(F))$ is the \emph{range space} induced by $\mathcal Q$ and $F$.
\end{Definition}

We extract the following lemma from \cite[Theorem 31]{FSS2020}, which is based on works of \cite{FL2011, BFL2016}:s

\begin{Theorem}[Range Space Approximation to Coresets, Theorem 31, \cite{FSS2020}]\label{thm:fss2020-thm31}
Let $F$ be a finite set of $n = \abs*{F}$ functions from a set $\mathcal Q$ to $[0,\infty)$, and let $\eps\in(0,1/2)$. Let
\[
    \tilde\bfsigma_f \geq \sup_{Q\in\mathcal Q}\frac{f(Q)}{\sum_{h\in F} h(Q)}, \qquad \tilde{\mathfrak{S}} = \sum_{f\in F} \tilde\bfsigma_f.
\]
Furthermore, assume that for each $f\in F$, $\tilde\bfsigma_f \geq 1/n$ and that $\tilde\bfsigma_f$ is an integer power of $2$, so that for some $n^*\in\mathbb N$, $n^*\cdot \tilde\bfsigma_f\in\mathbb N$ for all $f\in F$. Let $F'$ be obtained by replacing each $f\in F$ by $n_f \coloneqq n^*\cdot \tilde\bfsigma_f$ copies of $f / n_f$. Suppose $S\subseteq F'$ is an $(\eta,\eps/2)$-approximation for $\eta = 1/\tilde{\mathfrak{S}}$ to the range space $\mathfrak R_{\mathcal Q, F'}$. Then, for all $Q\in \mathcal Q$
\[
    \abs*{\frac{\abs*{F'}}{\abs*{S}}\sum_{f\in S}f(Q) - \sum_{f\in F}f(Q)} \leq \eps \sum_{h\in F}h(Q)
\]
\end{Theorem}

The proof of \cite[Theorem 31]{FSS2020} uses the following result to obtain an $(\eta,\eps)$-approximation of $\mathfrak R_{\mathcal Q,F'}$:

\begin{Theorem}[Theorem 5, \cite{LLS2001}]\label{thm:lls2001-thm5}
Let $\mathfrak R = (F,\ranges)$ with finite $F \neq\varnothing$ be a range space with VC-dimension $d$, $\eta > 0$, and $\eps,\delta\in(0,1)$. There is a universal constant $c>0$ such that a sample of
\[
    s \geq \frac{c}{\eta\eps^2}\cdot\parens*{d\log\frac1\eta + \log\frac1\delta}
\]
elements drawn independently and uniformly at random from $F$ is an $(\eta,\eps)$-approximation for $(F,\ranges)$ with probability at least $1-\delta$.
\end{Theorem}

We first note that the uniform sampling with replacement can be replaced by uniform sampling without replacement, at a sacrifice of $s^2/n$ in the success probability.

\begin{Corollary}[\cite{LLS2001} by Sampling without Replacement]\label{cor:uniform-wor}
Let $\mathfrak R = (F,\ranges)$ with finite $F \neq\varnothing$ be a range space with VC-dimension $d$, $\eta > 0$, and $\eps,\delta\in(0,1)$. There is a universal constant $c>0$ such that a sample of
\[
    s \geq \frac{c}{\eta\eps^2}\cdot\parens*{d\log\frac1\eta + \log\frac1\delta}
\]
\emph{distinct} elements drawn uniformly at random from $F$ is an $(\eta,\eps)$-approximation for $(F,\ranges)$ with probability at least $1-\delta-s^2/n$.
\end{Corollary}
\begin{proof}
The probability that a uniform sample with replacement selects an item twice is less than $s^2 / n$ by the union bound. Then with probability at least $1 - s^2 / n$, the uniform sample with replacement selects $s$ unique elements. Conditioned on this event, by symmetry, the sample is drawn as a uniformly random subset of size $s$. Now suppose that the conclusion is false. Let good denote the event that uniform sampling with replacement is successful, and let unique denote the event that uniform sampling with replacement selects a unique set of $s$ elements. Then,
\begin{align*}
    \Pr(\text{good}) &= \Pr(\text{good}\mid \text{unique})\Pr(\text{unique}) + \Pr(\text{good}\mid \neg\text{unique})\Pr(\neg\text{unique}) \\
    &< \Pr(\text{good}\mid \text{unique}) + \frac{s^2}{n} \\
    &< 1 - \delta - \frac{s^2}{n} + \frac{s^2}{n} \\
    &= 1 - \delta,
\end{align*}
which contradicts the conclusion of Theorem \ref{thm:lls2001-thm5}. 
\end{proof}

We now switch to sampling each of the $n^*\cdot \tilde{\mathfrak S}$ items independently with probability $(1+\eps)s/n$. Note then that with probability at least $1-\delta$, we sample at least $s$ and at most $(1+\eps)^2s$ elements by Chernoff bounds, since $s \geq c\eps^{-2}\log\frac1\delta$. Furthermore, conditioned on the number $\abs*{S}$ of elements sampled, the sample $S$ is a uniformly random subset of size $\abs*{S}$ by symmetry. Furthermore, conditioned on each of these, we obtain a good approximation with probability at least $1 - \delta - \abs*{S}^2 / n$ by Corollary \ref{cor:uniform-wor}. This yields the following:

\begin{Corollary}[\cite{LLS2001} by Independent Sampling]\label{cor:indep-sample}
Let $\mathfrak R = (F,\ranges)$ with finite $F \neq\varnothing$ be a range space with VC-dimension $d$, $\eta > 0$, and $\eps,\delta\in(0,1)$. There is a universal constant $c>0$ such that if
\[
    s \geq \frac{c}{\eta\eps^2}\cdot\parens*{d\log\frac1\eta + \log\frac1\delta},
\]
then a sample $S$ obtained by independently sampling each $f\in F$ with probability $s/n$ is an $(\eta,\eps)$-approximation for $\mathfrak R$ with probability at least $1-\delta-4s^2/n$.
\end{Corollary}

Now by combining Corollary \ref{cor:indep-sample} with Theorem \ref{thm:fss2020-thm31} yields the following:

\begin{Theorem}[\cite{FL2011, BFL2016, FSS2020} by Independent Sampling]\label{thm:fl2011-indep}
Let $F$ be a finite set of $n = \abs*{F}$ functions from a set $\mathcal Q$ to $[0,\infty)$, and let $\delta,\eps\in(0,1/2)$. Let
\[
    \tilde\bfsigma_f \geq \sup_{Q\in\mathcal Q}\frac{f(Q)}{\sum_{h\in F} h(Q)}, \qquad \tilde{\mathfrak{S}} = \sum_{f\in F} \tilde\bfsigma_f.
\]
Let $\mathcal S \geq \tilde{\mathfrak S}$ be an upper bound on the total sensitivity $\tilde{\mathfrak S}$. 
Furthermore, assume that for each $f\in F$, $\tilde\bfsigma_f \geq 1/n$ and that $\tilde\bfsigma_f$ is an integer power of $2$, so that for some $n^*\in\mathbb N$, $n^*\cdot \tilde\bfsigma_f\in\mathbb N$ for all $f\in F$. Choose this $n^*$ so that $n^* \geq c'\cdot n^2/\delta$ for a sufficiently large constant $c'$.

Let $F'$ be obtained by replacing each $f\in F$ by $n_f \coloneqq n^*\cdot \tilde\bfsigma_f$ copies of $f / n_f$. Let $d$ be the VC-dimension of $\mathfrak R_{\mathcal Q, F'}$. Let
\[
    s \geq c\cdot \frac{\mathcal S}{\eps^2}\parens*{d\log\mathcal S + \log\frac1\delta}
\]
for a sufficiently large constant $c$. Let $S$ be a random sample obtained by sampling each $f\in F'$ with probability $s / (n^* \cdot \mathcal S)$. Then, with probability at least $1-\delta$, simultaneously for all $Q\in \mathcal Q$
\[
    \abs*{\frac{\abs*{F'}}{s}\sum_{f\in S}f(Q) - \sum_{f\in F}f(Q)} \leq O(\eps) \sum_{h\in F}h(Q)
\]
\end{Theorem}
\begin{proof}
Note first that we can handle the fact that we only have an upper bound $\mathcal S$ instead of $\tilde{\mathfrak S}$ by adding an empty function at the end of $F$ with a sensitivity overestimate $\mathcal S - \tilde{\mathfrak S}$, so the previous results assuming the knowledge of an exact sum $\tilde{\mathfrak S}$ apply. Second, note that by Chernoff bounds, $\abs*{S} = (1\pm\eps) s$. Thus, $\abs*{F'}/\abs*{S} = (1\pm\eps)\abs*{F'}/s$ which yields the conclusion. Finally, the success probability is $1-O(\delta)$ since we either select $n$ elements, in which case we deterministically get an exact approximation, or $s^2 / n^* \leq \delta$, in which case the failure probability is as claimed by Corollary \ref{cor:indep-sample}.
\end{proof}
\begin{Remark}
Note that splitting $f$ into $n_f = n^*\cdot\tilde\bfsigma_f$ copies of $f / n_f$ and sampling each with probability $p_f = s/(n^*\cdot\mathcal S)$ can be combined into just sampling a weight $w_f \sim \Binomial(n_f, p_f)$.
\end{Remark}

\subsection{Improved Feldman--Langberg Framework for Clustering}

Next, we adapt the improved \cite{FL2011} argument for clustering which achieves a linear dependence on $k$, from Theorem 15.5. We first set up some notation:

\begin{Definition}
Let $\bfA\in\mathbb R^{n\times d}$, let $p\geq 1$, and let $\eps\in(0,1)$. Let $B\subseteq\mathbb R^d$ be a set of centers (not necessarily of size $k$) such that
\[
    \sum_{i=1}^n d(\bfa_i, B)^p \leq \gamma\cdot\min_{C\subseteq\mathbb R^d, \abs*{C}\leq k} \sum_{i=1}^n d(\bfa_i, C)^p
\]
for some $\gamma>0$. Let $\bfa_i' \coloneqq \argmin_{\bfb\in B}\norm*{\bfa_i - \bfb}_2$ be the center in $B$ closest to $\bfa_i$, and let $P_i\subseteq[n]$ denote the set of indices belonging to the set cluster as $i\in[n]$.
\end{Definition}

We sample each row with probability
\begin{equation}\label{eq:cluster-sample-def}
    p_i \geq \min\braces*{1, \max\braces*{\beta_1\cdot\frac1{\eps^{p+1}}\frac{\norm*{\bfa_i - \bfa_i'}_2^p}{\sum_{j=1}^n d(\bfa_i,B)^p}, \beta_2\cdot\frac1{\eps^2}\frac1{\abs*{P_i}}}}
\end{equation}
and set the weight to be $\bfw_i = 1/p_i$ if we sample the row and $0$ otherwise, for oversampling parameters $\beta_1, \beta_2 \geq 1$.

First, we show that the above sampling algorithm preserves the sizes of the clusters. 
\begin{Lemma}[Preserving Size of Clusters]\label{lem:preserve-clusters}
Let $P\subseteq[n]$ be all the rows clustered to a given center in $B$. Then,
\[
    \sum_{i\in P}\bfw_i = (1\pm\eps)\abs*{P}
\]
with probability at least $1 - 2\exp(-\beta_2/3)$.
\end{Lemma}
\begin{proof}
Note that $\bfw_i \leq \eps^2\abs*{P} / \beta_2$ for all $i\in P$, so $\beta_2 \bfw_i / (\eps^2\abs*{P}) \in[0, 1]$, and
\[
    \E\bracks*{\sum_{i\in P}\bfw_i} = \abs*{P}.
\]
Then by Chernoff bounds, 
\[
    \Pr\braces*{\abs*{\sum_{i\in P}\bfw_i - \abs*{P}} \geq \eps\abs*{P}} \leq 2\exp\parens*{-\frac13\eps^2\beta_2\frac{\abs*{P}}{\abs*{P}}} \leq 2\exp(-\beta_2/3).\qedhere
\]
\end{proof}

By taking $\beta_2 \geq 3\log(\abs*{B}/\delta)$, the above holds for all centers $B$ with probability at least $1 - 2\delta$. We condition on this event for the rest of this section.

We will need the following relaxed triangle inequality:
\begin{Lemma}[Relaxed Triangle Inequality, Corollary A.2, \cite{MMR2019}]
    For vectors $\bfu, \bfv, \bfw$ and $\eps>0$ and $p\geq 1$, we have
    \[
        \norm*{\bfu-\bfw}_2^p \leq (1+\eps)\norm*{\bfu-\bfv} + \parens*{\frac{1+\eps}{\eps}}^{p-1}\norm*{\bfv-\bfw}_2^p.
    \]
\end{Lemma}

Next, we give a net argument to show that it suffices to preserve the cost of centers $C$ in a set of size at most $\poly(n)^{dk}$. 

\begin{Lemma}[Net argument]\label{lem:net-cluster}
Let $\bfA\in\mathbb R^{n\times d}$. Let $\eps\in(0,1/4)$ with $\eps\geq1/\poly(n)$. Let $\bfw_i$ be sampled as in \eqref{eq:cluster-sample-def} with $\beta_2 \geq 3\log(\abs*{B}/\delta)$. With probability at least $1-\delta$, here is a set $\mathcal N\subseteq(\mathbb R^d)^k$ with $\abs*{\mathcal N}\leq \poly(n)^{dk}$ such that if
\begin{equation}\label{eq:net-guarantee}
    \mbox{for all $C\in\mathcal N$},\qquad\sum_{i=1}^n d(\bfa_i, C)^p = (1\pm\eps)\sum_{i=1}^n \bfw_i d(\bfa_i, C)^p
\end{equation}
then we have that
\[
    \mbox{for all $C\subseteq\mathbb R^d, \abs*{C}\leq k$},\qquad\sum_{i=1}^n d(\bfa_i, C)^p = (1\pm10\eps)\sum_{i=1}^n \bfw_i d(\bfa_i, C)^p.
\]
\end{Lemma}
\begin{proof}
Note that all the $\bfa_i$ are within a distance of $\gamma^{1/p}\OPT^{1/p}$ of $B$, and a distance of $2\gamma^{1/p}\OPT^{1/p}$ of every point in the same cluster $P_i$. Now suppose that $\bfa_i$ is assigned to some center in a cluster $C$ with $d(\bfa_i, C)^p \geq 3^p\gamma\OPT / \eps^p$. Then, for any $j\in P_i$,
\[
    d(\bfa_j, C) \geq d(\bfa_i, C) - \norm*{\bfa_i - \bfa_j}_2 \geq 2\gamma^{1/p}\OPT^{1/p}/\eps
\]
so
\begin{align*}
    d(\bfa_j, C) &= d(\bfa_j', C) \pm \norm*{\bfa_j' - \bfa_i}_2 \\
    &= d(\bfa_j', C) \pm 2\gamma^{1/p}\OPT^{1/p} \\
    &= d(\bfa_j', C) \pm \eps d(\bfa_j, C)
\end{align*}
so $d(\bfa_j', C) = (1\pm\eps)d(\bfa_j,C)$. Then by Lemma \ref{lem:preserve-clusters},
\begin{align*}
    \sum_{j\in P_i}\bfw_j d(\bfa_j, C)^p = (1\pm\eps)\sum_{j\in P_i}\bfw_j d(\bfa_j', C)^p = (1\pm\eps)^2\sum_{j\in P_i} d(\bfa_j', C)^p = (1\pm\eps)^3\sum_{j\in P_i} d(\bfa_j, C)^p.
\end{align*}
Thus, it suffices to consider centers with $d(\bfa_i, C)^p \leq \alpha\OPT \coloneqq 3^p\gamma\OPT/\eps^p$. 

Let $C^*$ be an optimal solution achieving $\OPT$, and consider a ball $\mathcal B_l$ of radius $(3\gamma^{1/p}/\eps+1)\OPT^{1/p}$ centered at each of the $k$ centers $l\in[k]$. Note that every point $\bfa_i$ is within a distance of $\OPT^{1/p}$ of one of the centers, so if a $k$-tuple of centers $C$ has any point outside of these $k$ balls, then by the triangle inequality, that point alone already has a cost of at least $\alpha\OPT$ and thus we need not consider it.

We now consider an $\eps'$-net $\mathcal N_l$ for each $\mathcal B_l$, for $\eps' = \OPT^{1/p}/\poly(n)$. Note that each net has size at most $\poly(\alpha n)^d$. We then set $\mathcal N$ to be the $k$ tuples of the union of these $k$ nets. 

Now let $C$ be any set of $k$ centers satisfying within a cost of $\alpha\OPT$. Then, as argued previously, each point lies in one of the $\mathcal B_l$, so we can choose a $C'\in \mathcal N$ such that for each point in $C$, there is a point in $C'$ at a distance of at most $\OPT^{1/p} / \poly(n)$. Then by the relaxed triangle inequality,
\[
    d(\bfa_i, C)^p = (1\pm\eps) d(\bfa_i, C')^p \pm \parens*{\frac{1+\eps}{\eps}}^{p-1}\frac{\OPT}{\poly(n)} = (1\pm\eps)d(\bfa_i, C')^p \pm \frac{\eps}{\poly(n)}\OPT
\]
so summing over $i\in[n]$ gives
\[
    \sum_{i=1}^n d(\bfa_i, C)^p = (1\pm\eps)\sum_{i=1}^n d(\bfa_i, C')^p \pm \eps\OPT
\]
and
\[
    \sum_{i=1}^n \bfw_i d(\bfa_i, C)^p = (1\pm\eps)\sum_{i=1}^n \bfw_i d(\bfa_i, C')^p \pm \eps\OPT
\]
since $\bfw_i$ is bounded by $\poly(n)$. Note also that
\[
    \sum_{i=1}^n \bfw_i d(\bfa_i, C')^p \geq (1-\eps)\sum_{i=1}^n d(\bfa_i, C')^p \geq (1-\eps)\OPT
\]
so
\[
    \sum_{i=1}^n \bfw_i d(\bfa_i, C)^p \geq \sum_{i=1}^n \bfw_i d(\bfa_i, C')^p - \eps\OPT \geq (1-2\eps)\OPT
\]
Then,
\begin{align*}
    \sum_{i=1}^n d(\bfa_i, C)^p &= (1\pm\eps)\sum_{i=1}^n d(\bfa_i, C')^p \pm \eps\OPT \\
    &= (1\pm2\eps)\sum_{i=1}^n d(\bfa_i, C')^p \\
    &= (1\pm2\eps)(1\pm\eps)\sum_{i=1}^n \bfw_i d(\bfa_i, C')^p \\
    &= (1\pm2\eps)(1\pm\eps)\bracks*{\sum_{i=1}^n \bfw_i d(\bfa_i, C)^p \pm \eps\OPT} \\
    &= (1\pm2\eps)(1\pm\eps)\parens*{1\pm\frac{\eps}{1-2\eps}}\sum_{i=1}^n \bfw_i d(\bfa_i, C)^p \\
    &= (1\pm10\eps)\sum_{i=1}^n \bfw_i d(\bfa_i, C)^p
\end{align*}
for $\eps$ sufficiently small.
\end{proof}

With the net argument in hand, we seek to prove \eqref{eq:net-guarantee}. We will need the following lemma of \cite{HV2020}, which fixes a claim in \cite[Theorem 15.5]{FL2011}. 
\begin{Lemma}[Lemma 8.1, \cite{HV2020}]\label{lem:hv2020-lem8}
If
\[
    \abs*{d(\bfa_i,C)^p - d(\bfa_i',C)^p} \geq \frac{p\norm*{\bfa_i-\bfa_i'}_2^p}{\eps^{p-1}},
\]
then
\[
    \abs*{d(\bfa_i,C)^p - d(\bfa_i',C)^p} \leq p\eps\cdot \max\braces*{d(\bfa_i,C)^p, d(\bfa_i',C)^p}
\]
so $d(\bfa_i, C) = (1\pm\eps)d(\bfa_i', C)$. 
\end{Lemma}

Now, the idea of \cite[Theorem 15.5]{FL2011} is that since it is easy to preserve the costs $d(\bfa_i', C)$ using, for example, Lemma \ref{lem:preserve-clusters}, it now suffices to preserve the \emph{difference} between $d(\bfa_i, C)^p$ and $d(\bfa_i', C)$. This will either be bounded by $\norm*{\bfa_i-\bfa_i'}_2^p / \eps^{p-1}$ or be larger than it; in the former case, we have a good bound for use in a Bernstein bound, while in the latter, $d(\bfa_i, C)^p$ must be close to $d(\bfa_i', C)^p$ by Lemma \ref{lem:hv2020-lem8}.

\begin{Lemma}[Bernstein Bounds]\label{lem:bernstein-cluster}
Fix $C\subseteq\mathbb R^d$ with $\abs*{C}\leq k$. Let $\bfw_i$ be sampled as in \eqref{eq:cluster-sample-def}. Then, with probability at least $1 - 2\exp(-\beta_1/2^{p+2}) - 2\abs*{B}\exp(-\beta_2)$, we have that
\[
    \abs*{\sum_{i=1}^n(\bfw_i-1)d(\bfa_i, C)^p} \leq O(\gamma \eps)\sum_{i=1}^n d(\bfa_i, C)^p
\]
\end{Lemma}
\begin{proof}
We partition the rows $[n]$ into two parts, $G$ and $B = [n]\setminus G$, where
\[
    G = \braces*{i\in[n] : \abs*{d(\bfa_i, C)^p - d(\bfa_i', C)^p} \leq \frac{\norm*{\bfa_i-\bfa_i'}_2^p}{\eps^{p-1}}}.
\]
For indices in $G$, we use Bernstein bounds to bound
\[
    \abs*{\sum_{i\in G}(\bfw_i - 1)(d(\bfa_i, C)^p - d(\bfa_i', C)^p)}
\]
For $i\in G$, we have that
\[
    \bfw_i \abs*{d(\bfa_i, C)^p - d(\bfa_i', C)^p} \leq \bfw_i \frac{\norm*{\bfa_i-\bfa_i'}_2^p}{\eps^{p-1}} \leq \frac1{\beta_1} \eps^2 \sum_{j=1}^n d(\bfa_j, B)^p \leq \frac1{\beta_1}\gamma\eps^2\sum_{j=1}^n d(\bfa_j, C)^p
\]
and thus the variance is bounded by
\begin{align*}
    \sum_{i\in G}p_i \cdot \parens*{\bfw_i \abs*{d(\bfa_i, C)^p - d(\bfa_i', C)^p}}^2 &\leq \frac1{\beta_1}\eps^2\parens*{\sum_{i\in G}\abs*{d(\bfa_i, C)^p - d(\bfa_i', C)^p}}\parens*{\sum_{j=1}^n d(\bfa_j, B)^p} \\
    &\leq \frac1{\beta_1}\eps^2\parens*{\sum_{i\in G}(2^{p-1}+1)d(\bfa_i, C)^p + 2^{p-1}\norm*{\bfa_i - \bfa_i'}_2^p}\parens*{\sum_{j=1}^n d(\bfa_j, B)^p} \\
    &\leq \frac{2^p}{\beta_1}\eps^2\gamma^2\parens*{\sum_{j=1}^n d(\bfa_j, C)^p}^2
\end{align*}
Then Bernstein bounds give that
\begin{align*}
    &\Pr\braces*{\abs*{\sum_{i\in G}(\bfw_i-1) (d(\bfa_i, C)^p - d(\bfa_i', C)^p)} \geq \gamma\eps \sum_{j=1}^n d(\bfa_i, C)^p} \\
    \leq~&2\exp\parens*{-\frac12\frac{\gamma^2\eps^2\parens*{\sum_{j=1}^n d(\bfa_i, C)^p}^2}{2^p\eps^2\gamma\parens*{\sum_{j=1}^n d(\bfa_j, C)^p}^2 + \eps^3\gamma^2\parens*{\sum_{j=1}^n d(\bfa_j, C)^p}^2/3}\beta_1} \leq 2\exp\parens*{-\frac{\beta_1}{2^{p+2}}}.
\end{align*}
Now condition on the complement of this event, as well as the event of Lemma \ref{lem:preserve-clusters}. Then,
\begin{align*}
    \abs*{\sum_{i\in G}(\bfw_i-1)d(\bfa_i, C)^p} &= \abs*{\sum_{i\in G}(\bfw_i-1)d(\bfa_i, C)^p - (\bfw_i-1)d(\bfa_i', C)^p + (\bfw_i-1)d(\bfa_i', C)^p} \\
    &\leq \abs*{\sum_{i\in G}(\bfw_i-1)d(\bfa_i, C)^p - (\bfw_i-1)d(\bfa_i', C)^p} + \abs*{\sum_{i\in G}(\bfw_i-1)d(\bfa_i', C)^p} \\
    &= \abs*{\sum_{i\in G}(\bfw_i-1)(d(\bfa_i, C)^p - d(\bfa_i', C)^p)} + \abs*{\sum_{i\in G}(\bfw_i-1)d(\bfa_i', C)^p} \\
    &\leq \gamma\eps \sum_{j=1}^n d(\bfa_i, C)^p + \eps \sum_{i\in G}d(\bfa_i', C)^p \\
    &\leq \gamma\eps \sum_{j=1}^n d(\bfa_i, C)^p + 2^{p-1}\eps \sum_{i\in G}d(\bfa_i, C)^p + \norm*{\bfa_i-\bfa_i'}_2^p \\
    &\leq (\gamma + (\gamma+1) 2^{p-1})\eps \sum_{j=1}^n d(\bfa_i, C)^p \\
    &\leq (2^p+1)\gamma \eps\sum_{j=1}^n d(\bfa_i, C)^p. 
\end{align*}

For indices in $B$, we have that
\begin{align*}
    \sum_{i\in B}\bfw_i d(\bfa_i, C)^p &= (1\pm\eps)\sum_{i\in B}\bfw_i d(\bfa_i', C)^p && \text{Lemma \ref{lem:hv2020-lem8}} \\
    &= (1\pm\eps)^2\sum_{i\in B}d(\bfa_i', C)^p && \text{Lemma \ref{lem:preserve-clusters}} \\
    &= (1\pm\eps)^3\sum_{i\in B}d(\bfa_i, C)^p && \text{Lemma \ref{lem:hv2020-lem8}}.
\end{align*}
Altogether, we conclude that
\[
    \abs*{\sum_{i=1}^n(\bfw_i-1)d(\bfa_i, C)^p} \leq O(\gamma \eps)\sum_{i=1}^n d(\bfa_i, C)^p.\qedhere
\]
\end{proof}

Finally, we are ready to put all the pieces together.

\begin{Theorem}\label{thm:fl-coreset}
Let $\bfw_i$ be drawn as in \eqref{eq:cluster-sample-def}, with
\begin{align*}
    \beta_1 &= O\parens*{dk\log n + \log\frac1\delta} \\
    \beta_2 &= O\parens*{\log\abs*{B} + \log\frac1\delta} \\
\end{align*}
Then, for all $C\subseteq\mathbb R^d$ with $\abs*{C}\leq k$, we have that
\[
    \abs*{\sum_{i=1}^n(\bfw_i-1)d(\bfa_i, C)^p} \leq O(\gamma \eps)\sum_{i=1}^n d(\bfa_i, C)^p.
\]
\end{Theorem}
\begin{proof}
For the given choice of $\beta$, by Lemma \ref{lem:bernstein-cluster} with, we can union bound over a net of size $\poly(n)^{dk}$ from Lemma \ref{lem:net-cluster}, with probability $1-\delta$. The guarantee of the net argument then gives the lemma.
\end{proof}

In particular, we achieve a coreset of size
\[
    O\parens*{\frac1{\eps^{p+1}}\parens*{dk + \log\frac1\delta} + \frac{\abs*{B}}{\eps^2}\log\abs*{B}}
\]
by summing over $i\in[n]$ in \eqref{eq:cluster-sample-def}. For constant $\delta$, $\abs*{B} = O(k)$, and $d = \log(k/\eps)/\eps^2$ using a terminal embedding after a first $\poly(k/\eps)$-sized coreset, this is size
\[
    \tilde O\parens*{\frac{k}{\eps^{p+3}}}.
\]

\section{Online Euclidean \texorpdfstring{$(k,p)$}{(k,p)}-Clustering}\label{sec:lss}

We show that online $k$-means clustering algorithm of \cite{LSS2016} immediately generalizes to Euclidean $(k,p)$ clustering. 

\LSS*

The algorithm and analysis of \cite{LSS2016} for the case of $p = 2$ generalizes with little obstructions. We work out the details below.

\begin{algorithm}
\caption{Online Euclidean $(k,p)$-Clustering \cite{LSS2016}}
\textbf{input:} $\bfA\in\mathbb R^{n\times d}$, number of clusters $k$, cost lower bound $w^*$. \\
\textbf{output:} Online clusters.
\begin{algorithmic}[1] 
    \State $C\gets \varnothing$
    \State $r\gets 1$; $q_1 \gets 0$; $f_1\gets w^* / k\log n$
    \For{$i\in[n]$}
        \State $p_i\gets \min(1, d(\bfa_i, C)^p / f_r)$\label{line:open-cluster}
        \State $C\gets C\cup\{\bfa_i\}$ and $q_r \gets q_r + 1$ with probability $p_i$
        \If{$q_r \geq 3k(1+\log_2 n)$}
            \State $r\gets r+1$; $q_r\gets 0$; $f_r\gets 2\cdot f_{r-1}$
        \EndIf
        \State Assign $\bfa_i$ to the closest center in $C$
    \EndFor
\end{algorithmic}\label{alg:online-k-clustering}
\end{algorithm}

Let $S_1^*, \dots, S_k^*\subseteq[n]$ be an optimal partition with centers $\bfc_1^*, \dots, \bfc_k^*$, and let
\[
    W_i^* = \sum_{j\in S_i^*} \norm*{\bfa_j - \bfc_i}_2^p, \qquad W^* = \sum_{i=1}^k W_i^*
\]
be the costs of the $i$th cluster and all the clusters, respectively. Let $A_i^* = W_i^* / \abs*{S_i^*}$ be the average cost of a vector in the $i$th cluster. Also define rings
\begin{align*}
    S_{i,0}^* &= \braces*{j\in S_i^* : \norm*{\bfa_j - \bfc_i^*}_2^p \leq A_i^*} \\
    S_{i,\tau}^* &= \braces*{j\in S_i^* : \norm*{\bfa_j - \bfc_i^*}_2^p \in \left(2^{\tau-1}A_i^*, 2^\tau A_i^*\right]} && 1 \leq \tau \leq \log_2 n
\end{align*}

\begin{Lemma}[Bound on Number of Clusters]\label{lem:bound-num-clusters}
The expected number of clusters formed by Algorithm \ref{alg:online-k-clustering} is at most
\[
    \E[\abs*{C}] = O\parens*{k(\log n)\log\frac{W^*}{w^*}}
\]
\end{Lemma}
\begin{proof}
The proof is nearly identical to Theorem 1 of \cite{LSS2016}. Let $r'$ be the first round $r$ such that
\[
    f_{r'} \geq \frac{W^*}{k\log_2 n}
\]
There are at most $\log\frac{f_{r'}}{f_1}$ rounds before $r'$, so the number of clusters opened before round $r'$ is at most $O\parens*{k(\log n)\log\frac{W^*}{w^*}}$. It suffices to bound the number of clusters opened after round $r'$. 

Fix a ring $\tau$, and define $S_{i,\tau,r}^*$ to be the points in $S_{i,\tau}^*$ encountered in round $r$. The first point $\ell$ from $S_{i,\tau}^*$ chosen to be a center contributes $1$ towards the number of clusters opened, while the rest open
\[
    \sum_{r\geq r'}\frac{2^{p}\cdot 2^\tau A_i^*}{f_r}\abs*{S_{i,\tau,r}^*}
\]
in expectation, since the probability $p_j$ of opening a new cluster in Line \ref{line:open-cluster} is bounded by
\[
    \frac{d(\bfa_j,C)^p}{f_r} \leq \frac{d(\bfa_j,\bfa_\ell)^p}{f_r} \leq \frac{2^{p-1}\cdot(\norm*{\bfa_j - \bfc_i^*}_2^p + \norm*{\bfc_i^* - \bfa_\ell}_2^p)}{f_r} \leq \frac{2^{p-1}\cdot 2\cdot 2^\tau A_i^*}{f_r} = \frac{2^p 2^\tau A_i^*}{f_r}.
\]
Then, summing over $\tau\geq 0$ gives
\begin{align*}
    \sum_{\tau\geq 0}\parens*{1 + \sum_{r\geq r'}\frac{2^{p}\cdot 2^\tau A_i^*}{f_r}\abs*{S_{i,\tau,r}^*}} &\leq O(\log n) + \sum_{\tau\geq 0}\sum_{r\geq r'}\frac{2^{p}\cdot 2^\tau A_i^*}{f_{r'}}\abs*{S_{i,\tau,r}^*} \\
    &\leq 1 + \log_2 n + \frac{2^{p+1}}{f_{r'}}\sum_{\tau\geq 0}2^{\tau-1} A_i^*\abs*{S_{i,\tau}^*} \\
    &\leq 1 + \log_2 n + \frac{2^{p+1}}{f_{r'}}\parens*{A_i^*\abs*{S_i^*} + \sum_{\tau\geq 1}2^{\tau-1} A_i^*\abs*{S_{i,\tau}^*}} \\
    &\leq 1 + \log_2 n + \frac{2^{p+1}}{f_{r'}}\parens*{W_i^* + W_i^*} \\
    &\leq 1 + \log_2 n + 2^{p+2}k(\log_2 n)\frac{W_i^*}{W^*}.
\end{align*}
Then, summing over $i\in[k]$ and using that $\sum_{i=1}^k W_i^* = W^*$ then gives a bound of $(2^{p+2} + 1)k(1+\log_2 n)$. 
\end{proof}

\begin{Lemma}[Lemma 1, \cite{LSS2016}]\label{lem:seq-exp}
Let $p_i \geq \min\{A_i/B, 1\}$ be probabilities, for $A_i \geq 0$ and $B\geq 0$. Let $t$ be the number of sequential unsuccessful experiments. Then,
\[
    \E\bracks*{\sum_{i=1}^t A_i} \leq B.
\]
\end{Lemma}

\begin{Lemma}\label{lem:expected-fR}
Let $R$ denote the random variable representing the total number of rounds. Then,
\[
    \E[f_R] = O\parens*{\frac{W^*}{k\log n}}
\]
\end{Lemma}
\begin{proof}
We now estimate $\E[f_R]$. Let $r''$ be the first phase $r$ such that
\[
    f_{r''} \geq \frac{2^{p+4} W^*}{k(1+\log_2 n)}.
\]
Following the proof of Lemma \ref{lem:bound-num-clusters}, at most 
\[
    k(1+\log_2 n) + \frac{2^{p+2}}{f_{r''}} \leq \frac54 k(1 + \log_2 n)
\]
clusters are opened after round $r'$ in expectation. By Markov's inequality, the probability of opening more than $3k(1+\log_2 n)$ clusters is at most $4/9$. Then with probability at least $5/9$, the algorithm concludes while at round $r''$. Now let $q$ the probability that the algorithm terminates before round $r''$. We then have
\begin{align*}
    \E[f_R] &\leq q\cdot f_{r''-1} + (1-q)\sum_{r \geq r''} f_r\cdot \frac{5}{9} \cdot \parens*{\frac49}^{r - r''} \\
    &\leq f_{r''} + f_{r''}\sum_{i\geq 0}2^i\parens*{\frac49}^i = O(f_{r''})
\end{align*}
as claimed.
\end{proof}

\begin{Lemma}[Bound on Expected Cost]
The cost of the clustering in Algorithm \ref{alg:online-k-clustering} is at most $O(W^*)$ in expectation.
\end{Lemma}
\begin{proof}
The proof is nearly identical to Theorem 2 of \cite{LSS2016}. 

Fix a cluster $S_i^*$ and a ring $\tau$. We first consider the cost of points in $S_{i,\tau}^*$ before the first point from $S_{i,\tau}^*$ was chosen as a center. Note that each $j\in S_{i,\tau}^*$ is chosen with probability at least $p_j \geq \min\{d(\bfa_j, C)^p / f_R, 1\}$, where $C$ is the set of centers chosen by time $j$. Then by Lemma \ref{lem:seq-exp}, the sum of these costs is bounded by $f_R$, and summing over all $i$ and $\tau$ gives a bound of $O(f_R k\log n)$. By Lemma \ref{lem:expected-fR}, this is $O(W^*)$. 

We next consider the cost of points in $S_{i,\tau}^*$ after the first point from $S_{i,\tau}^*$ was chosen as a center. Then as shown in the proof of Lemma \ref{lem:bound-num-clusters}, the cost of these points is at most $O(W^*)$.
\end{proof}

\end{document}